\documentclass[review,12pt]{elsarticle}
\usepackage[margin=1in,footskip=0.25in]{geometry}
\usepackage{times}
\usepackage{amssymb}
\usepackage{graphicx}
\usepackage{amsmath,amsthm,mathtools}
\usepackage[version=4]{mhchem}
\usepackage{siunitx}
\usepackage{float,subcaption}
\usepackage{longtable,tabularx}
\usepackage{multirow}
\usepackage{xcolor}
\usepackage{hyperref}
\hypersetup{colorlinks, breaklinks, citecolor=blue, linkcolor=blue, urlcolor=blue}
\usepackage[nameinlink,noabbrev,sort&compress]{cleveref}
\Crefname{equation}{Eq.}{Eqs.}
\Crefname{figure}{Fig.}{Figs.}
\crefname{figure}{Figure}{Figures}
\newtheorem{theorem}{Theorem}

\newtheorem{problem}{Problem}

\newtheorem{assumption}{Assumption}

\newcommand{\diag}{\mathrm{diag}}

\biboptions{numbers,sort&compress}

\makeatletter
\def\ps@pprintTitle{%
	\let\@oddhead\@empty
	\let\@evenhead\@empty
	\let\@oddfoot\@empty
	\let\@evenfoot\@empty
}
\makeatother
\begin{document}
	
	\begin{frontmatter}

		\title{Trajectory tracking control of USV with actuator constraints in the presence of disturbances} %% Article title
		
		\author[rmv]{Ram Milan Kumar Verma\corref{cor1}} %% Author name
		\cortext[cor1]{Corresponding author}
		\ead{rmverma@aero.iitb.ac.in}
		\author[rmv]{Shashi Ranjan Kumar}
		\ead{srk@aero.iitb.ac.in}
		\author[rmv]{Hemendra Arya}
		\ead{arya@aero.iitb.ac.in}
		%% Author affiliation
		\affiliation[rmv]{organization={Indian Institute of Technology Bombay},%Department and Organization
			addressline={Powai}, 
			city={Mumbai},
			postcode={400076}, 
			state={Maharashtra},
			country={India}}
		
		%% Abstract
		\begin{abstract}
			All practical systems often pose a problem of finite control capability, which can notably degrade the performance if not properly addressed. 
			Since actuator input bounds are typically known, integrating actuator saturation considerations into the control law design process can lead to enhanced performance and more precise trajectory tracking. Also, the actuators cannot provide the demanded forces or torques instantaneously; hence, there is a limitation on the rate of magnitude. 
			This work proposes nonlinear feedback controller designs developed using the Lyapunov stability and backstepping method while actively considering the actuator magnitude and rate constraints.
			The system dynamics are augmented with a smooth control input saturation model. 
			Additionally, an observer is incorporated to estimate the disturbance vector. Through Lyapunov stability analysis, we demonstrate the system's stability under the proposed controller for the Uncrewed Surface Vessel (USV), ensuring adherence to actuator constraints provided their initial values fall within the prescribed bounds. Extensive numerical simulations performed by considering various trajectories and multiple initial conditions demonstrate the effectiveness of the controller in maintaining tracking performance without violating actuator constraints. This work also relaxes the assumption of equally capable actuators to be used to control the motion of USVs, affirming the viability of the controller in practical applications.
		\end{abstract}

		
	\end{frontmatter}
	\section{Introduction}\label{sec:introduction}
	The rapid growth of emerging marine applications calls for advanced solutions capable of working in real-world scenarios. 
	Surface Vessels are being extensively utilised in many such applications, ranging from scientific exploration to defense and industrial monitoring. 
	The surface vessels can be used to autonomously navigate to the offshore oil and gas mining stations, dock at the stations, and come back. 
	The other application could be to act as a support ship to other underwater vehicles, which may involve coordination between multiple vehicles to achieve the mission objectives.
	A more detailed overview of key application areas and research projects has been presented in \cite{zereik2018challenges}. 
	Most of the applications require motion control algorithms such that the vehicle can be made to follow a designated path or trajectory.
	Uncrewed surface vessels (USVs) are equipped with thrusters and control fins that can change the speed and heading of the vehicle to follow arbitrary desired paths or trajectories. Application of trajectory tracking for USVs is an active area of exploration, particularly when addressing the challenges posed by the actuator input saturation and unknown disturbances \cite{yin2024fixed,mu2019adaptive,chen2020}.
	
	As the world is sailing towards achieving greater autonomy, the idea of endowing these vehicles with autonomous capabilities to perform mission objectives, such as trajectory tracking and path following, is widely regarded \cite{liu2020robust}. 
	However, developing stable controllers for these vehicles to perform these tasks accurately is fraught with several challenges. 
	All practical systems operate under bounded control inputs due to the physical limits imposed by the actuators driving the system. Hence, addressing the problem of actuator saturation while designing stable controllers is of paramount importance; if not dealt with adequately, it may lead to performance degradation to an unacceptable limit or even cause the system's instability \cite{xia2021solution}. 
	The nonlinear and coupled dynamics also add to the difficulty in designing suitable and stable controllers.
	% The other challenges associated with USVs are that the dynamics are highly nonlinear, with multiple degrees of freedom (DOF) coupled, making it difficult to design suitable and stable controllers.
	Apart from these, commonly, the USVs are underactuated; that is, there are fewer independent control actuators than degrees of freedom. Hence, it cannot generate linear and angular accelerations in all directions, leading to limited motion capabilities.
	Additionally, model parameters are not known precisely. The unmodeled dynamics and external disturbances, such as ocean currents, wind, and wave forces, introduce significant challenges to achieving accurate and reliable control. Addressing these issues is essential for the development of practical and safe autonomous marine systems.
	
	To tackle these challenges, various approaches based on different techniques were developed. In \cite{zhao2014, he2017, WEI2019}, the authors used a neural network-based approach to deal with parametric uncertainties for fully-actuated surface vessels. In \cite{godhavn1996, LI2007, sun2018robust}, the authors developed a robust nonlinear trajectory tracking method for underactuated marine surface vessels.  The authors in \cite{Neto2021_paramEstim, mu2019adaptive} presented methods for identifying the dynamics and uncertainties due to unmodelled dynamics and external disturbances. The authors in \cite{yang2014_dyamics, chen2020adaptive, 8836553} designed a disturbance observer to estimate the unknown disturbances. 
	
	The challenges pertaining to the modeling uncertainties and nonlinearity have been dealt with adequately in the literature, but the incorporation of actuator input saturation at the controller design stage itself, requires further attention. The trajectory tracking and path-following problem with input saturation has been investigated in \cite{QIN2020, wang2019, ghommam2010global, do2010practical, li2021, dong2023eventtrig, ZHENG2016158}. There are many methods in the literature to handle actuator input saturation. Among them, the common method is to directly limit the actuator input; that is, when the control demand exceeds the bounds, the maximum or minimum limit is applied. This approach loses stability guarantee as well as poor tracking \cite{liu2020robust}. A hyperbolic tangent function is also used to incorporate the actuator constraint, which maps the input to a bounded range dictated by actuator bounds \cite{zheng2018, QIN2020}. 
	% The authors in \cite{reyhanoglu1997exponential} studied exponentially stable feedback law for set-point control of the underactuated surface vessels.
	% The authors in \cite{ZHENG2016158} investigated the path following of USVs with the input saturation. 
	The authors in \cite{sun2018robust} proposed an adaptive trajectory tracking algorithm using proportional-integral (PI) sliding mode control and the backstepping method. Many other trajectory tracking approaches used in the literature included artificial potential field-based, moving horizon optimal control, reference governor, and model predictive control to respect the actuator constraints imposed. These works were briefly summarized in \cite{zhao2014}. 
	The other methods used to represent the input saturation were the dead zone-based method \cite{mousavi2016dead} and the Gaussian error function \cite{ma2014adaptive, liu2020robust, zhu2020single}. 
	Using these to represent input saturation non-linearity results in a model that is non-affine in the model input, complicating the direct development of the control law. 
	% gong2023fixed
	A simple barrier Lyapunov function-based approach is also used to deal with constraints. The works in \cite{ngo2004, ngo2005, zhao2014, he2017, swati2024_blf, kpt2009_time_varying} implemented the barrier Lyapunov function-based controller design to incorporate the constraints on states of the system. In \cite{kpt2009}, the authors proposed a control design technique for a general single-input, single-output (SISO) nonlinear system in the strict feedback form to constrain the output states within their bounds. However, this approach is commonly employed when constraints are applied to the states of the system. 
	% In \cite{QIN2020}, the authors proposed finite-time trajectory control with state constraints in the presence of input saturation. 
	Apart from actuator magnitude constraints, it is also critical to take into account bounds on their rate. It is impractical to expect the actuators to respond fast and supply the demanded force or torque instantaneously due to their limited bandwidth \cite{xia2021solution}.
	The authors in \cite{wang2019,gong2023fixed,yin2024fixed} used an auxiliary system to address input saturation as well as their rate. A two-nested saturation system is used in \cite{xia2021solution}
	to model the actuator with magnitude limitation and rate limitation. The major difficulty with these approaches is the complicated controller design due to the resulting model being non-affine in the model input. In this work, we present an elegant way to address the actuator magnitude and its rate constraints. Therefore, in light of the above discussions, the contributions of this work are summarized next.
	% The authors in \cite{QIN2020, mu2018tracking} addressed the trajectory tracking problem under actuator saturation and disturbance uncertainties. 
	
	% \textcolor{red}{Write about the limitations of these techniques in brief.}
	
	% Most of the above-discussed works designed the controllers without considering the bounds on the surface vessel or using hyperbolic tan functions to approximate the input saturation, which is non-affine in model input, making the control development more involved.
	
	This work proposes a nonlinear closed-loop controller for a USV, based on the Lyapunov function and backstepping technique, enabling precise trajectory tracking while respecting the constraints posed due to input saturation bounds. 
	To take actuator constraints into consideration, we employ a smooth input saturation model which results in a dynamical system with affine in model input, unlike in \cite{ZHENG2016158, mu2019adaptive, zou2012neural}. 
	This proposed scheme significantly simplifies the controller design process while also ensuring the stability of the system under the designed control law.
	The application of saturation bounds in an ad-hoc manner may lead to poor tracking performance, and at the same time, there is no mathematical guarantee for stability. Additionally, the proposed model generates smooth control demand, which helps in preventing damage to the actuators, unlike the ad-hoc implementation that has corner points leading to the development of jerk and is harmful to the actuators, decreasing their lifespan and hence increasing the maintenance cost.
	
	In this design, we incorporate the asymmetric input magnitude saturation constraints. In real-life scenarios, the wear and tear in the actuators mounted on a system could be at different levels, and that can cause different capacity availability from each of them. Hence, it is important to deal with asymmetric actuator bounds.  Also, this allows us to use actuators with different capacities in different directions, helping reduce the cost. In addition to this, the proposed controller also accommodates the constraint on the rate of input magnitude, which is the case in real-life scenarios, as the bandwidth of the actuators is limited. These considerations make the proposed law more suitable for real-life applications.
	
	Besides these, the proposed scheme uses an observer to estimate the unknown disturbances and compensates for their adverse effect on the system performance by incorporating them.
	% Owing to the nonlinear framework used for deriving control inputs, it remains applicable even in situations where the deviations of system states are significantly large, and the control methods based on the linearization technique might fail. 
	
	The paper is organized as follows: The next section presents the problem formulation and overviews the vehicle dynamics. In Section~\ref{sec:main results}, we derive the necessary controller for the USV and present the relevant stability proof. Section~\ref{sec:simulation results} showcases the simulation results that validate the efficacy of the proposed control design. Finally, section~\ref{sec:conclusion} concludes the paper and indicates potential future research directions.
	\section{Problem Formulation}\label{sec: Problem}
	This section begins by describing the kinematics and dynamics of the USV and subsequently formulates the main problem addressed in this paper. 
	In general, the marine surface vessel experiences motion in 6-DOF, namely surge, sway, heave, roll, pitch, and yaw. 
	% The dynamics are generally described as a 6-degrees of freedom (DOF) equations of motion, manifesting as ordinary differential equations. 
	% These DOFs are referred to as the surge, sway, and heave for the position in three-dimensional space and the roll, pitch, and yaw to describe the vehicle's attitude. 
	Here, we assume that the motion of the USV is constrained to a plane, that is, the vessel can only translate on the water surface and rotate about the axis perpendicular to the surface. This is a reasonable assumption because the USVs are metacentrically stable with small amplitudes of roll, pitch angles, and their rates. Hence, the roll and pitch dynamics can be safely neglected for all practical applications. Similarly, the dynamics in the heave direction can also be ignored as the vessel is floating on the water surface. Thus, it is adequate to represent the dynamics of the USV as a simplified 3-DOF system. 
	\begin{figure}[ht]
		\centering
		\includegraphics[width=.5\linewidth]{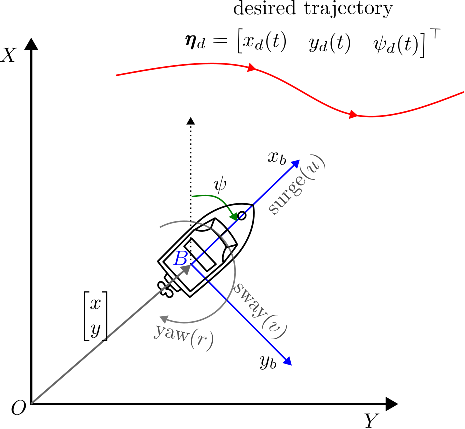}
		\caption{Two dimensional motion of the USV.} 
		\label{fig:ref_frame}
	\end{figure}
	For this purpose, we consider two frames of reference, namely, the Earth-fixed frame $(OXY)$ and the body-fixed frame $(Bx_{b}y_{b})$, to represent the equations of motion of the USV. As shown in \Cref{fig:ref_frame}, the points $O$ and $B$ denote the origin of the respective coordinate frames. The axes $OX$ and $OY$ are towards the North and the East, respectively. The vector $Bx_b$ is pointing from the aft to the fore, and $By_b$ is towards the starboard side. The rotation matrix,  $J$, that relates these two frames is given by
	\begin{align}
		\pmb{J}(\psi) = \begin{bmatrix}
			\cos\psi & -\sin \psi &0\\
			\sin\psi & \cos\psi & 0 \\
			0 & 0 & 1
		\end{bmatrix}.
	\end{align}
	% A vector represented in the body-fixed frame when premultiplied with $J(\psi)$ splits the components of the same vector in the Earth-fixed frame. 
	The vector $\pmb{\eta} = [x, y, \psi]^\top$ denotes the position $(x, y)$ and the heading $(\psi)$ of the USV in the Earth-fixed frame. The vector $\pmb{\nu} = [u, v, r ]^\top$ denotes the velocity vector components of the USV in the body-fixed frame. The variables $u$, $v$, and $r$ denote the linear velocity components in the surge and sway, and the yaw rate of the USV, respectively.  Now, by applying Newton's second law,  the 3-DOF nonlinear equations of motion can be expressed, \cite{ fossen2021handbook}, as
	\begin{align}\label{eqn: kinematics}
		\dot{\pmb{\eta}}  &= \pmb{J}(\psi) \pmb{\nu},
		\\
		\label{eqn: dynamics}
		\dot{\pmb{\nu}} &= \pmb{M}^{-1}\left[\pmb{\tau} - \pmb{C}(\pmb{\nu})\pmb{\nu} - \pmb{D}(\pmb{\nu})\pmb{\nu}  + \pmb{b}\right],
	\end{align}
	where $\pmb{\tau} = \left[
	\tau_1,\, \tau_2,\, \tau_3    
	\right]^\top=\left[
	\tau_x,\, \tau_y,\, \tau_\psi  
	\right]^\top$ is the control input in the body-fixed frame, $\pmb{M}\in \mathbb{R}^{3\times3}$ is a positive definite matrix, $\pmb{C}\in \mathbb{R}^{3\times3}$ is the Coriolis matrix, $\pmb{D}\in \mathbb{R}^{3\times3}$ is the damping matrix, and $\pmb{b} = \left[
	b_1(t) ,\, b_2(t) ,\, b_3(t)
	\right]^\top$ is the disturbance vector components in the body-fixed frame from the environment due to wind, waves, and ocean currents, which are time-dependent quantities. The matrices $\pmb{M}, \pmb{C}$ and $\pmb{D}$ are given by
	\begin{align}\label{eqn: M}
		\pmb{M} &= 
		\begin{bmatrix}
			m_{11} &0 &0\\
			0 &m_{22} &m_{23} \\
			0  &m_{32} &m_{33}
		\end{bmatrix}, \quad
		% \\
		% \label{eqn: C}
		\pmb{C}(\pmb{\nu}) = 
		\begin{bmatrix}
			0 &0 &-m_{22}\nu-m_{23}r\\
			0 &0 &m_{11}u \\
			m_{22}\nu+m_{23}r  &-m_{11}u &0
		\end{bmatrix},
		\\
		\label{eqn: D}
		\pmb{D}(\pmb{\nu}) &= 
		\begin{bmatrix}
			-X_u - X_{|u|u}|u| &0 &0\\
			0 &-Y_v - Y_{|v|v}|v| - Y_{|r|v}|r| &-Y_r - Y_{|v|r}|v| - Y_{|r|r}|r| \\
			0  &-N_v -N_{|v|v}|v| - N_{|r|v}|r| &-N_r - N_{|v|r}|v| - N_{|r|r}|r|
		\end{bmatrix}. 
	\end{align}
	The expressions for the elements of the matrices used in \Cref{eqn: M} are given by
	\begin{subequations}
		\begin{align}
			m_{11} &= \rm{m} - X_{\dot{u}},
			m_{22} = \rm{m} -  Y_{\dot{v}}, m_{33} = I_z -  N_{\dot{r}},\\
			m_{23} &= \rm{m}x_g -  Y_{\dot{r}},
			m_{32} = \rm{m}x_g -  N_{\dot{v}},
			% d_{11}(u) = -X_u - X_{|u|u}|u|,\\
			% d_{22}(v,r) &= -Y_v - Y_{|v|v}|v| - Y_{|r|v}|r|,\\
			% d_{23}(v,r) &= -Y_r - Y_{|v|r}|v| - Y_{|r|r}|r|,\\
			% d_{32}(v,r) &= -N_v -N_{|v|v}|v| - N_{|r|v}|r|,\\
			% d_{33}(v,r) &= -N_r - N_{|v|r}|v| - N_{|r|r}|r|,
		\end{align}  
	\end{subequations}
	where m is the mass of the surface vessel, $x_g$ is the distance of the geometric center of the surface vessel from the center of gravity (CG), and $I_z$  is the yaw moment of inertia of the surface vessel. The other symbols are related to hydrodynamic derivatives and have the usual meanings, as per the SNAME convention \cite{SNAME1950nomenclature}. 
	
	Let the desired trajectory be represented by $\pmb{\eta}_d =\left[
	x_d(t),\,  y_d(t),\,  \psi_d(t)
	\right]^\top,$
	and the desired surge, sway, and yaw rates are denoted by $\pmb{\nu_d}(t) = 
	\left[
	u_d(t),\, v_d(t) ,\,r_d(t)
	\right]^\top$. The desired body rates can be related to the inertial rates as
	$\pmb{\nu}_{d} = \pmb{J}(\pmb{\eta}_d)^\top \dot{\pmb{\eta}}_d$.
	% \textcolor{red}{So, $-\pmb{k}_a<\pmb{\eta} - \pmb{\eta}_d <\pmb{k}_a$; where $\pmb{k}_a$ is positive constant vectors. Note that vector inequalities are taken component-wise. IS THIS NECESSARY IN THIS PAPER?} 
	\begin{assumption}
		The system is fully actuated, and the system dynamics are known. The desired trajectory $\pmb{\eta_d}(t)$ is smooth and bounded, that is $ \underline{\pmb{A}}_0 \leq \pmb{\eta_d}(t) \leq \overline{\pmb{A}}_0$. Also, the first and second derivatives of $\pmb{\eta}_d$ are bounded for all time.
	\end{assumption}
	\begin{assumption}
		The unknown time-varying disturbance vector acting on the system is such that their rate of change is bounded by some positive value, $b_M$. Mathematically it can be described as $ ||\dot{\pmb{b}}(t)|| \leq b_M < \infty$
	\end{assumption}
	
	Given the assumptions above, we now state the problem statement addressed in this work. 
	% \textcolor{red}{Rephrase problem statement as we discussed}
	\begin{problem}
		The primary objective of this work is to develop a stable feedback controller for a surface vessel whose dynamics are characterized by \Cref{eqn: kinematics,eqn: dynamics}. The goal is to ensure that the vessel accurately follows a specified smooth trajectory, $\pmb{\eta}_d$ while adhering to the asymmetric input constraints dictated by the physical limitations of the actuators. Mathematically, the problem can be formulated as  
		$$ \pmb{\eta}(t) \rightarrow \pmb{\eta}_d(t) ~~\mathrm{as} ~~ t \rightarrow \infty  \quad \text{while maintaining} ~~ -\tau_{im}<{\tau_i} < \tau_{iM} ~(i \in \{1,2,3\}) ~ \forall t \geq 0,$$
		where $\tau_{im}$ and $\tau_{iM}$ denote the input bound for the actuator in positive and negative directions, respectively. 
		At the same time, it is also ensured that all closed-loop signals remain bounded.
	\end{problem}
	\begin{problem}
		The second objective of the work is to develop another stable feedback controller for trajectory tracking of a USV whose dynamics are governed by \Cref{eqn: kinematics,eqn: dynamics} while adhering to the bounds on actuators' input as well as their rates. Mathematically, the problem can be described as $\pmb{\eta}(t) \rightarrow \pmb{\eta}_d(t) ~~\mathrm{as} ~~ t \rightarrow \infty  \quad \text{while satisfying}$
		\begin{align*}
			-\tau_{iM}<{\tau_i} < \tau_{iM}, ~\rm{and}~-\tau_{idM}<\dot{\tau}_i < \tau_{idM}&  ~(i \in \{1,2,3\}) ~ \forall t \geq 0,
			% -\tau_{iM}<{\tau_i} < \tau_{iM},& \\
			% -\tau_{idM}<\dot{\tau}_i < \tau_{idM}&  ~(i\in \{1,2,3\}) ~ \forall t \geq 0,
		\end{align*} where $\tau_{iM}$ is the maximum available control input and $\tau_{idM}$ is the maximum rate of change of the control input available.
		
	\end{problem}
	The design of the proposed controller adheres to the nonlinear framework, eschewing any reliance on linearizations or approximations. This ensures the controller's efficacy even when there are substantial deviations from nominal trajectories.

	\section{Main Results}\label{sec:main results}
\begin{figure}[!ht]
		\centering
		\includegraphics[width=0.75\linewidth]{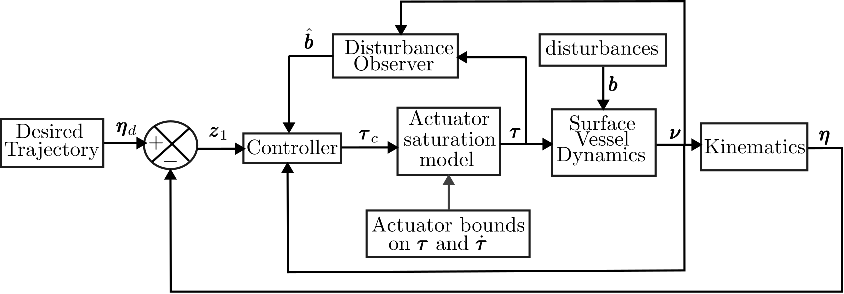}
		\caption{Schematic of trajectory tracking control of the USV under actuator constraints.}
		\label{fig:control_law}
	\end{figure}

	%\section{Preliminaries}\label{sec:preliminaries}
	This section presents the controller design to fulfill the objectives mentioned in the previous section. We provide a rigorous methodology to leverage a Lyapunov function and a backstepping-based approach \cite{100933}, both widely acknowledged in the literature for designing stable controllers. 
	% Before proceeding to the actual controller design process, we present foundational results crucial for deriving the required control commands. 
	% \begin{lemma}\cite{kpt2009}\label{lem1} \textcolor{red}{put diff lemma from KD kumar paper}
		%  For any positive constants $k_{a_1}, k_{b_1}$,
		%  let $Z_1 := \left\{z_1 \in \mathbb{R}: -k_{a_1}<z_1 <k_{b_1}\right\}\subset \mathbb{R}$ and $\mathbb{N} := \mathbb{R}^l \times \mathbb{Z}_1 \subset \mathbb{R}^{l+1}$ be open sets. Consider the system $$\dot \eta= h(t,\eta),$$ where $\eta := [w,~~z_1]^T\in \mathbb{N}$, and $h : \mathbb{R}_+ \times \mathbb{N} \to \mathbb{R}^{l+1}$ is piece-wise continuous in $t$ and locally Lipschitz in $\eta$, uniformly in $t$, on $\mathbb{R}_+ \times \mathbb{N}$. Suppose that there exist functions $U: \mathbb{R}^l \to \mathbb{R}_+$ and $\mathcal{V}_1: Z_1 \to \mathbb{R}_+$, continuously differentiable and positive definite in their respective domains, such that $$V_1(z_1)\to\infty~\text{as}~ z_1\to -k_{a_1}~\text{or}~z_1\to k_{b_1}, ~~ \gamma_1(||w||)\le \mathcal{U}(w)\le \gamma_2(||w||),$$ where $\gamma_1$ and $\gamma_2$ are class $K_{\infty}$ functions. Let $$\mathcal{V}(\eta) := \mathcal{V}_1(z_1) + \mathcal{U}(w),$$ and $z_1(0)$ belong to the set $z_1 \in (-k_{a_1}, k_{b_1})$. If the inequality holds:  $$\dot{\mathcal{V}} = \dfrac{\partial\mathcal{V}}{\partial \eta}h\le 0,$$ then $z_1(t)$ remains in the open set $z_1 \in (-k_{a_1}, k_{b_1})~\forall\,\,t \,\in [0,\infty)$.
		% \end{lemma}
	The closed-loop trajectory tracking control of the surface vessel consists of a disturbance observer, controller, and actuator input saturation model.  The interconnection and flow of information between different elements have been illustrated in the \Cref{fig:control_law}. The desired trajectory information comes from the mission. The control command is computed based on the deviation from the desired state, which then passes through an actuator saturation model before being applied to the vehicle.
	\subsection{Disturbance Observer}
	Before proceeding to the next section, we briefly describe the disturbance observer as presented in \cite{yang2014_dyamics, do2010practical}. The disturbance observer has been constructed using the surface vessel dynamics for estimating the disturbance term $\pmb{b(t)}$ in \Cref{eqn: dynamics}. 
	\begin{align}
		\label{eqn:b_hat}\hat{\pmb{b}} &= \pmb{z} + \pmb{K}_0 \pmb{M} \pmb{\nu}, \\
		\label{eqn:z_dot}\dot{\pmb{z}}  &= -\pmb{K}_0 \pmb{z} - \pmb{K}_0\left[\pmb{\tau} - \pmb{C}(\pmb{\nu})\pmb{\nu} - \pmb{D}(\pmb{\nu})\pmb{\nu}  + \pmb{K}_0\pmb{M}\pmb{\nu}\right],
	\end{align}
	where $\pmb{z}$ is an internal state variable and $\hat{\pmb{b}}$ is the estimate for the unknown, bounded, and slowly time-varying disturbance. The above observer has been designed using the Lyapunov function method and was shown to be exponentially stable in \cite{do2010practical,yang2014_dyamics}.  To design a nonlinear disturbance observer-based controller, a general procedure as described in \cite{chen2004disturbance}, is to first design a stable controller assuming that the disturbance is measurable and then replace the disturbance vector in the control law by the estimate provided by the designed disturbance observer. 
	Let us define the error in the estimate of the disturbance vector as $\pmb{b}_e = \pmb{b} - \hat {\pmb{b}}$, which on taking the time derivative and using \Cref{eqn:b_hat,eqn:z_dot}, with further manipulation, lead to
	\begin{align}
		\label{eqn:b_e_dot}\dot{\pmb{b}}_e = \dot{\pmb{b}}- \pmb{K}_0\pmb{b}_e.
	\end{align}
	Now, consider the Lyapunov function candidate $\mathcal{V}_{be} = \dfrac{1}{2}\pmb{b}_e^\top \pmb{b}_e$. Differentiating it with respect to time and substituting $\dot{\pmb{b}}_e$ from \Cref{eqn:b_e_dot} along with the use of the inequality  $|\pmb{b}_e^\top\dot{\pmb{b}}| \leq  \varepsilon_1 \pmb{b}_e^\top\pmb{b}_e + \dfrac{1}{4\varepsilon_1}\dot{\pmb{b}}^\top \dot{\pmb{b}},$
	% \begin{equation}\label{eqn:b_e_inequality}
		%     |\pmb{b}_e^\top\dot{\pmb{b}}| \leq  \varepsilon_1 \pmb{b}_e^\top\pmb{b}_e + \dfrac{1}{4\varepsilon_1}\dot{\pmb{b}}^\top \dot{\pmb{b}},
		% \end{equation}
	gives \begin{equation}
		\dot{\mathcal{V}}_{be} \leq -\pmb{b}_e^\top \pmb{K}_0 \pmb{b}_e + \varepsilon_1 \pmb{b}_e^\top\pmb{b}_e + \dfrac{1}{4\varepsilon_1}\dot{\pmb{b}}^\top \dot{\pmb{b}}
		\leq -2[\lambda_{\rm{min}}(\pmb{K}_0) -\varepsilon_1]\mathcal{V}_{be} + \dfrac{b_M^2}{4\varepsilon_1}\label{eqn:V_dot_be},
	\end{equation}
	where $\lambda_{\rm{min}}(\cdot)$ is the minimum eigenvalue and $\varepsilon_1 > 0$ such that $\lambda_{\rm{min}}(\pmb{K}_0) -\varepsilon_1 >0$. On solving \Cref{eqn:V_dot_be}, we find that $\mathcal{V}_{be}$ diminishes exponentially to a ball of radius $b_M^2/[8\varepsilon_1(\lambda_{\rm{min}}(\pmb{K}_0) -\varepsilon_1)]$. Hence, it can be concluded that $||\pmb{b}_e(t)||$ converges to ball of radius $\sqrt{b_M^2/[8\varepsilon_1(\lambda_{\rm{min}}(\pmb{K}_0) -\varepsilon_1)]}$ around origin exponentially. The size of the ball can be made arbitrarily small by carefully selecting the gains, $\pmb{K}_0$.\\
	Several other disturbance observers exist in the literature; many of which are described in \cite{chen2004disturbance}, but the focus of this work is to address the challenge of incorporating the actuator capability into controller design. Hence, now we are shifting our focus to the controller design, as stated in the problem statement. 
	\subsection{Input saturation model with asymmetric bounds}
	Most real-world systems suffer from actuator saturation as they may not always have ample capabilities. Also, the actuator capability in the positive direction and the negative direction might be different. For example, thrusters used in USVs generally have different forward thrust and reverse thrust capabilities. Hence, it is more favorable to account for asymmetric actuator saturation bounds in the design. Most of the work in the literature tries to address the symmetric actuator saturation bounds. It is also commonly assumed that all the thrusters used in the USVs are identical, meaning the same saturation bounds. But this may not be economical in practical applications. To increase safety and reduce cost, we face the challenge of using actuators of different capabilities for actuation in different directions. In \cite{10.1049/iet-cta.2016.1097}, the authors proposed a simple symmetric input saturation model with the same actuator bounds in all directions. However, taking inspiration from the same, in our work, we propose another actuator input saturation model, which is given by
	\begin{equation}\label{eqn: actuator}
		\dot{\pmb{\zeta}} = \pmb{Q}(\pmb{I} - \pmb{G}_M) \pmb{\tau}_c - \pmb{\rho} \pmb{\zeta} + (\pmb{I}-\pmb{Q})(\pmb{I} - \pmb{G}_m) \pmb{\tau}_c, \quad \pmb{\zeta}(0) = \pmb{0}, \quad \pmb{\tau} = \pmb{\zeta},
	\end{equation}
	where  $\pmb{\tau}_{c}$, $\pmb{\tau}$, and $\pmb{\zeta} \in \mathbb{R}^3$ are the input, the output, and the state of the actuator saturation model, respectively. The term 
	$\pmb{\rho} = \diag \begin{bmatrix}
		\rho_1 & \rho_2 & \rho_3
	\end{bmatrix} $ where
	$\rho_1, \rho_2, \rho_3 > 0 $ are the design constants. The matrices $\pmb{I},\pmb{G}_M, \pmb{G}_m \in \mathbb{R}^{3 \times 3} $, where $\pmb{I}$ is the identity matrix,  $\pmb{G}_M$ is a diagonal matrix with $\pmb{G}_M=\diag \begin{bmatrix}
		g_{1M} & g_{2M} & g_{3M}
	\end{bmatrix}$ where the terms $g_{iM} = \left(\dfrac{\tau_i}{\tau_{iM}}\right)^n$  $\forall~i\in \{1,2,3\}$, and  $\pmb{G}_m$ is a diagonal matrix with $\pmb{G}_m=\diag \begin{bmatrix}
		g_{1m} & g_{2m} & g_{3m}
	\end{bmatrix}$ where the terms $g_{im} = \left(\dfrac{\tau_i}{\tau_{im}}\right)^n$  $\forall~i\in\{1,2,3\}$ and $n\geq 2$ being an even integer. The matrix $\pmb{Q} = \diag \begin{bmatrix}
		q_1 & q_2& q_3
	\end{bmatrix}$ helps in enforcing the asymmetric actuator saturation bounds, where 
	\begin{equation}\label{eqn:q_zeta}
		q_i (\zeta_i) = \begin{cases}
			1, & \text{if $\zeta_i$} > 0\\
			0, & \text{$\zeta_i$} \leq 0
		\end{cases}.
	\end{equation}
	The actuator saturation bounds in forward and reverse directions are $\tau_{iM}$ and $\tau_{im}$, respectively. It should also be noted that each actuator can have different asymmetric actuator saturation bounds; that is, the assumption of using the equally capable actuator for actuation in all directions is relaxed. 
	Now, we show how the proposed model in \Cref{eqn: actuator} enforces the actuator bounds. To do so, we have the following result that ensures the boundedness of the control inputs within their bounds. 
	\begin{theorem}\label{th:input_sat}
		Consider the actuator input saturation model as described by \Cref{eqn: actuator}. If the designed model input $\pmb{\tau}_c(t)$ satisfies 
		\begin{equation} \label{eqn:tauc_M}
			\|\pmb{\tau}_c(t)\| \leq \tau_{cM} ~(\forall~ t \geq 0), 
		\end{equation} 
		where $\tau_{cM}$ is some positive constant,  then the model output, $\tau_i$ will always respect the saturation bounds. Equivalently, $-\tau_{i0m} \leq\tau_i(t) \leq \tau_{i0M}~~\forall~~i \in \{1,2,3\})$ for all time $t \geq 0$, where the terms $\tau_{i0M}$ and $\tau_{i0m}$ are given by 
		\begin{equation}
			\label{eqn:tau_i0M} \tau_{i0M} = \tau_{iM}\left[ \dfrac{\tau_{cM}}{\tau_{cM} + \rho_i \tau_{iM}}\right]^{1/n} < \tau_{iM}, \quad 
			\tau_{i0m} = \tau_{im}\left[ \dfrac{\tau_{cM}}{\tau_{cM} - \rho_i \tau_{im}}\right]^{1/n} < \tau_{im}.
		\end{equation}
		% \begin{subequations}
			%     \begin{align}
				%     \label{eqn:tau_i0M}\tau_{i0M} = \tau_{iM}\left[ \dfrac{\tau_{cM}}{\tau_{cM} + \rho_i \tau_{iM}}\right]^{1/n} < \tau_{iM},\\
				%     \label{eqn:tau_i0m}\tau_{i0m} = \tau_{im}\left[ \dfrac{\tau_{cM}}{\tau_{cM} - \rho_i \tau_{im}}\right]^{1/n} < \tau_{im}.
				% \end{align}
			% \end{subequations}
	\end{theorem}
	
	\begin{proof}
		Here, the proof is done in two sub-steps. First, it is proved that when the model state value reaches the maximum actuator bound, its value is guaranteed to decrease. Then, it is shown that the output is strictly less than the maximum actuator bound at all times. Now, we further divide the proof into two cases: when the actuator is saturated in the forward direction and in the reverse direction.\\
		\emph{Case1: $\zeta_i > 0$}, where $\zeta_i$ is an element of the state of the actuator saturation model, which is equal to the actuator input in the $i$\textsuperscript{th} channel. This means we will first prove the saturation of the actuator in the positive direction of the actuator.
		From the smooth input saturation model in \Cref{eqn: actuator}, we write the element-wise equation as 
		\begin{equation}
			\label{eqn: zeta_i} \dot{\zeta}_i = q_i(\zeta_i)(1-g_{iM})\tau_{ic} - \rho_i \zeta_i +(1- q_i(\zeta_i))(1-g_{im})\tau_{ic}.
		\end{equation}
		From \Cref{eqn:q_zeta}, we can see that for  $\zeta_i >0$, $q_i = 1$, which reduces \Cref{eqn: zeta_i} to
		\begin{equation}
			\label{eqn: zeta_i_M} \dot{\zeta}_i = \left[ 1- \left(\dfrac{\zeta_i}{\tau_{iM}}\right)^n\right]\tau_{ic} - \rho_i \zeta_i.
		\end{equation}
		Let $ \zeta_i  = \tau_{iM}$, then on substituting this in \Cref{eqn: zeta_i_M}, we get $\dot{\zeta}_i = -\rho_i \zeta_i$, which on integrating gives $\zeta_i = \tau_{iM}e^{-\rho_it}$. Since, $\rho_i$ and $\tau_{iM}$ are positive, $\zeta_i$ will decrease.  This clearly indicates that $\tau_i(t) = \zeta_i(t) \leq \tau_{iM} \,\forall t\geq 0$.
		
		Now, as the second sub-step, our objective is to show that $\tau_{i}(t) = \zeta_i(t) \leq \tau_{i0M}$ and $\tau_{i0M} < \tau_{iM} \,\forall t \geq 0$. 
		As we already proved that $\zeta_i  \leq \tau_{iM}$, which implies that $\left(\dfrac{\zeta_i}{\tau_{iM}}\right)^n \leq 1$. Thus, the coefficients of $\tau_{ic}$ and $\zeta_i$ in \Cref{eqn: zeta_i_M} are both non-negative. Now, it is obvious that the only way the magnitude of $\zeta_i$ can increase is when both $\zeta_i(t)$ and $\tau_{ic}(t)$ are of the same sign, that is, $\zeta_i(t) \tau_{ic}(t) \geq 0$. Therefore, for this case, without loss of generality, we can assume that $\tau_{ic} \geq0 $ as  $\zeta_i(t) > 0$.
		Now using the inequality (\ref{eqn:tauc_M}) and taking $\tau_{cM}$ as common, \Cref{eqn: zeta_i_M} can be written as
		\begin{equation}
			\dot{\zeta}_i \leq \left[ 1- \left(\dfrac{\zeta_i}{\tau_{iM}}\right)^n\right] \tau_{cM}- \rho_i \zeta_i = \tau_{cM}\left[ 1- \left(\dfrac{\zeta_i}{\tau_{iM}}\right)^n - \dfrac{\rho_i \tau_{iM}}{\tau_{cM}}\times \dfrac{\zeta_i}{\tau_{iM}} 
			\right]
		\end{equation}
		Since, $-\dfrac{\zeta_i}{\tau_{iM}} \leq - \left(\dfrac{\zeta_i}{\tau_{iM}}\right)^n$, the above inequality can be expressed as 
		\begin{equation} \label{eqn:zetai_ineq_}
			\dot{\zeta}_i \leq \tau_{cM}\left[ 1- \left(\dfrac{\zeta_i}{\tau_{iM}}\right)^n - \dfrac{\rho_i \tau_{iM}}{\tau_{cM}}\times \left(\dfrac{\zeta_i}{\tau_{iM}}\right)^n 
			\right]
			\leq \tau_{cM}\left[1- \left( 1+ \dfrac{\rho_i \tau_{iM}}{\tau_{cM}}\right) \left( \dfrac{\zeta_i}{\tau_{iM}}\right)^n\right].
		\end{equation}
		From \Cref{eqn:zetai_ineq_}, as $\tau_{cM}>0$, we can see that  
		$\dot{\zeta}_i \leq 0$, if \begin{equation}
			\left[1- \left( 1+ \dfrac{\rho_i \tau_{iM}}{\tau_{cM}}\right) \left( \dfrac{\zeta_i}{\tau_{iM}}\right)^n\right] \leq 0 \implies \zeta_i \geq \tau_{iM} \left[ \dfrac{\tau_{cM}}{\tau_{cM} + \rho_i \tau_{iM}}\right]^{1/n} = \tau_{i0M}.
		\end{equation} 
		Therefore, $\zeta_i(t)$ can not cross $\tau_{i0M}$, otherwise $\dot{\zeta}_i$ will become non-positive, leading to decrease of $\zeta_i$ immediately. 
		Now, we have $$\tau_{i0M} = \tau_{iM} \left[ \dfrac{\tau_{cM}}{\tau_{cM} + \rho_i \tau_{iM}}\right]^{1/n}  =  \tau_{iM} \left[ \dfrac{1}{1 + \rho_i \tau_{iM}/\tau_{cM}}\right]^{1/n} < \tau_{iM}.$$
		Hence, $\tau_i(t) \leq \tau_{i0M} < \tau_{iM} (i \in \{1,2,3\}), \forall~ t \geq 0.$ \\
		\emph{Case2: $\zeta_i \leq 0$}, that is, when the actuator is acting in the reverse direction.\\
		From \Cref{eqn:q_zeta}, we can see that for  $\zeta_i \leq 0$, $q_i = 0$, which reduces the element-wise equation of the smooth input saturation model given by \Cref{eqn: zeta_i} to
		\begin{equation}\label{eqn: zeta_i_m}
			\dot{\zeta}_i = (1-g_{im})\tau_c - \rho_i \zeta_i 
			= \left[ 1- \left(\dfrac{\zeta_i}{\tau_{im}}\right)^n\right]\tau_{ic} - \rho_i \zeta_i.
		\end{equation}
		Suppose the actuator input reaches the saturation bound of the actuator in the reverse direction that is, $ \zeta_i  = -\tau_{im}$, then on substituting this in \Cref{eqn: zeta_i_m}, we get $\dot{\zeta}_i = -\rho_i \zeta_i$. On performing integration, we get, $\zeta_i = -\tau_{im}e^{-\rho_it}$. Since $\rho_i$ and $\tau_{im}$ are positive, $\zeta_i$ will become less negative, that is, increase.  This clearly indicates that $-\tau_{im} \leq  \tau_i(t)  =\zeta_i(t) \,\forall t\geq 0$.
		
		Now, as the second sub-step for this case, our objective is to show that $-\tau_{i0m} \leq \tau_{i}(t) = \zeta_i(t) $ and $\tau_{i0m} < \tau_{im} \forall t \geq 0$. 
		As we already proved that $-\tau_{im} \leq \zeta_i  $, which implies that $\left(\dfrac{\zeta_i}{\tau_{im}}\right)^n \leq 1$. 
		Now, as pointed out earlier, that the magnitude of $\zeta_i(t)$ can only increase when both $\zeta_i(t)$ and $\tau_{ic}(t)$ are of the same sign, that is, $\zeta_i(t) \tau_{ic}(t) \geq 0$. Since $\zeta_i(t) \leq 0$, without loss of generality, we can assume that $\tau_{ic} <0 $. Now using the inequality (\ref{eqn:tauc_M}) and taking $\tau_{cM}$ as common, \Cref{eqn: zeta_i_m} can be written as
		\begin{align}
			\dot{\zeta}_i \leq \left[ 1- \left(\dfrac{\zeta_i}{\tau_{im}}\right)^n\right] \tau_{cM}- \rho_i \zeta_i = \tau_{cM}\left[ 1- \left(\dfrac{\zeta_i}{\tau_{im}}\right)^n - \dfrac{\rho_i (-\tau_{im})}{\tau_{cM}}\times \dfrac{\zeta_i}{-\tau_{im}} 
			\right]
		\end{align}
		Now, since, $-\left(\dfrac{\zeta_i}{-\tau_{im}}\right) \leq -\left(\dfrac{\zeta_i}{-\tau_{im}}\right)^n$, the above inequality can be rewritten as 
		\begin{align}
			\dot{\zeta}_i \leq \tau_{cM}\left[1- \left( 1- \dfrac{\rho_i \tau_{im}}{\tau_{cM}}\right) \left( \dfrac{\zeta_i}{\tau_{im}}\right)^n
			\right],
		\end{align}
		which indicates that 
		$\dot{\zeta}_i \textcolor{blue}{\geq} 0$, if $$\zeta_i \leq -\tau_{im} \left[ \dfrac{\tau_{cM}}{\tau_{cM} - \rho_i \tau_{im}}\right]^{1/n} = -\tau_{i0m}.$$ 
		Therefore, $\zeta_i(t)$ can not go below $-\tau_{i0m}$, otherwise $\dot{\zeta}_i$ will become non-negative, leading to increase of $\zeta_i$ immediately. Also, similar to \emph{case 1}, it can be shown that $\tau_{im}>\tau_{i0m}$
		Hence,  $-\tau_{im} < -\tau_{i0m} \leq \tau_i(t) \leq \tau_{i0M} < \tau_{iM} (i \in \{1,2,3\}), \forall ~t \geq 0$, and this satisfies \Cref{eqn:tau_i0M}. This completes the proof.
	\end{proof} 
	Next, we proceed to design the control law by augmenting the input saturation model in \Cref{{eqn: actuator}} with the system kinematics and dynamics.  Instead of designing $\pmb{\tau}$, we will design the input, $\pmb{\tau}_c$ of the augmented system. As a result, the model output $\pmb{\tau}$ will always respect the actuators' bounds, preventing them from overburdening.   It is to be noted that augmenting the governing equations of the USV with the input saturation model transformed the constrained input problem into an unconstrained input problem where the model is in affine form with respect to the model input. This allows us to conveniently apply the Lyapunov stability and backstepping approach to design the control law. 
	\begin{theorem}\label{th:input_sat_ctrl_design}
		Consider the dynamics of the USV described by \Cref{eqn: kinematics,eqn: dynamics}, augmented with the smooth asymmetric input saturation model in \Cref{eqn: actuator}. If the input to the augmented system is chosen as 
		\begin{equation}\label{eqn: tau_c}
			% \pmb{\tau}_c &= (\pmb{I}-\pmb{G})^{-1} (\pmb{\rho} \pmb{\tau} + \dot{\pmb{\alpha}}_2) - \pmb{K}_3\left( \pmb{\tau} -\pmb{C}(\pmb{\nu})\pmb{\nu} - \pmb{D}(\pmb{\nu})\pmb{\nu} - \pmb{M}\dot{\pmb{\alpha}}_1 + J^\top(\pmb{\eta})\pmb{z}_1\right) + (\pmb{K}_2\pmb{K}_3-\pmb{I})\left(\pmb{\nu} -  \pmb{J}(\pmb{\eta})^T \left( \dot{\pmb{\eta}}_d -  \pmb{K}_{1}\pmb{z}_1 \right)\right)
			\pmb{\tau}_c = \left[\pmb{Q}(\pmb{I} - \pmb{G}_M) + (\pmb{I}-\pmb{Q})(\pmb{I} - \pmb{G}_m) \right]^{-1} (\pmb{\rho} \pmb{\tau} + \dot{\pmb{\alpha}}_2) - \pmb{K}_3 \pmb{z}_3 -\pmb{z}_2,
		\end{equation} 
		where $\pmb{K}_i = \diag (\begin{bmatrix}
			k_{i1} & k_{i2} & k_{i3}
		\end{bmatrix})$ $(i \in \{1,2,3\})$ are the gain parameters, 
		\begin{subequations}
			\begin{align}
				\pmb{z}_3 &= \pmb{\tau} - \pmb{\alpha}_2,\\
				\label{eqn: alpha2_tau}
				\pmb{\alpha}_2 
				&= \pmb{C}(\pmb{\nu})\pmb{\nu} + \pmb{D}(\pmb{\nu})\pmb{\nu} + \pmb{M}\dot{\pmb{\alpha}}_1 -\pmb{K}_2\pmb{z}_2 - J^\top(\pmb{\eta})\pmb{z}_1 - \hat{\pmb{b}}, \\ 
				\label{eqn: z2}
				\pmb{z}_{2} &= \pmb{\nu} - \pmb{\alpha}_1 =\left[
				z_{21},\, z_{22},\, z_{23}
				\right]^\top, \\
				\label{eqn: alpha1}
				\pmb{\alpha}_1 &= \pmb{J}(\pmb{\eta})^T \left( \dot{\pmb{\eta}}_d -  \pmb{K}_{1}\pmb{z}_1 \right), \\
				\pmb{z}_1 &= \pmb{\eta} - \pmb{\eta_d} = \left[
				z_{11},\, z_{12},\, z_{13}
				\right]^\top = \left[
				x-x_d ,\,  y-y_d ,\,  \psi-\psi_d
				\right]^\top,
			\end{align}
		\end{subequations} 
		and output $\pmb{\tau}$ of the augmented system is applied to the USV, then the vehicle's states will track the desired trajectory. Moreover, the actuator constraints are guaranteed to be satisfied at all times.
	\end{theorem}
	\begin{proof}
		We augment the system dynamics with a smooth input saturation model to be able to apply the Lyapunov and backstepping techniques to design a stable controller even in the presence of input magnitude bounds.
		Now, to derive the necessary controller to satisfy the objectives of this work, let us define the output tracking errors as 
		\begin{align}\label{eqn:z1z2}
			\pmb{z}_1 = \pmb{\eta} - \pmb{\eta_d} , \quad \pmb{z}_{2} = \pmb{\nu} - \pmb{\alpha}_1,
		\end{align}
		where $\pmb{\alpha}_1$ is a stabilising function to be designed. 
		Consider a Lyapunov function candidate as $\mathcal{V}_1 = \dfrac{1}{2} \pmb{z}_1^\top \pmb{z}_1$, whose time derivative is given by
		\begin{equation}\label{eqn: v1_dot1}
			\dot{\mathcal{V}}_1 = \pmb{z}_1^\top \dot{\pmb{z}}_1.
		\end{equation}
		% \begin{align}\label{eqn:V_1}
			%   \mathcal{V}_1 = \dfrac{1}{2} \pmb{z}_1^\top \pmb{z}_1.
			% \end{align}
		Now, since $\dot{\pmb{z}}_1 = \dot{\pmb{\eta}} - \dot{\pmb{\eta}}_d$, and using \Cref{eqn: kinematics,eqn:z1z2}, one can express $\dot{\pmb{z}}_1$ as
		\begin{align}\label{eqn: z1dot}
			\dot{\pmb{z}}_1 = \pmb{J}(\pmb{\eta}) \pmb{\nu} - \dot{\pmb{\eta}}_d = J(\pmb{\eta}) (\pmb{z}_2 + \pmb{\alpha}_1) - \dot{\pmb{\eta}}_d.
		\end{align}
		% \begin{equation}\label{eqn:e1_dot}
			%     \dot{\pmb{z}}_1 = \pmb{J}(\pmb{\eta}) \pmb{\nu} - \dot{\pmb{\eta}}_d.
			% \end{equation}
		% To design the controller using the backstepping method, let us define another tracking error in $\pmb{\nu}$ as 
		% \begin{equation}\label{eqn: z2}
			%     \pmb{z}_{2} = \pmb{\nu} - \pmb{\alpha}_1, \\
			% %     = \begin{bmatrix}
				% %     z_{21}& z_{22}& z_{23}
				% % \end{bmatrix}^\top,
			% \end{equation}
		% where \(\pmb{\alpha}_1\) is the stabilizing function to be designed. 
		% Using \Cref{eqn: z2}, the expression for $\dot{\pmb{z}}_1$ can be can be written in terms of variables $\pmb{z}_2$ and $\pmb{\alpha}_1$ as 
		% \begin{equation}\label{eqn: z1dot}
			%     \dot{\pmb{z}}_1 = J(\pmb{\eta}) (\pmb{z}_2 + \pmb{\alpha}_1) - \dot{\pmb{\eta}}_d.
			% \end{equation}
		% In order to design the stabilizing function $\pmb{\alpha}_{1}$, differentiating the chosen Lyapunov function candidate in \Cref{eqn:V_1} with respect to time and 
		By substituting $\dot{\pmb{z}}_1$ from \Cref{eqn: z1dot} into \Cref{eqn: v1_dot1}, we obtain,
		\begin{equation}\label{eqn: v1dot}
			\dot{\mathcal{V}}_1 = \pmb{z}_1^\top \dot{\pmb{z}}_1 = \pmb{z}_1^\top \left[J(\pmb{\eta}) (\pmb{z}_2 + \pmb{\alpha}_1) - \dot{\pmb{\eta}}_d \right].
		\end{equation}
		Upon choosing the stabilizing function $\pmb{\alpha}_1$, as given by 
		\Cref{eqn: alpha1} and substituting the value of chosen $\pmb{\alpha}_1$ into \Cref{eqn: v1dot}, one may arrive at
		% obtain the time derivative of the Lyapunov function candidate, $\mathcal{V}_1$ as
		\begin{equation}
			\dot{\mathcal{V}}_1 
			% =& -\sum_{i=1}^{3} k_{1i}z_{1i}^2 + 
			% \sum_{i=1}^{3} \left[ z_{1i}J_i(\pmb{\eta})\pmb{z}_2
			% \right],\\
			=-\pmb{z}_1^\top \pmb{K}_1 \pmb{z}_1 + \pmb{z}_1^\top J(\pmb{\eta})\pmb{z}_2,
		\end{equation}
		where $\pmb{z}_1^\top J(\pmb{\eta})\pmb{z}_2$ is the coupling term, which will be canceled in the next step.
		% where $J_i(\pmb{\eta})$ represents the $i^{\rm th}$ row of the $\pmb{J}$ matrix.
		To again apply the backstepping, we define another tracking error as
		\begin{equation}\label{eqn: z3}
			\pmb{z}_{3} = \pmb{\tau} - \pmb{\alpha}_2,
			%     =\begin{bmatrix}
				%     z_{31}& z_{32}& z_{33}
				% \end{bmatrix}^\top,
		\end{equation} 
		where $\pmb{\alpha}_2$ is the stabilizing function to be designed next.
		Towards that consider a Lyapunov function candidate by augmenting $\mathcal{V}_1$ with the quadratic function in terms of $\pmb{z}_2$ and $\pmb{b}_e$ as   
		\begin{equation}
			\mathcal{V}_2 = \mathcal{V}_1 + \dfrac{1}{2}\pmb{z}_2^\top\pmb{Mz}_2
			+ \dfrac{1}{2}\pmb{b}_e^\top\pmb{b}_e,
		\end{equation}
		where, $\pmb{M} \succ 0$, as described in \Cref{eqn: M}.
		On differentiating $\mathcal{V}_2$ with respect to time and using \Cref{eqn:b_e_dot}, we get,
		\begin{equation}
			\label{eqn:V2_dot} \dot{\mathcal{V}}_2 = \dot{\mathcal{V}}_1 + \pmb{z}_2^\top\pmb{M}\dot{\pmb{z}}_2 + \pmb{b}_e^\top\dot{\pmb{b}}_e = \dot{\mathcal{V}}_1 + \pmb{z}_2^\top\pmb{M}\dot{\pmb{z}}_2 + \pmb{b}_e^\top\dot{\pmb{b}}- \pmb{b}_e^\top\pmb{K}_0\pmb{b}_e.
			%+ \Tilde{\pmb{b}}^\top\Gamma^{-1}\dot{\Tilde{\pmb{b}}}.
		\end{equation}
		Next, differentiating \Cref{eqn: z2} to obtain $\dot{\pmb{z}}_2$ and substituting the value of $\dot{\pmb{\nu}}$ from \Cref{eqn: dynamics}, results in 
		\begin{equation*}
			\dot{\pmb{z}}_2 = \dot{\pmb{\nu}} - \dot{\pmb{\alpha}}_1 = \pmb{M}^{-1}\left[\pmb{\tau} + \pmb{b} - \pmb{C}(\pmb{\nu})\pmb{\nu} - \pmb{D}(\pmb{\nu})\pmb{\nu} \right] - \dot{\pmb{\alpha}}_1,
		\end{equation*}
		which can be rewritten with the help of \Cref{eqn: z3} as 
		\begin{equation}
			\label{eqn: z2dot}
			\dot{\pmb{z}}_2   =\pmb{M}^{-1}\left[\pmb{z}_3 + \pmb{\alpha}_2 + \pmb{b} - \pmb{C}(\pmb{\nu})\pmb{\nu} - \pmb{D}(\pmb{\nu})\pmb{\nu} \right] - \dot{\pmb{\alpha}}_1.
		\end{equation}
		Now, if we choose $\pmb{\alpha}_2$ as given by \Cref{eqn: alpha2_tau}
		and substitute the value $\dot{\pmb{z}}_2$ from \Cref{eqn: z2dot} into \Cref{eqn:V2_dot}, the time derivative of the Lyapunov function candidate becomes 
		\begin{align}\label{eqn: V2_dot_final}
			\dot{\mathcal{V}}_2 &= -\pmb{z}_1^\top \pmb{K}_1 \pmb{z}_1 -\pmb{z}_2^\top \pmb{K}_2 \pmb{z}_2  - \pmb{b}_e^\top\pmb{K}_0\pmb{b}_e + \pmb{z}_2^\top\pmb{z}_3 + \pmb{z}_2^\top\pmb{b}_e  + \pmb{b}_e^\top\dot{\pmb{b}}.
		\end{align}
		Here also, there is one coupling term, $\pmb{z}_2^\top\pmb{z}_3$, which will be canceled out in the next step. Otherwise, the leftover terms can be made into square form by using Young's inequality, and can be expressed as
		\begin{equation}
			\label{eqn:b_e_inequality}\pmb{b}_e^\top\dot{\pmb{b}} \leq \varepsilon_1 \pmb{b}_e^\top\pmb{b}_e + \dfrac{1}{4\varepsilon_1}\dot{\pmb{b}}^\top \dot{\pmb{b}}, \quad \pmb{z}_2^\top\pmb{b}_e \leq \varepsilon_2 \pmb{z}_2^\top\pmb{z}_2 + \dfrac{1}{4\varepsilon_2}\pmb{b}_e^\top \pmb{b}_e,
		\end{equation}
		where $\varepsilon_2$ is also a positive number like $\varepsilon_1$.
		At last, to design $\pmb{\tau}_c$, consider the Lyapunov function candidate by augmenting $\mathcal{V}_2$ with quadratic function in terms of $\pmb{z}_3$ as    
		\begin{align}\label{eqn:aV3}
			\mathcal{V}_3 =& \mathcal{V}_2 + \dfrac{1}{2}\pmb{z}_3^\top\pmb{z}_3.
		\end{align}
		On differentiating $\mathcal{V}_3$ with respect to time, we get 
		\begin{align}\label{eqn: V3_dot}
			\dot{\mathcal{V}}_3 =& \dot{\mathcal{V}}_2 + \pmb{z}_3^\top \dot{\pmb{z}}_3.
		\end{align}
		Using the \Cref{eqn: actuator,eqn: z3}, we can obtain $\dot{\pmb{z}}_3$ as
		\begin{align}\label{eqn: az3dot}
			\dot{\pmb{z}}_3 &= \dot{\pmb{\tau}} -\dot{\pmb{\alpha}}_2 = \pmb{Q}(\pmb{I} - \pmb{G}_M) \pmb{\tau}_c - \pmb{\rho} \pmb{\tau} + (\pmb{I}-\pmb{Q})(\pmb{I} - \pmb{G}_m) \pmb{\tau}_c -\dot{\pmb{\alpha}}_2 \\
			&= \left[\pmb{Q}(\pmb{I} - \pmb{G}_M) + (\pmb{I}-\pmb{Q})(\pmb{I} - \pmb{G}_m) \right]\pmb{\tau}_c - \pmb{\rho} \pmb{\tau}  -\dot{\pmb{\alpha}}_2.
		\end{align}
		Now, if the control input $\pmb{\tau}_c$ is chosen as
		\begin{align}\label{eqn: tau_c}
			\pmb{\tau}_c = \left[\pmb{Q}(\pmb{I} - \pmb{G}_M) + (\pmb{I}-\pmb{Q})(\pmb{I} - \pmb{G}_m) \right]^{-1} (\pmb{\rho} \pmb{\tau} + \dot{\pmb{\alpha}}_2) - \pmb{K}_3 \pmb{z}_3 -\pmb{z}_2,
		\end{align} 
		where $\pmb{K}_3 = \diag (\begin{bmatrix}
			k_{31} & k_{32} & k_{33}
		\end{bmatrix})$ such that $k_{3i}>0$ and substitute in \Cref{eqn: az3dot}, we get $\dot{\pmb{z}}_3 = -\pmb{K}_3 \pmb{z}_3 - \pmb{z}_2.$
		Finally, on substituting $\dot{\pmb{z}}_3$ in \Cref{eqn: V3_dot}, we get
		\begin{align}\label{eqn: aV3_dot_final}
			\dot{\mathcal{V}}_3 & =  -\left( \pmb{z}_1^\top \pmb{K}_1 \pmb{z}_1 + \pmb{z}_2^\top \pmb{K}_2 \pmb{z}_2 + \pmb{z}_3^\top \pmb{K}_3 \pmb{z}_3 + \pmb{b}_e^\top\pmb{K}_0\pmb{b}_e
			\right)  + \pmb{z}_2^\top\pmb{b}_e  + \pmb{b}_e^\top\dot{\pmb{b}}.
		\end{align}
		Now, using the inequalities from \Cref{eqn:b_e_inequality}, we can write
		\begin{align}\label{eqn:V_3_dot_inequality_1}
			\dot{\mathcal{V}}_3 & \leq  -\left( \pmb{z}_1^\top \pmb{K}_1 \pmb{z}_1 + \pmb{z}_2^\top \pmb{K}_2 \pmb{z}_2 + \pmb{z}_3^\top \pmb{K}_3 \pmb{z}_3 + \pmb{b}_e^\top\pmb{K}_0\pmb{b}_e \right)  
			+ \varepsilon_1 \pmb{b}_e^\top\pmb{b}_e + \dfrac{1}{4\varepsilon_1}\dot{\pmb{b}}^\top \dot{\pmb{b}}  
			+ \varepsilon_2 \pmb{z}_2^\top\pmb{z}_2 + \dfrac{1}{4\varepsilon_2}\pmb{b}_e^\top \pmb{b}_e.
		\end{align}
		Now, our goal is to express the term on the right-hand side of \Cref{eqn:V_3_dot_inequality_1}, in terms of $\mathcal{V}_3$, and for that, we need to mold the terms containing $\pmb{z}_2$ as
		\begin{equation}\label{eqn:z2_M_z2_inequality}
			- \pmb{z}_2^\top \pmb{K}_2 \pmb{z}_2 \leq -\lambda_{\rm{min}}(\pmb{K}_2\pmb{M}^{-1})\pmb{z}_2^\top \pmb{M} \pmb{z}_2, \quad \pmb{z}_2^\top \pmb{z}_2 \leq \lambda_{\rm{max}}(\pmb{M}^{-1})\pmb{z}_2^\top \pmb{M} \pmb{z}_2. 
		\end{equation}
		Therefore, using \Cref{eqn:z2_M_z2_inequality} and eigenvalue property of positive definite matrix, \Cref{eqn:V_3_dot_inequality_1} 
		can be rewritten as 
		\begin{align}
			\nonumber \dot{\mathcal{V}}_3 &  \leq \dfrac{b_M^2}{4\varepsilon_1} + -2\left[ \lambda_{\rm{min}}(\pmb{K}_1) \dfrac{1}{2}\pmb{z}_1^\top\pmb{z}_1  + 
			\left(\lambda_{\rm{min}}(\pmb{K}_2\pmb{M}^{-1}) -\varepsilon_2\lambda_{\rm{max}}(\pmb{M}^{-1})\right)\dfrac{1}{2}\pmb{z}_2^\top\pmb{M}\pmb{z}_2 \right.\\
			& \left. \quad
			+ \lambda_{\rm{min}}(\pmb{K}_3) \dfrac{1}{2}\pmb{z}_3^\top\pmb{z}_3  +\left(\lambda_{\rm{min}}(\pmb{K}_0)-\varepsilon_1 - \dfrac{1}{4\varepsilon_2}\right) \dfrac{1}{2}\pmb{b}_e^\top \pmb{b}_e
			\right], \label{eqn:v_3_dot_inequality}
		\end{align}
		where $\left(\lambda_{\rm{min}}(\pmb{K}_2) -\varepsilon_2\lambda_{\rm{max}}(\pmb{M}^{-1})\right)  > 0, \quad 
		\left(\lambda_{\rm{min}}(\pmb{K}_0)-\varepsilon_1 - \dfrac{1}{4\varepsilon_2}\right) >0.$
		Since the coefficients of $\pmb{z}_i^\top\pmb{z}_i~~ (i \in \{1,2,3\})$ and $\pmb{b}_e^\top\pmb{b}_e$ are all positive, all of it can be replaced with the minimum of these coefficients in the inequality (\ref{eqn:v_3_dot_inequality}). So, in order to neatly represent, let us define
		\begin{equation*}
			a_1:= \rm{min}\left\{ \lambda_{\rm{min}}(\pmb{K}_1),
			\left(\lambda_{\rm{min}}(\pmb{K}_2\pmb{M}^{-1}) -\varepsilon_2\lambda_{\rm{max}}(\pmb{M}^{-1})\right),
			\lambda_{\rm{min}}(\pmb{K}_3),
			\left(\lambda_{\rm{min}}(\pmb{K}_0)-\varepsilon_1 - \dfrac{1}{4\varepsilon_2}\right)
			\right\},
		\end{equation*}
		with the help of which the derivative of $\mathcal{V}_3$ can be expressed as
		\begin{align}
			\label{eqn:V_3_dot_compact} \dot{\mathcal{V}}_3(t) & \leq -2a_1\left( \dfrac{1}{2}\pmb{z}_1^\top\pmb{z}_1 + \dfrac{1}{2}\pmb{z}_2^\top\pmb{z}_2 + 
			\dfrac{1}{2}\pmb{z}_3^\top\pmb{z}_3 + \dfrac{1}{2}\pmb{b}_e^\top \pmb{b}_e
			\right) + \dfrac{b_M^2}{4\varepsilon_1} \leq -2a_1 \mathcal{V}_3(t) + \dfrac{b_M^2}{4\varepsilon_1}.
		\end{align}
		On integrating \Cref{eqn:V_3_dot_compact}, we get the solution as 
		\begin{equation}\label{eqn:V3_integrated}
			\mathcal{V}_3(t) \leq \dfrac{b_M^2}{8a_1\varepsilon_1} + \left( \mathcal{V}_3(0) - \dfrac{b_M^2}{8a_1\varepsilon_1}\right)e^{-2a_1t}.
		\end{equation}
		From \Cref{eqn:V3_integrated}, it can be concluded that the $\mathcal{V}_3$ is globally ultimately bounded and converges exponentially to a ball of radius $\dfrac{b_M^2}{8a_1\varepsilon_1}$. 
		% \textcolor{red}{conclusion about other states} 
		% Since the gains $k_{1i}$, $k_{2i}$, and $ k_{3i}$ are positive real numbers, the square of a real number is always positive.
		% Hence, $\dot{\mathcal{V}}_3$ is negative definite, that is, $\dot{\mathcal{V}}_3 < 0$. From $\dot{\mathcal{V}}_3 < 0$, it follows that $\mathcal{V}_3(t) < \mathcal{V}_3(0)$. 
		Hence, all the error signals $z_{1i}(t), z_{2i}(t), z_{3i}(t) ~\forall i\in(\{1,2,3\})$ remains bounded. 
	\end{proof}
	\subsection{Input magnitude and rate saturation model}
	Due to the limited capability of the actuators, which are generally known a priori, it is important to consider these constraints while designing the control law. Most of the work in the literature ignores the dynamics of the actuator. This is important because, in reality, the actuators will take some time to produce the output as demanded. It cannot provide whatever is demanded instantaneously. 
	% Some works in the literature consider first-order actuator dynamics. This can be taken into account by considering the bound on actuator rate as well.
	Thus, considering the bound on actuator input magnitude and its rate in the design represents wider applicability to practical scenarios.
	To do so, we propose a strategy inspired by \cite{8401917} to take this into account. The system dynamics in \Cref{eqn: kinematics,eqn: dynamics} is augmented with
	\begin{subequations}
		\begin{align}
			\label{eqn:zeta_dot}\dot{\pmb{\zeta}} &=  (\pmb{I} - \pmb{G}_1)\pmb{\tau}_c - \pmb{\rho}_1\dfrac{\tau_{dM}}{\tau_M}\pmb{\zeta}, \quad \pmb{\zeta}(0) = \pmb{0}, \quad \pmb{\tau}=\pmb{\zeta}\\
			\label{eqn:tauc_dot}\dot{\pmb{\tau}}_c &= (\pmb{I}-\pmb{G}_2)\pmb{\tau}_d - \pmb{\rho}_2 \pmb{\tau}_c, \quad \pmb{\tau}_c(0)=\pmb{0} 
		\end{align}
	\end{subequations}
	where $\pmb{\zeta} \in \mathbb{R}^3$, $\pmb{\tau}_c \in \mathbb{R}^3$ are the states of the magnitude and rate saturation model with $\pmb{\tau}_d \in \mathbb{R}^3$ as input and $\pmb{\tau} \in \mathbb{R}^3$ as output, $\pmb{I} \in \mathbb{R}^{3\times3}$ is identity matrix, $\pmb{G}_1 = \diag \begin{bmatrix}
		g_{11} & g_{12} & g_{13}
	\end{bmatrix}$ and $\pmb{G}_2 = \diag \begin{bmatrix}
		g_{21} & g_{22} & g_{23}
	\end{bmatrix}$. The elements of diagonal matrix $\pmb{G}_1$ are defined as $g_{1i} = \left(\dfrac{\zeta_i}{\tau_{iM}}\right)^n$ and the matrix $\pmb{G}_2$ are  $g_{2i} = \left[\dfrac{\tau_{ci}}{(1-\rho_{1i})\tau_{idM}}\right]^n $ for $(i \in \{1,2,3\})$, with $n \geq 2$ being an even integer.
	The constant matrices, $\pmb{\rho}_1 = \diag \begin{bmatrix}
		\rho_{11} &\rho_{12}  & \rho_{13}
	\end{bmatrix}$ and $\pmb{\rho}_2= \diag \begin{bmatrix}
		\rho_{21} &\rho_{22}  & \rho_{23}
	\end{bmatrix}$ are the design parameters whose elements satisfy, $0< \rho_{1i} < 1$ and $\rho_{2i} > 0$ for $(i\in \{1,2,3\})$. Now, we show how \Cref{eqn:zeta_dot,eqn:tauc_dot} prevents violation of the bounds of input magnitude and its rate in the next theorem.
	\begin{theorem}
		Consider the input magnitude and its rate saturation model described by \Cref{eqn:zeta_dot,eqn:tauc_dot}. If this model's input $\pmb{\tau}_d$ of the saturation model is bounded, that is, $|| \pmb{\tau}_d(t)|| <\mathcal{T}_M ~\forall ~t \geq 0$), then 
		\begin{enumerate}
			\item the output $\pmb{\tau}$ of the model satisfies the constraints on its magnitude and rate, that is, $|\tau_i(t)| < \tau_{iM}$ and $|\dot{\tau}_i(t)| < \tau_{idM} \, (i\in \{1,2,3\}), \, \forall ~t\geq0$.
			\item $|\tau_{ci}(t)| \leq \tau_{c0i} \leq (1-\rho_{1i})\tau_{idM} ~~(i\in \{1,2,3\}) ~\forall ~t \geq 0$, where \\
			$\tau_{c0i} = \left[ \dfrac{\mathcal{T}_M}{\mathcal{T}_M + \rho_{2i}(1-\rho_{1i})\tau_{idM}} \right]^{(1/n)}(1-\rho_{1i})\tau_{idM}.$
		\end{enumerate}
	\end{theorem}
	\begin{proof}
		The proof of this theorem is similar to the \Cref{th:input_sat}. Now, since the model's input $\pmb{\tau}_d(t)$ is bounded, that is,  $|| \pmb{\tau}_d(t)|| <\mathcal{T}_M ~\forall ~t \geq 0$), and if $|\tau_{ci}| = (1-\rho_{1i})\tau_{idM}$, then $ \dot{\tau}_{ci} = -\rho_{2i}\tau_{ci}$. This implies that the magnitude of $\tau_{ci}$ is always going to decrease once $|\tau_{ci}| = (1-\rho_{1i})\tau_{idM}$. Hence, $|\tau_{ci}| \leq (1-\rho_{1i})\tau_{idM} ~ \forall ~t\geq 0$. Similarly, if $|\zeta_i| = \tau_{iM}$, then we have $\dot{\zeta}_i = -\rho_{1i}\dfrac{\tau_{idM}}{\tau_{iM}} \zeta_i$. Since, $\rho_i, \tau_{idM}$, and $\tau_{iM}$ are all positive constants, the magnitude of $\zeta_i$ is always going to decrease for all time, once it becomes equal to $\tau_{iM}$. Hence, $|\zeta_i| \leq \tau_{iM}$, which means that $\tau_i$ is bounded by $\tau_{iM}$. Now, on writing \Cref{eqn:zeta_dot} in elementwise form, we get, $\dot{\zeta}_i = (1-g_{1i})\tau_{ci} - \rho_{1i}\dfrac{\tau_{idM}}{\tau_{iM}} \zeta_i$, which can be converted to satisfy the inequality
		\begin{equation}\label{eqn:zeta_i_dot_ineq}
			\dot{\zeta}_i \leq |\tau_{ci}| + \rho_{1i}\dfrac{\tau_{idM}}{\tau_{iM}}|\zeta_i|,
		\end{equation} by applying triangle inequality and using the fact that $(1-g_{1i})$ will always be less than or equal to $1$.
		Now, since, $\zeta_i \leq \tau_{iM}$ and  $|\tau_{ci}| \leq (1-\rho_{1i})\tau_{idM}$, we can write \Cref{eqn:zeta_i_dot_ineq} as
		\begin{equation}
			\dot{\zeta}_i \leq (1-\rho_{1i})\tau_{idM} + \rho_{1i}\dfrac{\tau_{idM}}{\tau_{iM}}{\tau_{iM}} \leq \tau_{idM}.
		\end{equation}
		Now, in order to complete the proof, we need to prove that $|\tau_{ci}|$ cannot increase and will be less than $(1-\rho_{1i})\tau_{idM}$. Until now, we have shown that the input magnitude and rate constraints will be satisfied if this condition is maintained. \\
		On writing \Cref{eqn:tauc_dot} in element-wise form, we get
		\begin{equation}
			\dot{\tau}_{ci} = \left[1 - \left(\dfrac{\tau_{ci}}{(1-\rho_{1i})\tau_{idM}}\right)^n\right]\tau_{di} - \rho_{2i}\tau_{ci}.
		\end{equation}
		Now, since, $0 \leq1 - \left[\dfrac{\tau_{ci}}{(1-\rho_{1i})\tau_{idM}}\right]^n \leq 1$, and when $\tau_{ci} > 0$ and $\tau_{di}(t) <0$, then $\dot{\tau}_{ci} <0$. This implies that $\tau_{ci}$ is bound to increase. Similarly, when $\tau_{ci} < 0$ and $\tau_{di}(t) >0$, then $\dot{\tau}_{ci} >0$, which implies that $\tau_{ci}$ is now going to increase. From this discussion, we can conclude that the magnitude of $\tau_{ci}$ can only increase when $\tau_{ci}$ and $\tau_{di}(t)$ are of the same sign, that is, $\zeta_i U_i(t) \geq 0$. So, without loss of generality, we can conveniently assume $\tau_{ci} \geq 0$ and $\tau_{di}(t) \geq 0$ and write the inequality
		\begin{align}\label{eqn:tauci_dot_ineq}
			\dot{\tau}_{ci} \leq \left[1 - \left(\dfrac{\tau_{ci}}{(1-\rho_{1i})\tau_{idM}}\right)^n\right]\mathcal{T}_M - \rho_{2i}\tau_{ci},
		\end{align} as $||\pmb{\tau}_d|| < \mathcal{T}_M$, hence $\tau_{di} < \mathcal{T}_M$. Now, by taking $\mathcal{T}_M$ as common, the \Cref{eqn:tauci_dot_ineq} can be re-written as 
		\begin{align}
			\label{eqn:tauci_dot_1} \dot{\tau}_{ci} &\leq \left[1 - \left(\dfrac{\tau_{ci}}{(1-\rho_{1i})\tau_{idM}}\right)^n - \dfrac{\rho_{2i}\tau_{ci}}{\mathcal{T}_M}\right] \mathcal{T}_M \\  
			& \leq \left[1 - \left( 1+ \dfrac{\rho_{2i}\tau_{ci} \left[(1-\rho_{1i})\tau_{idM}\right]^n}{\tau_{ci}^n \mathcal{T}_M}\right) \dfrac{\tau_{ci}^n}{\left[(1-\rho_{1i})\tau_{idM}\right]^n}\right] \mathcal{T}_M.
		\end{align}
		
		% \begin{align}\label{eqn:tauci_dot_1}
			%     \dot{\tau}_{ci} & \leq \left[1 - \left( 1+ \dfrac{\rho_{2i}\tau_{ci} \left[(1-\rho_{1i})\tau_{idM}\right]^m}{\tau_{ci}^m \mathcal{T}_M}\right) \dfrac{\tau_{ci}^m}{\left[(1-\rho_{1i})\tau_{idM}\right]^m}\right] \mathcal{T}_M
			% \end{align}
		Also, we have assumed $\tau_{ci}\geq 0$ and already proved that $|\tau_{ci}| \leq (1-\rho_{1i})\tau_{idM}$ which lead us to 
		\begin{align}\label{eqn:tau_c_ineq}
			-\dfrac{(1-\rho_{1i})\tau_{idM}}{\tau_{ci}} \geq - \left[\dfrac{(1-\rho_{1i})\tau_{idM}}{\tau_{ci}}\right]^n.
		\end{align}
		Now, using \Cref{eqn:tau_c_ineq} into \Cref{eqn:tauci_dot_1}, we get 
		\begin{align}\label{eqn:tau_ci_dot}
			\dot{\tau}_{ci} = \mathcal{T}_M \left[ 1- \left( 1 + \dfrac{\rho_{2i}(1-\rho_{1i})\tau_{idM}}{\mathcal{T}_M}\right)\dfrac{\tau_{ci}^n}{\left[(1-\rho_{1i})\tau_{idM}\right]^n}\right].
		\end{align}
		From \Cref{eqn:tau_ci_dot}, we can get $\dot{\tau}_{ci} \leq 0$, if 
		\begin{align}
			\nonumber & \quad \quad \left[ 1 + \dfrac{\rho_{2i}(1-\rho_{1i})\tau_{idM}}{\mathcal{T}_M}\right]\dfrac{\tau_{ci}^n}{\left[(1-\rho_{1i})\tau_{idM}\right]^n} \geq 1 
		\end{align} implies
		\begin{align}
			\tau_{ci} &\geq  \left( 1 + \dfrac{\rho_{2i}(1-\rho_{1i})\tau_{idM}}{\mathcal{T}_M}\right)^{-1/n}(1-\rho_{1i})\tau_{idM}  \geq  \left(\dfrac{\mathcal{T}_M}{\mathcal{T}_M + \rho_{2i}(1-\rho_{1i})\tau_{idM} }\right)^{1/n}(1-\rho_{1i})\tau_{idM}. \label{eqn:tau_ci_ineq}
		\end{align}
		
		From \Cref{eqn:tau_ci_ineq}, we can conclude that the rate of change of $\tau_{ci}$ becomes negative as the $\tau_{ci}$ approaches the value on the right-hand side of the \Cref{eqn:tau_c_ineq}. Hence, 
		\begin{equation}
			|\tau_{ci}(t)| \leq \left(\dfrac{\mathcal{T}_M}{\mathcal{T}_M + \rho_{2i}(1-\rho_{1i})\tau_{idM} }\right)^{1/n}(1-\rho_{1i})\tau_{idM}.
		\end{equation} Since, $\left(\dfrac{\mathcal{T}_M}{\mathcal{T}_M + \rho_{2i}(1-\rho_{1i})\tau_{idM} }\right)^{1/n} <1, $  $~|\tau_{ci}(t)| \leq \tau_{c0i} <  (1-\rho_{1i})\tau_{idM} ~(i \in \{1,2,3\})~ \forall ~t \geq 0.$ This completes the proof.
	\end{proof}
	In the following theorem, we present the design of the model input, $\pmb{\tau}_d$. 
	\begin{theorem}
		Consider the USV dynamics given by \Cref{eqn: kinematics,eqn: dynamics}, augmented with input magnitude and rate saturation model described by \Cref{eqn:zeta_dot,eqn:tauc_dot}. If the model input, $\pmb{\tau}_d$ is designed as 
		\begin{equation}
			\pmb{\tau}_d = (\pmb{I}-\pmb{G}_2)^{-1} \left[\pmb{\rho}_2 \pmb{\tau}_c + \dot{\pmb{\alpha}}_3 - \pmb{K}_4\pmb{z}_4 - (\pmb{I}-\pmb{G}_1)\pmb{z}_3
			\right],
		\end{equation}
		where $\pmb{K}_4$ is diagonal matrix given by $\pmb{K}_4 = \diag \begin{bmatrix}
			k_{41} & k_{42} & k_{43}
		\end{bmatrix}$ such that $k_{4i}>0$ for $(i\in \{1,2,3\})$, and $\pmb{z}_4$ and $\pmb{\alpha}_3$ are defined as
		\begin{equation}
			\label{eqn:z4}\pmb{z}_4  =\pmb{\tau}_c - \pmb{\alpha}_3, \quad \pmb{\alpha}_3 = (\pmb{I}-\pmb{G}_1)^{-1} \left(\pmb{\rho}_1\dfrac{\tau_{dM}}{\tau_M}\pmb{\zeta} + \dot{\pmb{\alpha}}_2 - \pmb{K}_3 \pmb{z}_3 -\pmb{z}_2\right) ,
		\end{equation} and all other intermediate variables are defined in \Cref{th:input_sat,th:input_sat_ctrl_design},
		then all the constraints on input magnitude and rate saturation are met, that is, $|\tau_i(t)| < \tau_{iM}$ and $|\dot{\tau}_i(t)| < \tau_{idM} \, (i\in \{1,2,3\}), \, \forall t\geq0$.
	\end{theorem}
	\begin{proof}
		Again, we augment the actuator input magnitude and rate saturation model with the USV dynamics and proceed to design the unconstrained virtual input $\pmb{\tau}_d$, using the Lyapunov stability and backstepping method. 
		As we proceed to the controller design, the steps till \Cref{eqn:aV3} of \Cref{th:input_sat_ctrl_design} will essentially be repeated, hence, omitted from here and referred to appropriately. Moving on, consider a Lyapunov function candidate as in \Cref{eqn:aV3}, whose time differentiation is given in \Cref{eqn: V3_dot}.
		
		On differentiating \Cref{eqn: z3} with respect to time and substituting the value of $\dot{\pmb{\tau}}$ from \Cref{eqn:zeta_dot}, we obtain
		\begin{align}\label{eqn: z3dot_}
			\dot{\pmb{z}}_3 = \dot{\pmb{\tau}} -\dot{\pmb{\alpha}}_2=  (\pmb{I} - \pmb{G}_1)\pmb{\tau}_c - \pmb{\rho}_1\dfrac{\tau_{dM}}{\tau_M}\pmb{\zeta} -\dot{\pmb{\alpha}}_2.
		\end{align}
		Now, defining another tracking variable, $\pmb{z}_4 \coloneq \pmb{\tau}_c - \pmb{\alpha}_3$, where $\pmb{\alpha}_3$ is a stabilising function to be designed, we can write \Cref{eqn: z3dot_} as
		\begin{align} \label{eqn: z3dot}
			\dot{\pmb{z}}_3 = (\pmb{I} - \pmb{G}_1)(\pmb{z}_4 + \pmb{\alpha}_3) - \pmb{\rho}_1\dfrac{\tau_{dM}}{\tau_M}\pmb{\zeta} -\dot{\pmb{\alpha}}_2.
		\end{align}
		By choosing the stabilizing function $\pmb{\alpha_3}$ as
		\begin{equation}\label{eqn:alpha3}
			\pmb{\alpha}_3 = (\pmb{I}-\pmb{G}_1)^{-1} \left(\pmb{\rho}_1\dfrac{\tau_{dM}}{\tau_M}\pmb{\zeta} + \dot{\pmb{\alpha}}_2 - \pmb{K}_3 \pmb{z}_3 -\pmb{z}_2\right) ,
		\end{equation}
		where $\pmb{K}_3 = \diag (\begin{bmatrix}
			k_{31} & k_{32} & k_{33}
		\end{bmatrix})$ and substitute in \Cref{eqn: z3dot}, we get,
		\begin{align}\label{eqn:z3dot_1}
			\dot{\pmb{z}}_3 = (\pmb{I}-\pmb{G}_1)\pmb{z}_4-\pmb{K}_3 \pmb{z}_3 - \pmb{z}_2.
		\end{align}
		Finally, on substituting the value of $\dot{\pmb{z}}_3$ from \Cref{eqn:z3dot_1} in the expression of $\dot{\mathcal{V}}_3$ given by \Cref{eqn: V3_dot}, we get
		\begin{align}
			\nonumber \dot{\mathcal{V}}_3 & = -\left( \pmb{z}_1^\top \pmb{K}_1 \pmb{z}_1 + \pmb{z}_2^\top \pmb{K}_2 \pmb{z}_2 + \pmb{z}_3^\top \pmb{K}_3 \pmb{z}_3 + \pmb{b}_e^\top\pmb{K}_0\pmb{b}_e
			\right) + \pmb{z}_3^\top(\pmb{I}-\pmb{G}_1)\pmb{z}_4 \\
			& \quad + \varepsilon_1 \pmb{b}_e^\top\pmb{b}_e + \dfrac{1}{4\varepsilon_1}\dot{\pmb{b}}^\top \dot{\pmb{b}} 
			+ \varepsilon_2 \pmb{z}_2^\top\pmb{z}_2 + \dfrac{1}{4\varepsilon_2}\pmb{b}_e^\top \pmb{b}_e.\label{eqn: V3_dot_final}
		\end{align}
		The cross term $\pmb{z}_3^\top(\pmb{I}-\pmb{G}_1)\pmb{z}_4$ in \Cref{eqn: V3_dot_final} will be canceled subsequently by considering the Lyapunov function candidate as 
		\begin{equation}\label{eqn:V4}
			\mathcal{V}_4 = \mathcal{V}_3 + \dfrac{1}{2}\pmb{z}_4^\top\pmb{z}_4.
		\end{equation}
		Taking differentiation of \Cref{eqn:V4}, with respect to time, we obtain $ \dot{\mathcal{V}}_4 = \dot{\mathcal{V}}_3 + \pmb{z}_4^\top\dot{\pmb{z}}_4.$
		% \begin{equation}\label{eqn:V4_dot}
			%     \dot{\mathcal{V}}_4 = \dot{\mathcal{V}}_3 + \pmb{z}_4^\top\dot{\pmb{z}}_4
			% \end{equation}
		Now, on differentiating $\pmb{z}_4$ and using \Cref{eqn:tauc_dot}, we get
		\begin{equation}\label{eqn:z4_dot}
			\dot{\pmb{z}}_4 = \dot{\pmb{\tau}}_c - \dot{\pmb{\alpha}}_3 = (\pmb{I}-\pmb{G}_2)\pmb{\tau}_d - \pmb{\rho}_2 \pmb{\tau}_c - \dot{\pmb{\alpha}}_3
		\end{equation}
		% Using, \Cref{eqn:xi_dot}, we get
		% \begin{equation}\label{eqn:z4_dot}
			%     \dot{\pmb{z}}_4 = (\pmb{I}-\pmb{G}_2)\pmb{\tau}_d - \pmb{\rho}_2 \pmb{\tau}_c - \dot{\pmb{\alpha}}_3,
			% \end{equation}
		Thus, using \Cref{eqn:z4_dot,eqn: aV3_dot_final}, we can write the derivative of Lyapunov function candidate, $\mathcal{V}_4$ as,
		\begin{align}
			\nonumber \dot{\mathcal{V}}_4 &= \dot{\mathcal{V}}_3 + \pmb{z}_4^\top\left[ (\pmb{I}-\pmb{G}_2)\pmb{\tau}_d - \pmb{\rho}_2 \pmb{\tau}_c - \dot{\pmb{\alpha}}_3\right]\\
			&= -\left( \pmb{z}_1^\top \pmb{K}_1 \pmb{z}_1 + \pmb{z}_2^\top \pmb{K}_2 \pmb{z}_2 + \pmb{z}_3^\top \pmb{K}_3 \pmb{z}_3 + \pmb{b}_e^\top\pmb{K}_0\pmb{b}_e
			\right)  + \pmb{z}_2^\top\pmb{b}_e  + \pmb{b}_e^\top\dot{\pmb{b}} 
			+\pmb{z}_4^\top\left[ (\pmb{I}-\pmb{G}_2)\pmb{\tau}_d - \pmb{\rho}_2 \pmb{\tau}_c - \dot{\pmb{\alpha}}_3\right].\label{eqn:V4_dot1}
		\end{align}
		By choosing the model's control input $\pmb{\tau}_d$ as,
		\begin{equation}\label{eqn:tau_d}
			\pmb{\tau}_d = (\pmb{I}-\pmb{G}_2)^{-1} \left[\pmb{\rho}_2 \pmb{\tau}_c + \dot{\pmb{\alpha}}_3 - \pmb{K}_4\pmb{z}_4 - (\pmb{I}-\pmb{G}_1)\pmb{z}_3
			\right],
		\end{equation}
		and substituting in \Cref{eqn:V4_dot1}, we obtain
		\begin{align}
			\nonumber \dot{\mathcal{V}}_4 &= -\left( \pmb{z}_1^\top \pmb{K}_1 \pmb{z}_1 + \pmb{z}_2^\top \pmb{K}_2 \pmb{z}_2 + \pmb{z}_3^\top \pmb{K}_3 \pmb{z}_3 + \pmb{z}_4^\top \pmb{K}_4 \pmb{z}_4 + \pmb{b}_e^\top\pmb{K}_0\pmb{b}_e
			\right) \\
			& \quad + \varepsilon_1 \pmb{b}_e^\top\pmb{b}_e + \dfrac{1}{4\varepsilon_1}\dot{\pmb{b}}^\top \dot{\pmb{b}}  
			+ \varepsilon_2 \pmb{z}_2^\top\pmb{z}_2 + \dfrac{1}{4\varepsilon_2}\pmb{b}_e^\top \pmb{b}_e.\label{eqn:V4_dot2}
		\end{align}
		Now, using \Cref{eqn:z2_M_z2_inequality} and eigenvalue property of positive definite matrix, \Cref{eqn:V4_dot2} 
		can be written as 
		\begin{align}
			\nonumber \dot{\mathcal{V}}_4 &  \leq \dfrac{b_M^2}{4\varepsilon_1}  -2\left[ \lambda_{\rm{min}}(\pmb{K}_1) \dfrac{1}{2}\pmb{z}_1^\top\pmb{z}_1  + 
			\left(\lambda_{\rm{min}}(\pmb{K}_2\pmb{M}^{-1}) -\varepsilon_2\lambda_{\rm{max}}(\pmb{M}^{-1})\right)\dfrac{1}{2}\pmb{z}_2^\top\pmb{M}\pmb{z}_2  \right. \\ 
			&\left. \quad  + \lambda_{\rm{min}}(\pmb{K}_3) \dfrac{1}{2}\pmb{z}_3^\top\pmb{z}_3 +  \lambda_{\rm{min}}(\pmb{K}_4) \dfrac{1}{2}\pmb{z}_4^\top\pmb{z}_4 +
			\left(\lambda_{\rm{min}}(\pmb{K}_0)-\varepsilon_1 - \dfrac{1}{4\varepsilon_2}\right) \dfrac{1}{2}\pmb{b}_e^\top \pmb{b}_e
			\right], \label{eqn:V4_dot_inequality}
		\end{align}
		where $\left(\lambda_{\rm{min}}(\pmb{K}_2) -\varepsilon_2\lambda_{\rm{max}}(\pmb{M}^{-1})\right)  > 0, \quad 
		\left(\lambda_{\rm{min}}(\pmb{K}_0)-\varepsilon_1 - \dfrac{1}{4\varepsilon_2}\right) >0.$
		Since the coefficients of $\pmb{z}_i^\top\pmb{z}_i~~ (i \in \{1,2,3\})$ and $\pmb{b}_e^\top\pmb{b}_e$ are all positive, all of it can be replaced with the minimum of these coefficients in the inequality (\ref{eqn:V4_dot_inequality}). So, for clear representation, we proceed by defining
		\begin{align*}
			a_2\coloneq \rm{min}\left\{ \lambda_{\rm{min}}(\pmb{K}_1),~ 
			\left(\lambda_{\rm{min}}(\pmb{K}_2\pmb{M}^{-1}) -\varepsilon_2\lambda_{\rm{max}}(\pmb{M}^{-1})\right),~
			\lambda_{\rm{min}}(\pmb{K}_3),~\lambda_{\rm{min}}(\pmb{K}_4),~ \right. \\ \left. 
			\left(\lambda_{\rm{min}}(\pmb{K}_0)-\varepsilon_1 - \dfrac{1}{4\varepsilon_2}\right)
			\right\}.
		\end{align*}
		Finally, we can write
		\begin{equation}
			\label{eqn:V_4_dot_compact} \dot{\mathcal{V}}_4(t)  \leq -2\left( \dfrac{1}{2}\pmb{z}_1^\top\pmb{z}_1 + \dfrac{1}{2}\pmb{z}_2^\top\pmb{z}_2 + 
			\dfrac{1}{2}\pmb{z}_3^\top\pmb{z}_3 + \dfrac{1}{2}\pmb{z}_4^\top\pmb{z}_4 + \dfrac{1}{2}\pmb{b}_e^\top \pmb{b}_e
			\right) + \dfrac{b_M^2}{4\varepsilon_1}
			\leq -2a_2 \mathcal{V}_4(t) + \dfrac{b_M^2}{4\varepsilon_1}.
		\end{equation}
		On integrating \Cref{eqn:V_4_dot_compact}, we get the solution as 
		\begin{equation}\label{eqn:V4_integrated}
			\mathcal{V}_4(t) \leq \dfrac{b_M^2}{8a_1\varepsilon_1} + \left( \mathcal{V}_4(0) - \dfrac{b_M^2}{8a_2\varepsilon_1}\right)e^{-2a_2t}.
		\end{equation}
		From \Cref{eqn:V4_integrated}, it can be concluded that the $\mathcal{V}_4$ is globally ultimately bounded and converges exponentially to a ball of radius $\dfrac{b_M^2}{8a_2\varepsilon_1}$. 
		% \textcolor{red}{conclusion about other states} 
		% Since the gains $k_{1i}$, $k_{2i}$, and $ k_{3i}$ are positive real numbers, the square of a real number is always positive.
		% Hence, $\dot{\mathcal{V}}_3$ is negative definite, that is, $\dot{\mathcal{V}}_3 < 0$. From $\dot{\mathcal{V}}_3 < 0$, it follows that $\mathcal{V}_3(t) < \mathcal{V}_3(0)$. 
		Hence, all the error signals $z_{1i}(t),\, z_{2i}(t),\, z_{3i}(t),$ and $ z_{4i}(t) ~\forall \,i \in (\{1,2,3,4\})$ remains bounded. This completes the proof. 
	\end{proof}
	To support the analytical results, we performed simulations presented in the following section. 
	\section{Simulation Results}\label{sec:simulation results}
	In this section, we validate the proposed control strategy by conducting a variety of numerical simulations using MATLAB\textsuperscript{\textregistered}.
	% In this section, we perform the validation of the various controllers developed through MATLAB\textsuperscript{\textregistered } simulations. 
	The trajectory tracking performance of the proposed controllers for the USV exposed to unknown bounded time-varying disturbances is assessed. This work designs controllers for addressing two realistic challenges pertaining to actuator saturation. One of which incorporates the asymmetric bounds on the actuators' input magnitude and the corresponding controller is given by \Cref{eqn: tau_c,eqn: actuator}. The second one incorporates the constraint on the input magnitude as well as the rate, and the derived control for this case is described by \Cref{eqn:zeta_dot,eqn:tau_ci_dot,eqn:tau_d}.
	% We performed various numerical simulations to verify the efficacy of the 
	% First we present the results proposed control algorithms obtained by combining \Cref{eqn: tau_c,eqn: actuator} for addressing the actuator input saturation while \Cref{eqn:zeta_dot,eqn:tau_ci_dot,eqn:tau_d}.
	To test the performance of the designed strategies, we consider a scaled model of CyberShip II for simulations.
	% The parameters of the CyberShip II model were used for performing simulations. 
	The associated parameters of the CyberShip II were taken from \cite{skjetne2004modeling}.
	The model is fully actuated and has the dimensions $(1.255\,\rm{m} \times 0.29\,\rm{m})$. \Cref{tab:paramstable} tabulates all the relevant parameters of the model ship.
	\begin{table}[htpb]
		\centering
		\begin{tabular}{|c|c||c|c||c|c||c|c||c|c||c|c| } 
			\hline
			$\rm{m}$ & 23.800 & $X_{\dot{u}}$ &-2.0 &
			$X_u$ & -0.72253 & $Y_{v}$ &-2.0  & $Y_{|r|v}$ & -0.805 & $N_{|r|v}$ &0.130 \\ 
			$I_z $& 1.760 & $Y_{\dot{v}}$ &-10.0 &
			$X_{|u|u}$ & -1.32742 & $Y_{|v|v}$ &-36.47287 & $Y_{r}$ & -7.250 & $N_{r}$ &-1.900\\
			$x_g$ & 0.046 & $Y_{\dot{r}}$ &-0.0 &
			$X_{uuu}$ & -5.86643 & $N_{v}$ &0.03130 & $Y_{|v|r}$ & -0.845 & $N_{|v|r}$ &0.080\\
			$N_{\dot{v}}$ &-0.0&$N_{\dot{r}}$ &-0.0&
			&  & $N_{|v|v}$ &3.95645 & $Y_{|r|r}$ & -3.450 & $N_{|r|r}$ &-0.750\\
			\hline
		\end{tabular}
		\caption{Model parameters of the CyberShip II \cite{skjetne2004modeling}.}
		\label{tab:paramstable}
	\end{table}
	The various controller parameters used for the simulations are given in \Cref{tab:controller_params}.
	We now demonstrate the trajectory tracking performance of the USV by considering simple as well as more complicated desired trajectories.
	\begin{table}[h]
		\centering
		\begin{tabular}{|c|c|c|}
			\hline
			\multirow{2}{*}{\textbf{Parameters}} & \multicolumn{2}{c|}{\textbf{Values}} \\ \cline{2-3}
			& Asymmetric input saturation & Input magnitude and rate\\ \hline
			$\pmb{K}_1$ &  $\diag[4, 3, 0.5]$ & $\diag[0.02, 0.02, 0.05]$ \\  \hline
			$\pmb{K}_2$ &  $\diag[2, 3, 0.5]$ & $\diag[2, 3, 5]$  \\  \hline
			$\pmb{K}_3$ &  $\diag[2, 3, 0.5]$ & $\diag[2, 1, 5]$ \\  \hline
			$\pmb{K}_4$ &   & $\diag[0.2, 0.5, 0.1]$ \\  \hline
			$\pmb{\rho}$ &   $\diag[0.5, 0.5, 0.5]$ & \\  \hline
			$\pmb{\rho}_1$ &   & $\diag[0.2, 0.2, 0.2]$ \\  \hline
			$\pmb{\rho}_2$ &   & $\diag[2, 2, 2]$ \\  \hline
			$\tau_{cM}$ &  100 &  \\  \hline
			$\tau_{dM}$ &   & $\diag[4,4,4]$  \\  \hline
			$n$ &  2 &  2\\  \hline
			$\pmb{\tau}_m$ &   $(-4,-4,-3)$& $(-5,-5,-5)$  \\  \hline
			$\pmb{\tau}_M$ &   $(5,4.5,4)$& $(5,5,5)$  \\  \hline
		\end{tabular}
		\caption{Controller parameters.}
		\label{tab:controller_params}
	\end{table}
	We first discuss the results from tracking when an asymmetric input saturation constraint is considered. Secondly, the tracking in the presence of constraints on input magnitude as well as on their rate is illustrated.
	%==========asym input saturation constraint result===============
	\subsection{Tracking with asymmetric input saturation constraint}
	To illustrate the tracking performance of the designed algorithm given by \Cref{eqn: tau_c,eqn: actuator} under the asymmetric actuator input constraint, we consider an elliptical trajectory as well as an 8-shaped trajectory. 
	\subsubsection{Tracking of an elliptical trajectory}
	% To validate the proposed controller given by \Cref{eqn: tau_c}, MATLAB \textsuperscript{\textregistered} simulations have been performed. 
	At first,  we consider a simple, that is, an elliptical reference trajectory, which is given by $\pmb{\eta}_d = \left[
	x_d ,\, y_d ,\, \psi_d
	\right]^\top$, where
	\begin{equation} \label{eqn: ellipse}
		x_d = 4 \sin(0.02t),\quad y_d = 2.5(1- \cos(0.02t)), \quad \psi_{d} = 0.02 \sin(0.02t),
	\end{equation}
	and $\pmb{\nu}_{d} = \pmb{J}({\psi}_d)^\top \dot{\pmb{\eta}}_d$.  
	To assess the controller’s performance against different initial conditions, the USV is placed at three distinct positions and orientations, labeled as P1, P2, and P3 in the simulation figures.
	In the case P1, the vehicle starts at $-1$ m North and $0$ m East with a heading angle of $0.01$~rad. In the second case, P2, the USV is placed at $1$~m North and $1$~m West with the body axis aligned at $0.01$~rad from the North. Lastly, in the case of P3, the USV starts the mission from $1$ m South and $1$ m East with the same heading angle, $0.01$ rad. In all of the scenarios, the vehicle starts with rest, that is, $\pmb{\nu}(0) = \left[
	u(0),\, v(0),\, r(0)
	\right]^\top =\left[
	0\,\rm{m/s}, ~ 0\,\rm{m/s},~ 0\,\rm{rad/s}
	\right]^\top.$
	% $$\pmb{\eta}(0) = \begin{bmatrix}
		%     x(0)& y(0)& \psi(0)
		% \end{bmatrix}^\top = \begin{bmatrix}
		%     -1\,\rm{m}& 0\,\rm{m}& 0.01\,\rm{rad}]
		% \end{bmatrix}^\top$$ 
	\begin{figure*}[h!]
		\centering
		\begin{subfigure}{0.45\linewidth}
			\centering
			\includegraphics[width=\linewidth]{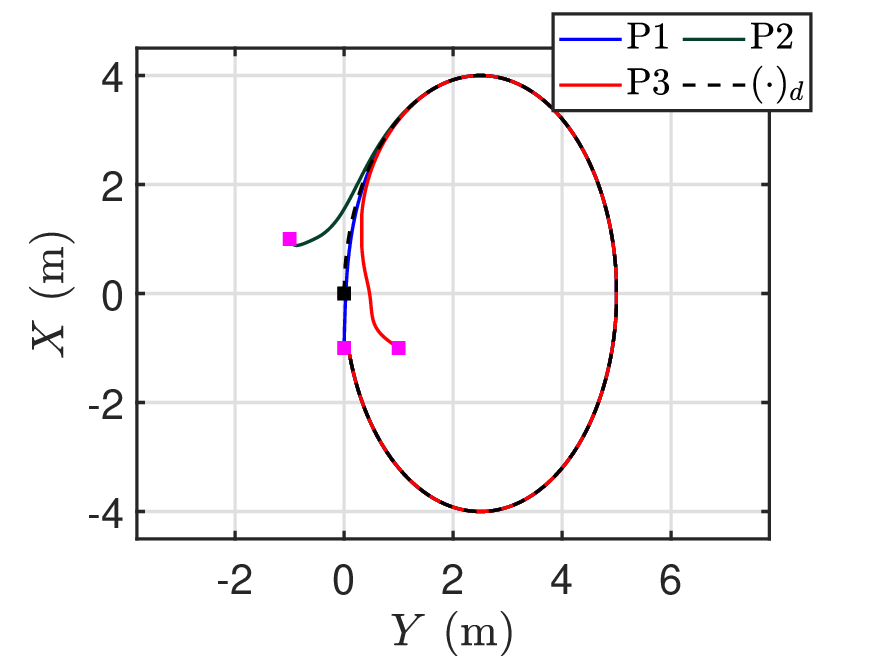}   
			\caption{Actual and desired trajectories.}
			\label{figac:xy_plot}
		\end{subfigure}%
		\begin{subfigure}{0.45\linewidth}
			\centering
			\includegraphics[width=\linewidth]{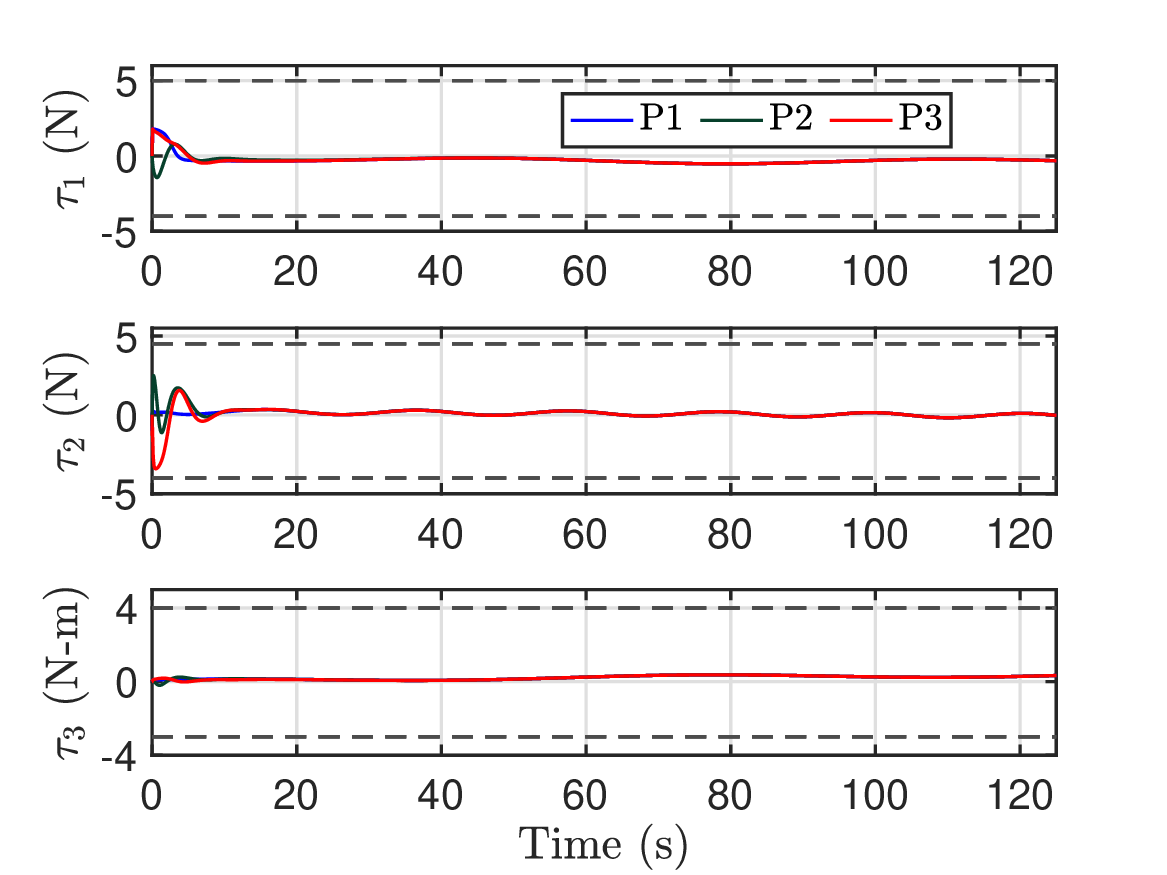}
			\caption{Control inputs.}
			\label{figac:tau}
		\end{subfigure}
		\begin{subfigure}{0.45\linewidth}
			\centering
			\includegraphics[width=\linewidth]{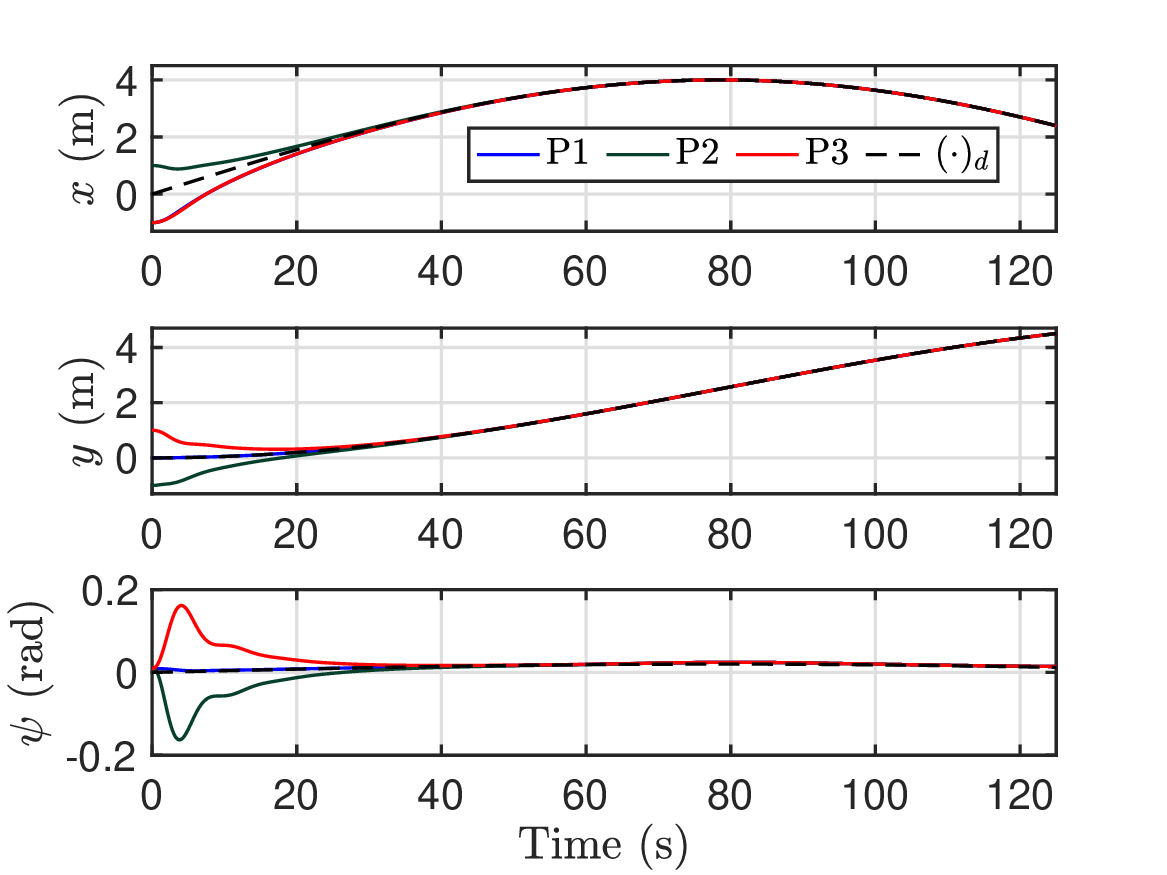}
			\caption{Position and heading of the vessel.}
			\label{figac:eta}
		\end{subfigure}%
		\begin{subfigure}{0.45\linewidth}
			\centering
			\includegraphics[width=\linewidth]{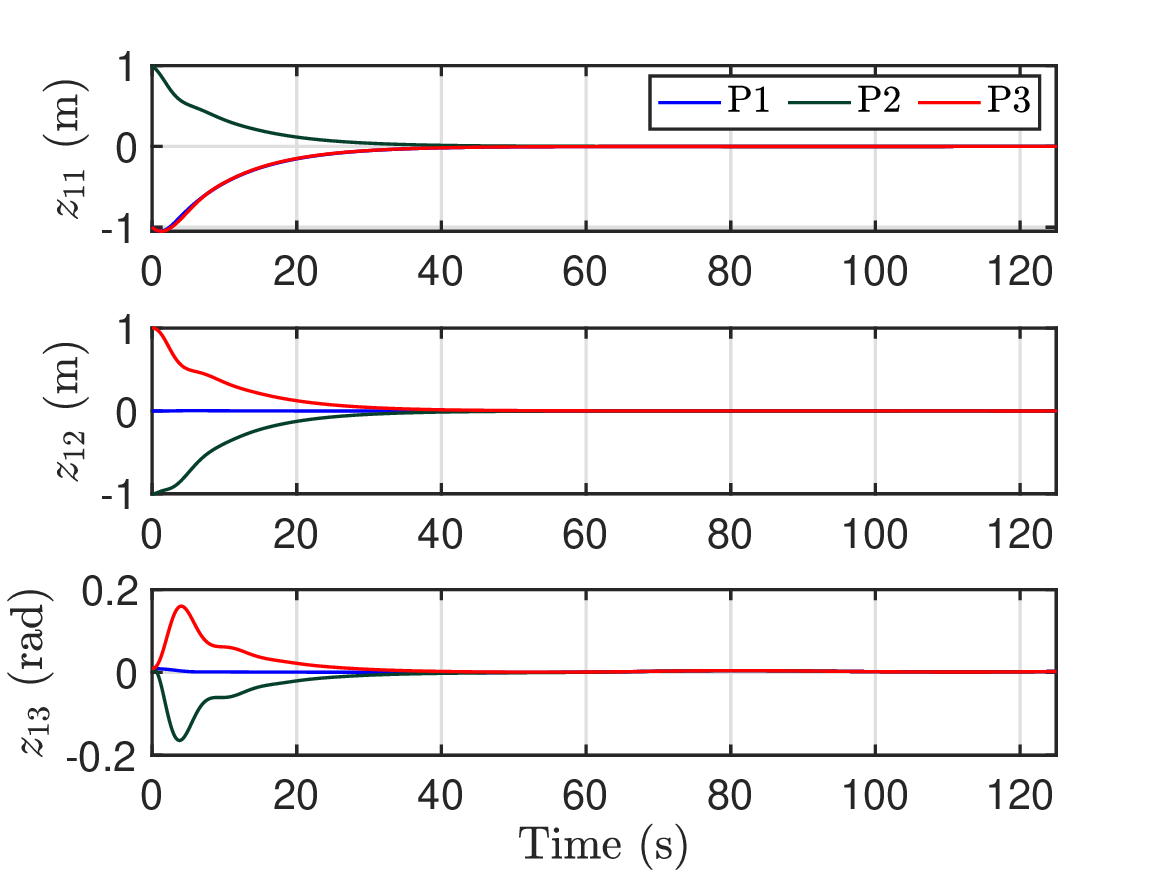}
			\caption{Errors in position and heading.}
			\label{figac:e1}
		\end{subfigure}
		\begin{subfigure}{0.45\linewidth}
			\centering
			\includegraphics[width=\linewidth]{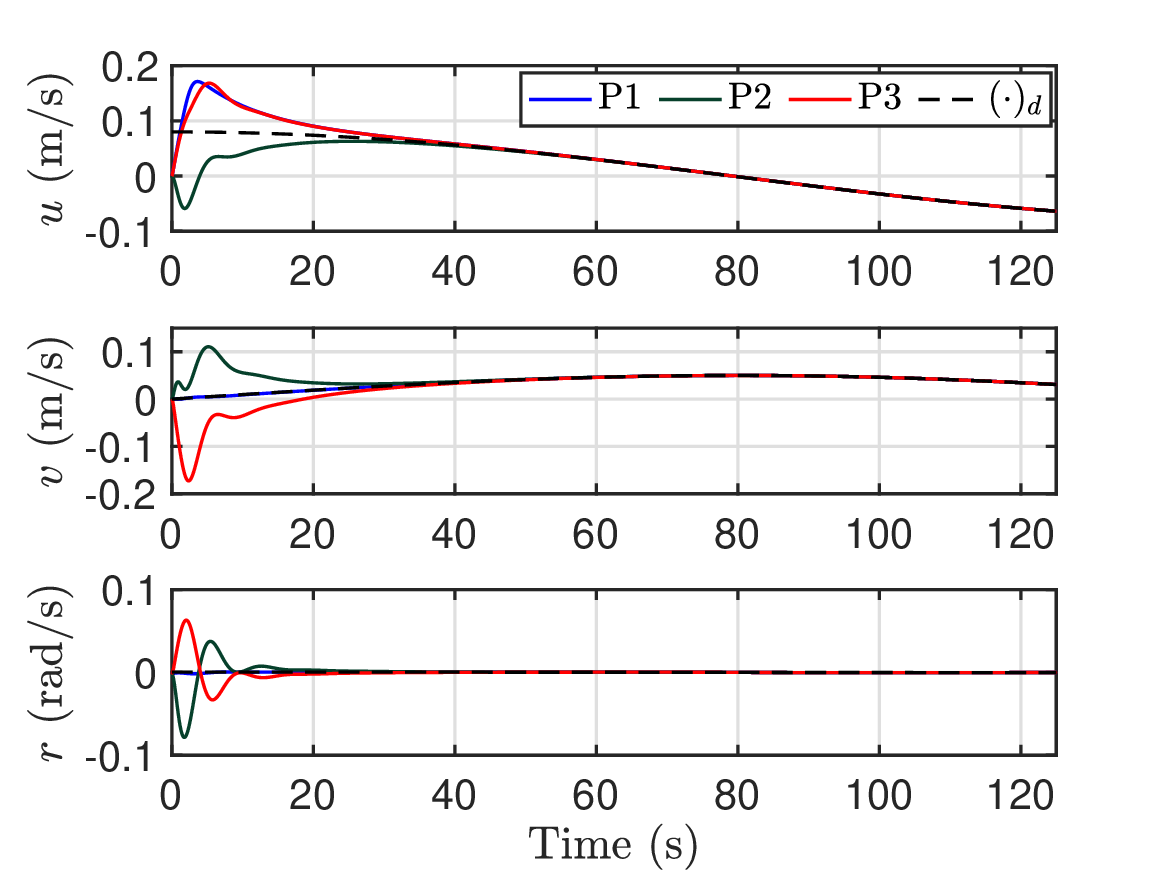}
			\caption{Surge, sway, and yaw rate.}
			\label{figac:nu}
		\end{subfigure}%
		\begin{subfigure}{0.45\linewidth}
			\centering
			\includegraphics[width=\linewidth]{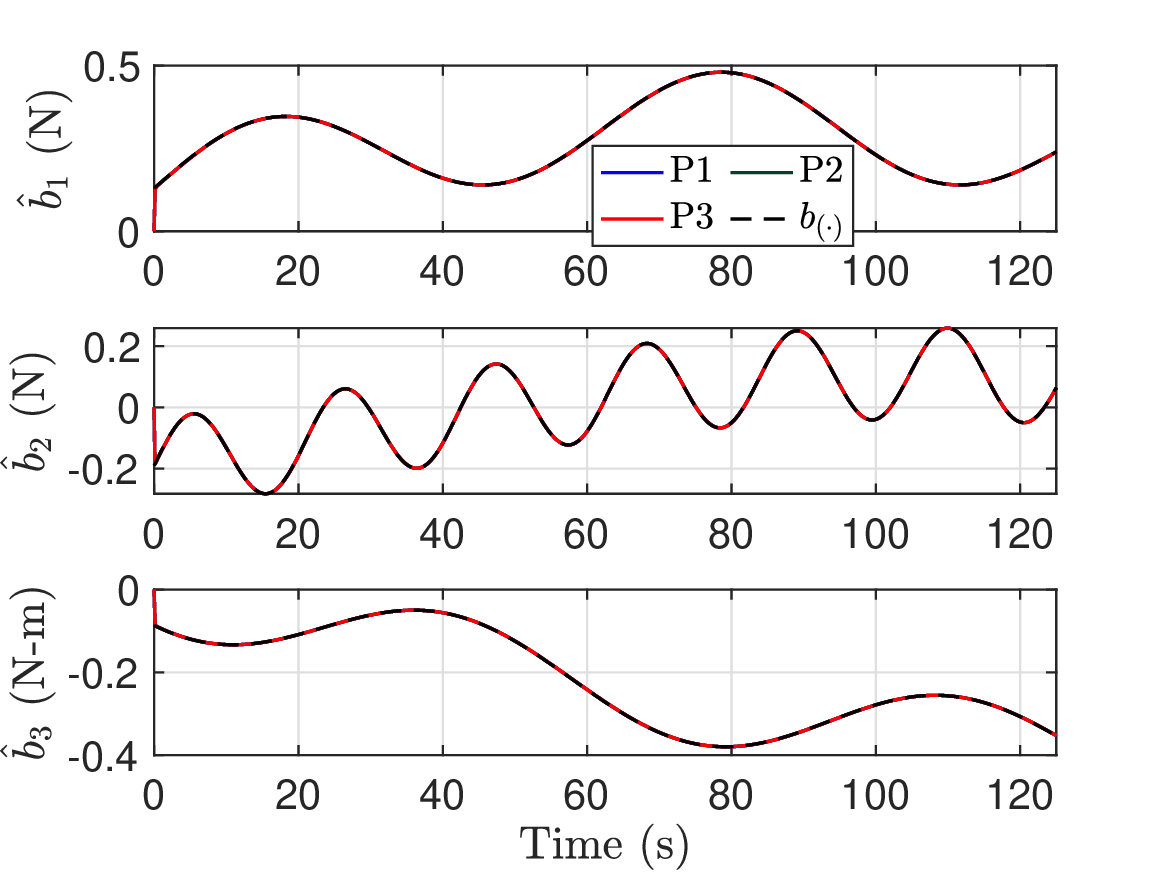}
			\caption{Estimation of disturbance vector.}
			\label{figac:d}
		\end{subfigure}
		\caption{Performance validation of proposed controller for a USV with elliptical trajectory under asymmetric input magnitude constraint.}
		\label{figcase: ac}
	\end{figure*}

	Performance results related to the elliptical trajectory tracking are depicted in \Cref{figcase: ac}, showcasing the tracking, input requirements, the time evolution of errors in position and heading, and the estimate of the disturbance vector.  \cref{figac:xy_plot} shows that the USV tracks the trajectory in all cases P1, P2, and P3, as evident from \Cref{figac:eta,figac:e1,figac:nu}. The errors in position and heading, initially large due to the initial placement of USV, near zero with time as seen from \Cref{figac:e1}. \cref{figac:tau} displays the control input demand, which is initially high but decreases as the error diminishes. Also, the actuator input remains within specified bounds throughout all cases P1, P2, and P3. It is to be noted that the actuator bounds are not symmetric in all three axes. The disturbance observer, as depicted in \Cref{figac:d}, is also able to estimate the time-varying external disturbances.  
	
	\subsubsection{Tracking of an 8-shaped trajectory}
	In this subsection, we test the controller's trajectory tracking performance with a more involved shape against three different initial conditions chosen at random. We considered an eight shaped reference trajectory given by 
	\begin{equation}\label{eqn: 8-shape}
		x_d = 4(\cos(0.05t)-1), \quad y_d = 2.5\sin(0.1t), \quad 
		\psi_{d} = \pi\sin(0.02t),
	\end{equation}
	and $\pmb{\nu}_{d} = \pmb{J}(\pmb{\eta})^T \dot{\pmb{\eta}}$. 
	The initial conditions for the case P1 are  $\pmb{\eta}(0) = \left[
	x(0),\, y(0),\, \psi(0)
	\right]^\top = \left[
	-1\,\rm{m},\, 0\,\rm{m},\, 0.01\,\rm{rad}
	\right]^\top$
	and $\pmb{\nu}(0) = \left[
	u(0),\, v(0),\, r(0)
	\right]^\top =\left[
	0\,\rm{m/s},\, 0\,\rm{m/s},\, 0\,\rm{rad/s}
	\right]^\top.$ In the second case P2, the USV is placed at 0.5~m North and 0.5~m West with a heading of 0.01~rad. In the third case P3, the USV is placed at 1~m South and 1~m East with the same heading as the previous case. 
	
	\begin{figure*}
		\centering
		\begin{subfigure}{0.45\linewidth}
			\centering
			\includegraphics[width=\linewidth]{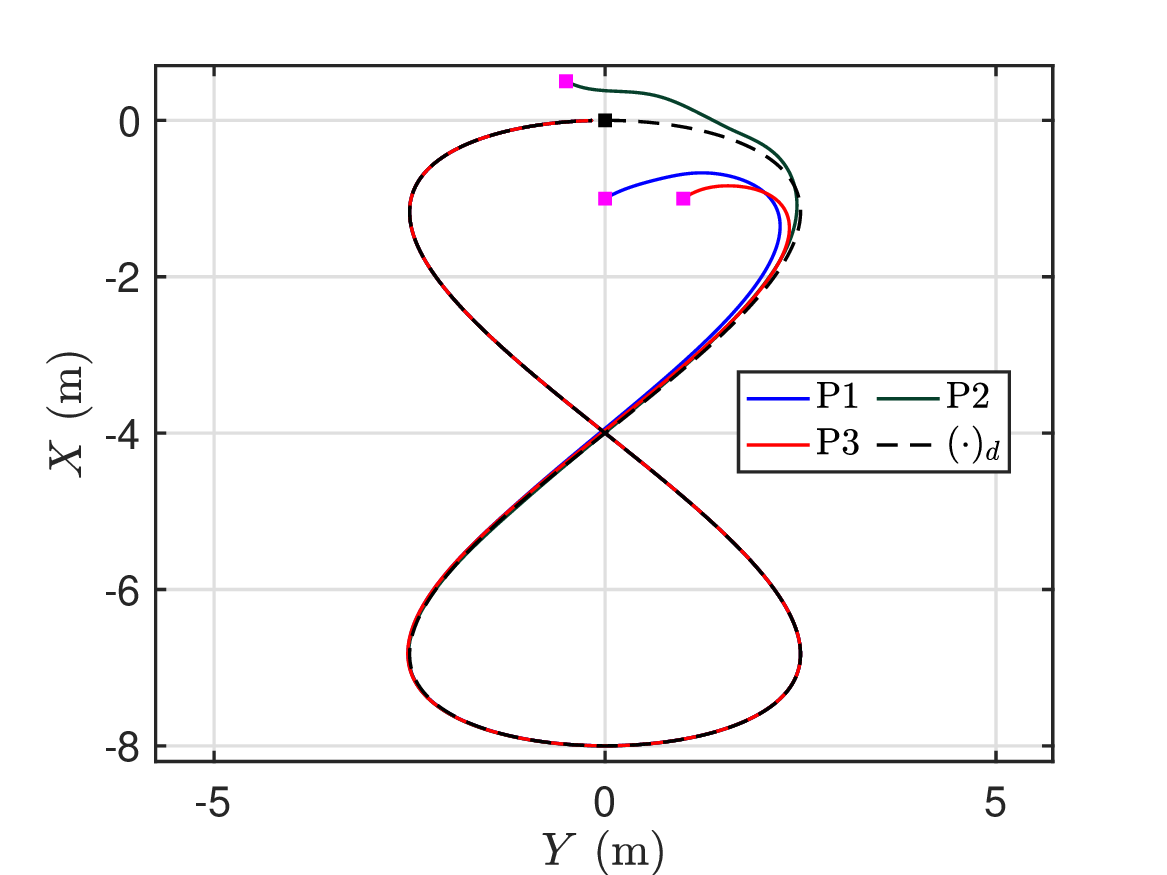}   
			\caption{Actual and desired trajectories.}
			\label{figa8:xy_plot}
		\end{subfigure}%
		\begin{subfigure}{0.45\linewidth}
			% \centering
			\includegraphics[width=\linewidth]{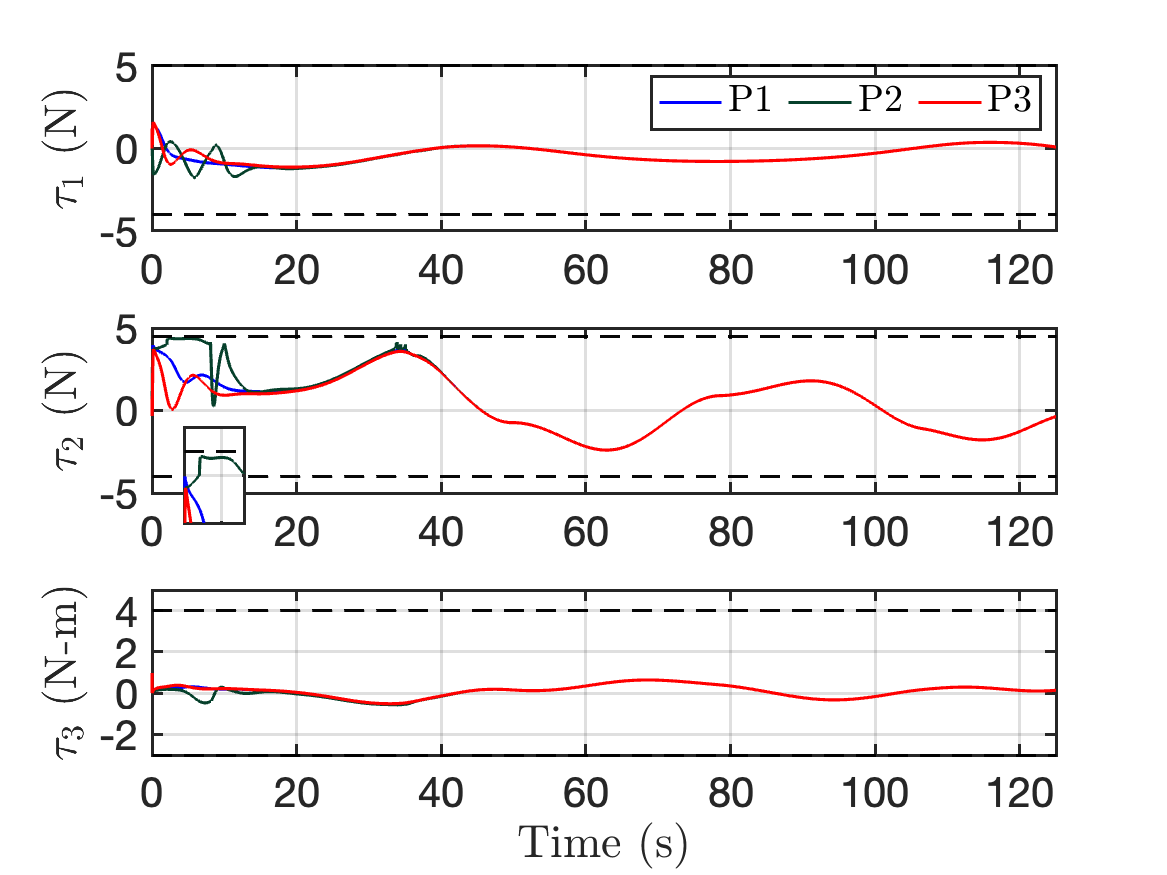}
			\caption{Control inputs.}
			\label{figa8:tau}
		\end{subfigure}
		\begin{subfigure}{0.45\linewidth}
			\centering
			\includegraphics[width=\linewidth]{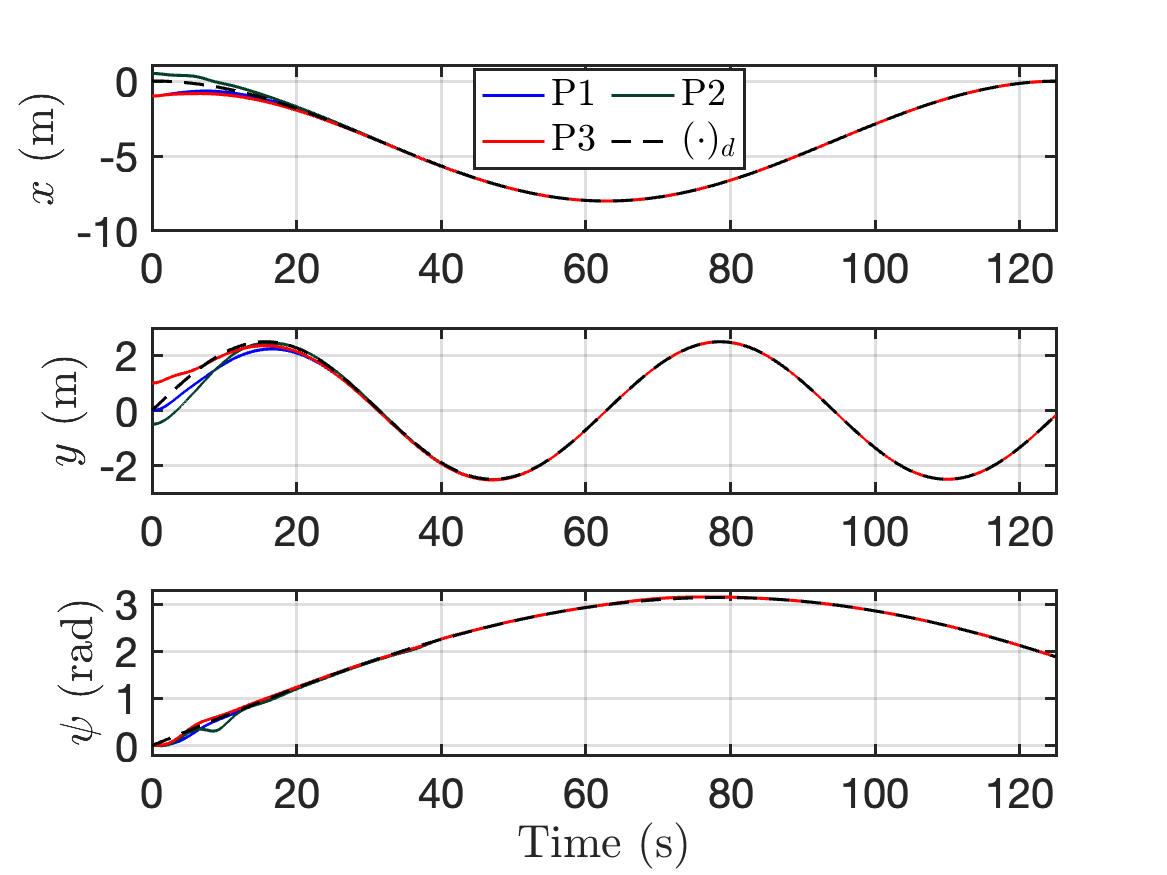}
			\caption{Position and heading of the USV.}
			\label{figa8:eta}
		\end{subfigure}%
		\begin{subfigure}{0.45\linewidth}
			% \centering
			\includegraphics[width=\linewidth]{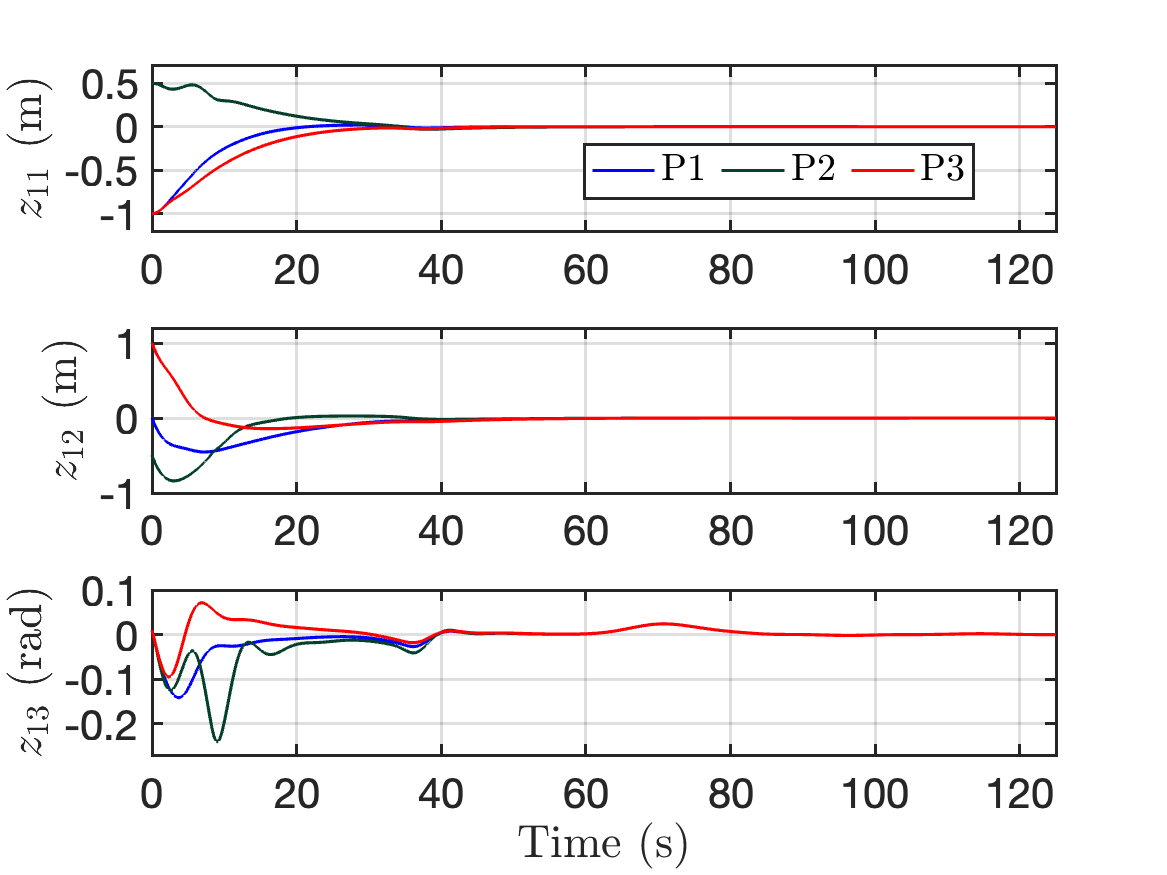}
			\caption{Error in position and heading.}
			\label{figa8:e1}
		\end{subfigure}
		\begin{subfigure}{0.45\linewidth}
			\centering
			\includegraphics[width=\linewidth]{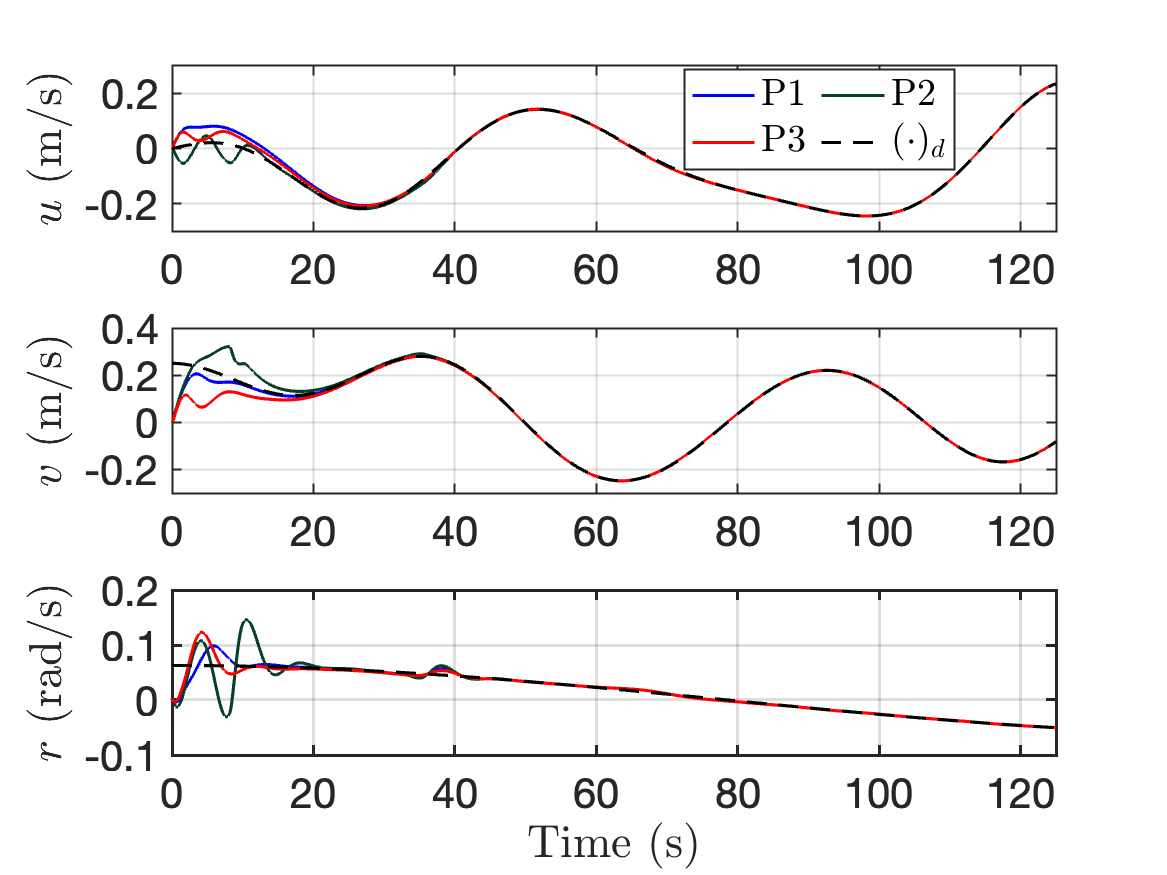}
			\caption{Surge, sway, and yaw rate.}
			\label{figa8:nu}
		\end{subfigure}%
		\begin{subfigure}{0.45\linewidth}
			\centering
			\includegraphics[width=\linewidth]{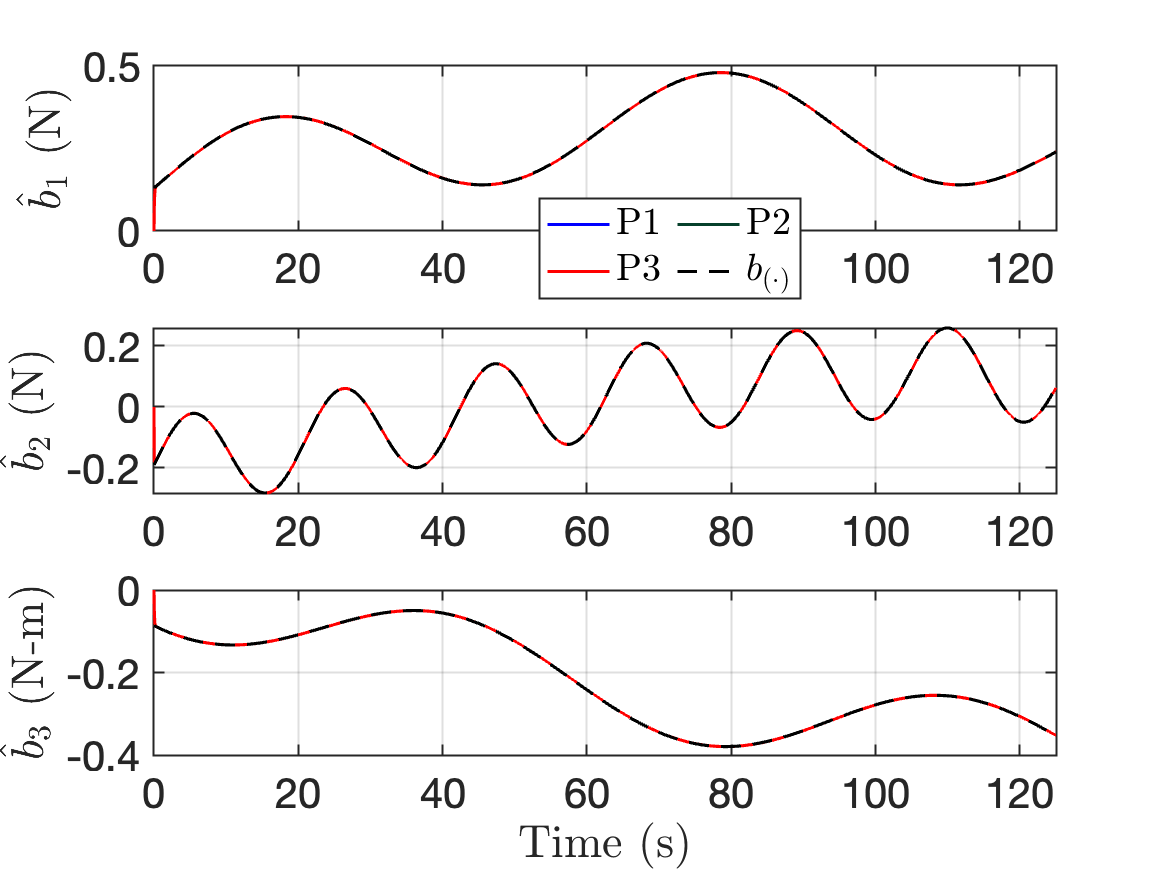}
			\caption{Estimation of disturbance vector.}
			\label{figa8:d}
		\end{subfigure}
		\caption{Performance validation of proposed controller for a surface vessel with 8-shaped trajectory under asymmetric input magnitude constraint.}
		\label{figcase: a8}
	\end{figure*}
	The simulation results for this 8-shaped trajectory are shown in \Cref{figcase: a8}. Again, for this trajectory also, there is an excellent agreement between the desired trajectory and the state of the USV without violating actuators' bounds, even in the presence of unknown time-varying disturbances. \Cref{figa8:xy_plot} shows that the USV follows an 8-shaped reference trajectory. The time variation of the positions, heading, and body rates are also compared with reference in \Cref{figa8:eta,figa8:nu}. It can be inferred from \Cref{figa8:e1} that the errors in position and heading go to zeros despite large initial deviations. As evident from \Cref{figa8:tau}, the controller never demands an input beyond the actuators' bounds. Hence, asymmetric input saturation constraints are always satisfied. Also, one can observe from \Cref{eqn: 8-shape}, the $x_d(t)~ \mathrm{and}~ \psi_d(t)$ varies slowly as compared to $y_d(t)$, resulting in relatively more control demand for $\tau_2$ in \Cref{figa8:tau}. Also, \Cref{figa8:d} shows that the disturbance vector is being estimated pretty well.
	% Similar conclusions to the previous case apply here as well. 
	\begin{figure*}
		\centering
		\begin{subfigure}{0.33\linewidth}
			% \centering
			\includegraphics[width=\linewidth]{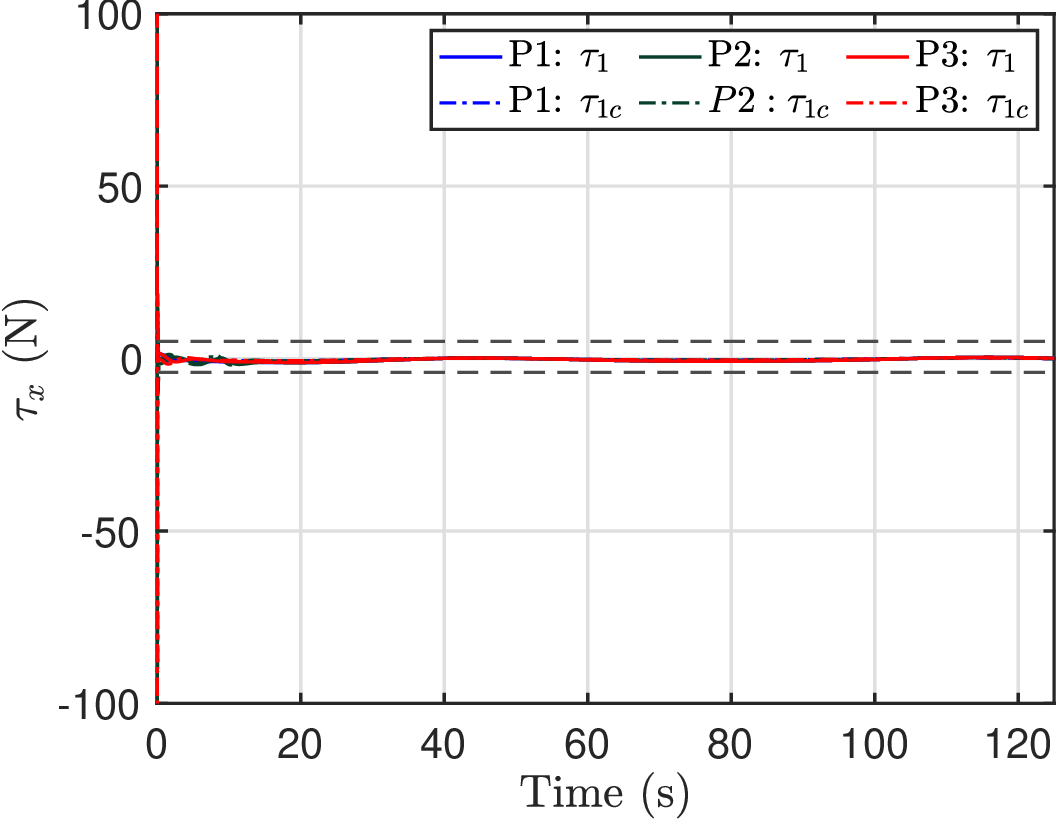}
			\caption{Effect  on $\pmb{\tau}_x$.}
			\label{figa8:tau_x}
		\end{subfigure}%
		\begin{subfigure}{0.33\linewidth}
			% \centering
			\includegraphics[width=\linewidth]{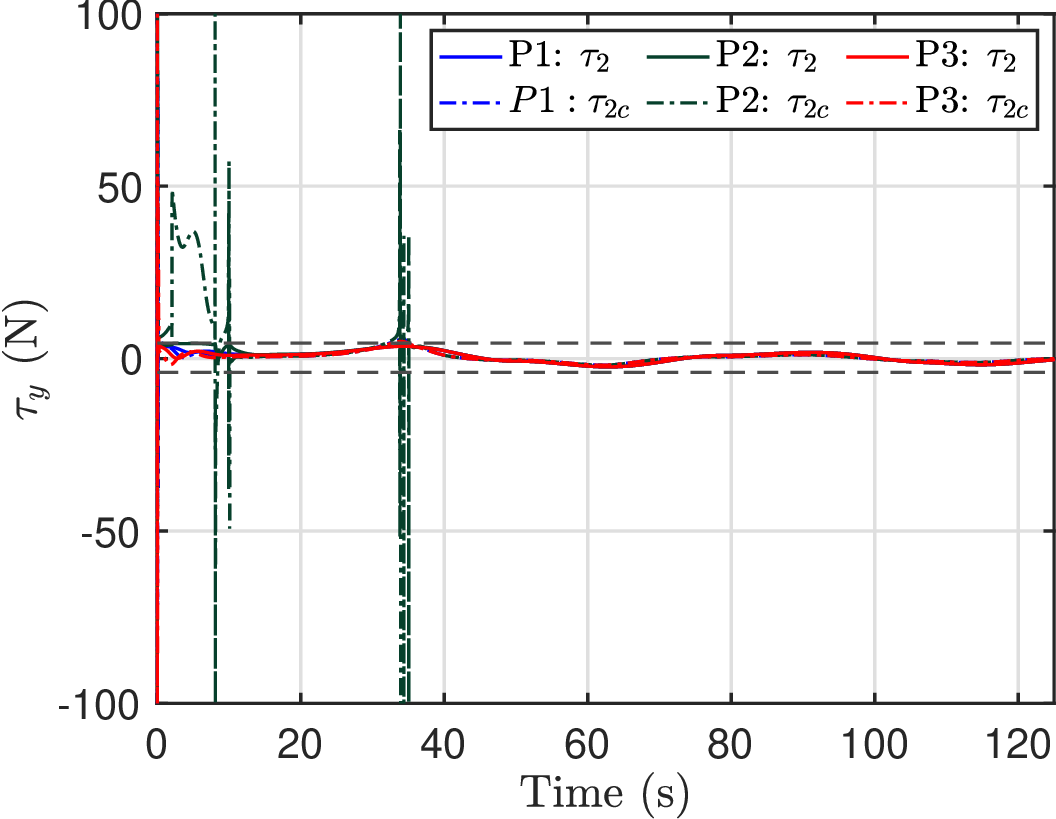}
			\caption{Effect on $\pmb{\tau}_y$.}
			\label{figa8:tau_y}
		\end{subfigure}
		\begin{subfigure}{0.33\linewidth}
			% \centering
			\includegraphics[width=\linewidth]{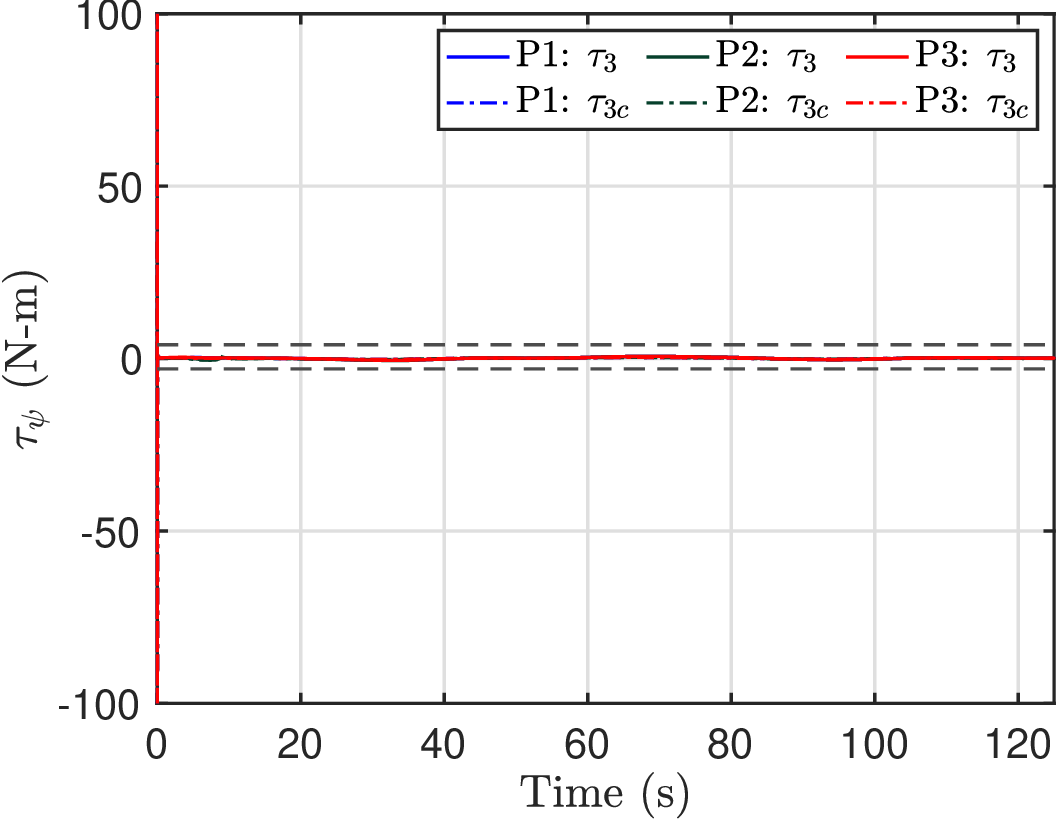}
			\caption{Effect on $\pmb{\tau}_\psi$.}
			\label{figa8:tau_psi}
		\end{subfigure}%
		\caption{Comparison of control input demand on applying the asymmetric input saturation model for 8-shaped trajectory.}
		\label{figcase: tauc_a8}
	\end{figure*}
	In \Cref{figcase: tauc_a8}, it is observed that, after incorporating the input saturation model, the overall control demand is less. It is also evident from \Cref{figa8:tau_y} that the control demand exceeds the actuator limits before applying the input saturation model. However, the proposed control law ensures that the control demand stays within permissible bounds while accurately tracking the trajectory.
	
	%%-------Comparison----------------------
	\subsubsection{Comparison across different methods}
	In this subsection, we performed a comparative study of the proposed controller with commonly applied approaches in the literature. The comparison is performed with three approaches. The first approach assumes ideal actuator conditions and disregards any bounds on the actuator while developing the controller. The second approach designs the controller, neglecting the bounds on the actuator, but applies the bounds on the controller output as per the actuators' abilities. The third approach is the one proposed in this work. These three approaches are indicated as ``Unbounded'', ``Adhoc'' and ``Proposed'', respectively, in the simulation figures. 
	\begin{figure*}
		\centering
		\begin{subfigure}{0.45\linewidth}
			\centering
			\includegraphics[width=\linewidth]{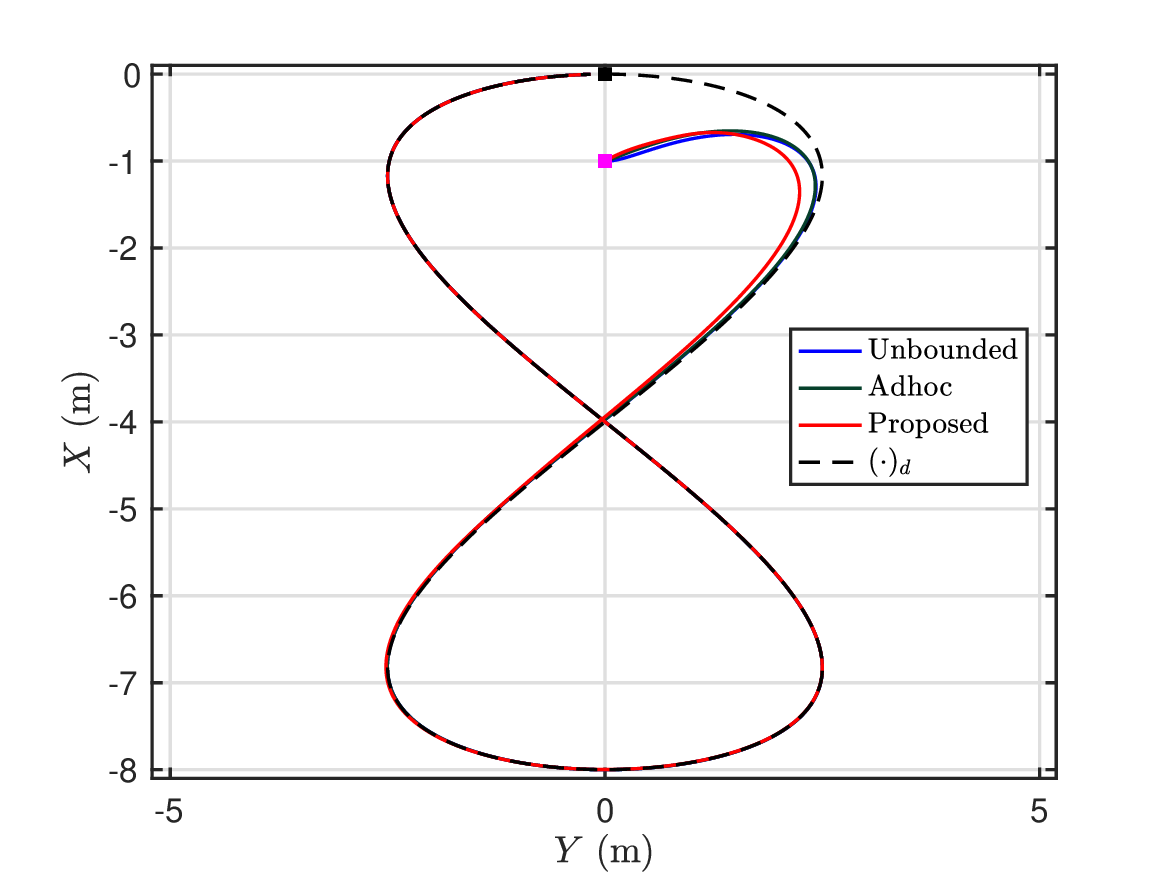}   
			\caption{Actual and desired trajectories.}
			\label{figa8comp:xy_plot}
		\end{subfigure}%
		\begin{subfigure}{0.45\linewidth}
			% \centering
			\includegraphics[width=\linewidth]{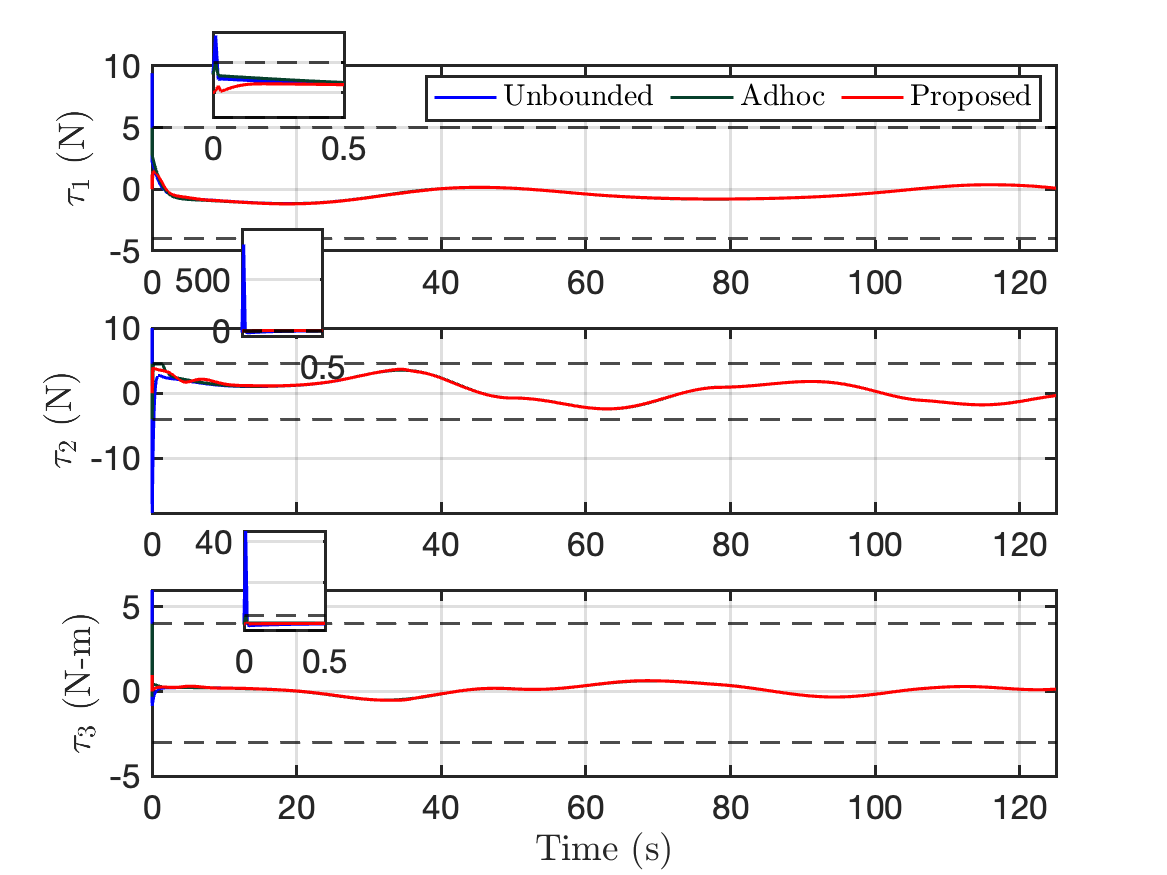}
			\caption{Control inputs.}
			\label{figa8comp:tau}
		\end{subfigure}
		\begin{subfigure}{0.45\linewidth}
			\centering
			\includegraphics[width=\linewidth]{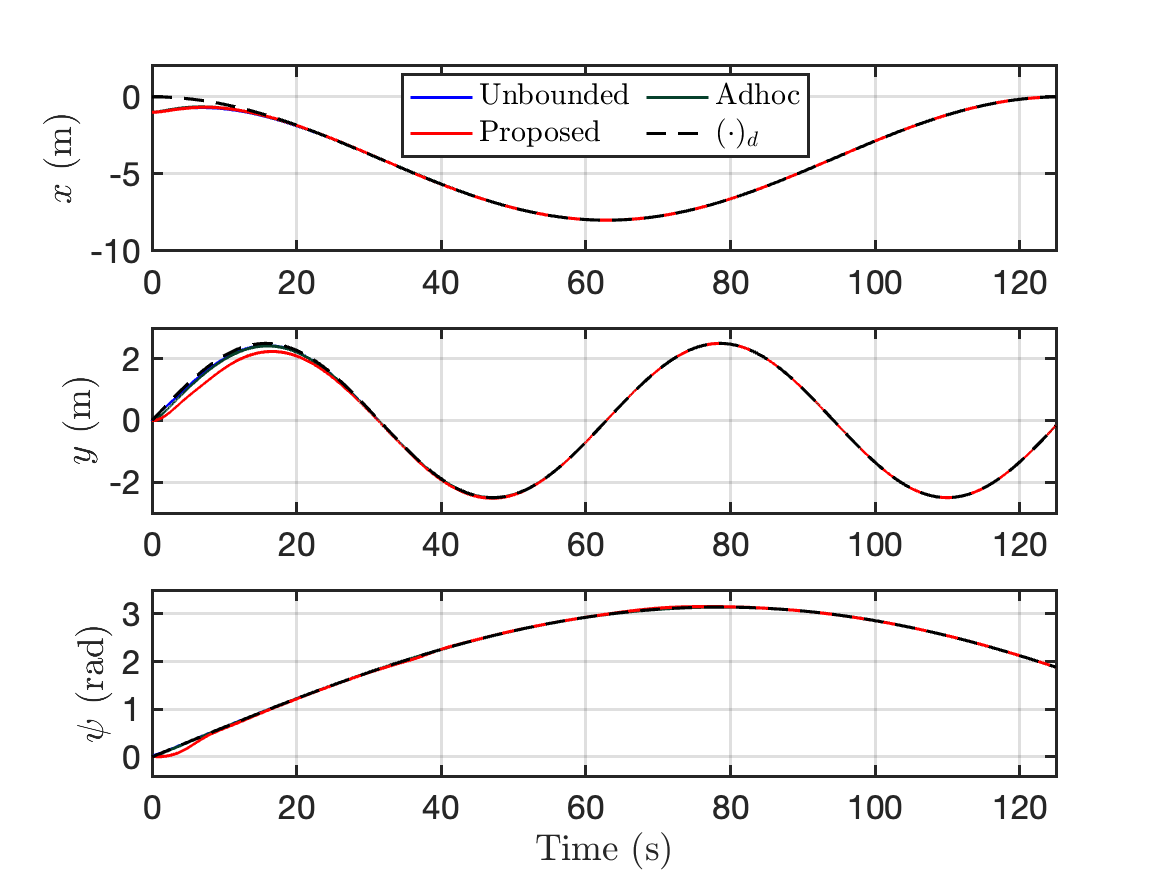}
			\caption{Position and heading of the vessel.}
			\label{figa8comp:eta}
		\end{subfigure}%
		\begin{subfigure}{0.45\linewidth}
			% \centering
			\includegraphics[width=\linewidth]{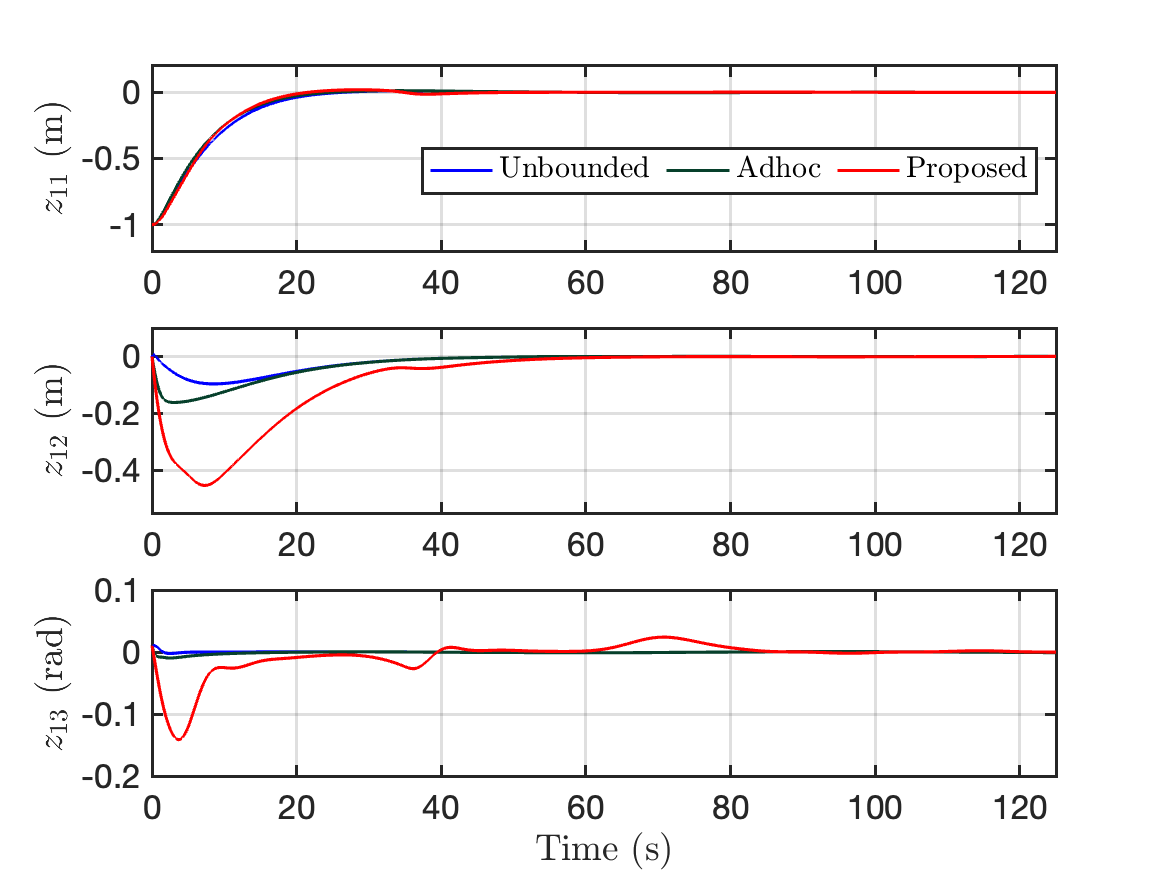}
			\caption{Error in position and heading.}
			\label{figa8comp:e1}
		\end{subfigure}
		\begin{subfigure}{0.45\linewidth}
			\centering
			\includegraphics[width=\linewidth]{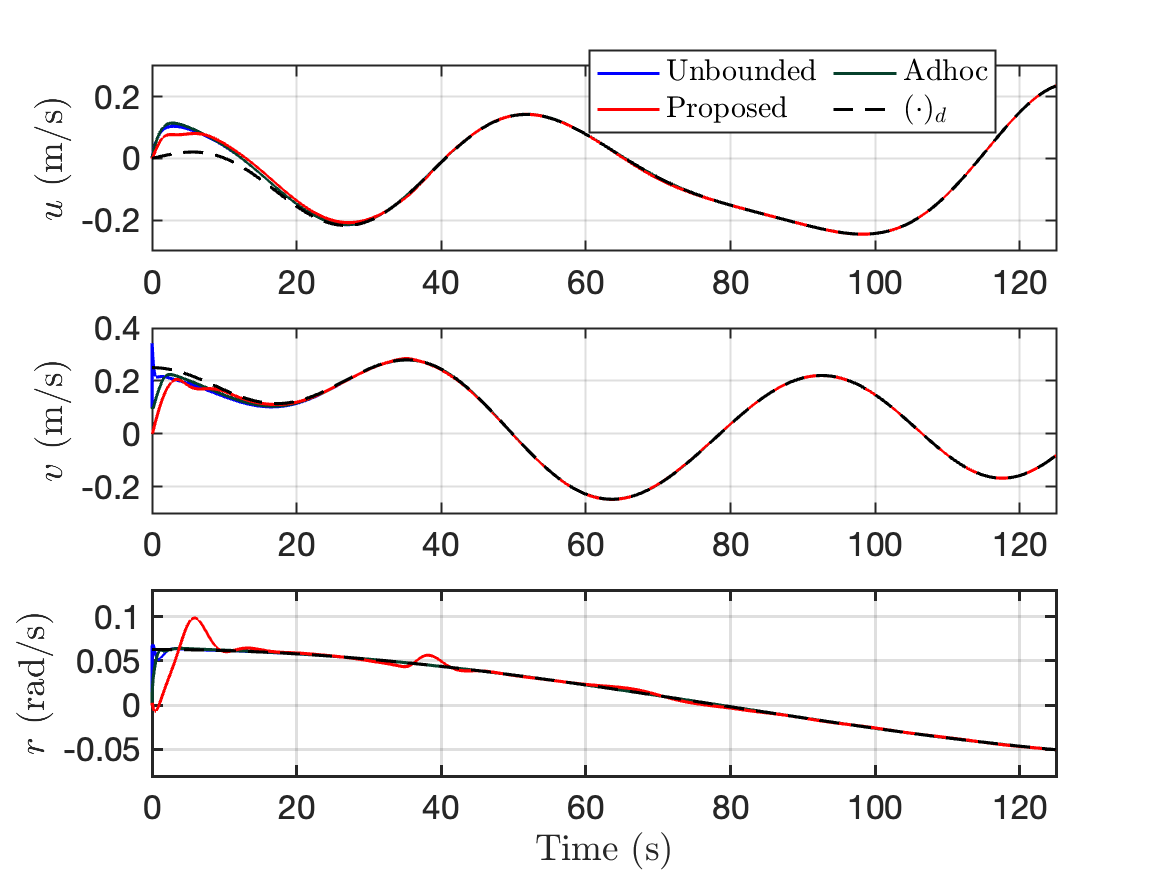}
			\caption{Surge, sway, and yaw rate.}
			\label{figa8comp:nu}
		\end{subfigure}%
		\begin{subfigure}{0.45\linewidth}
			\centering
			\includegraphics[width=\linewidth]{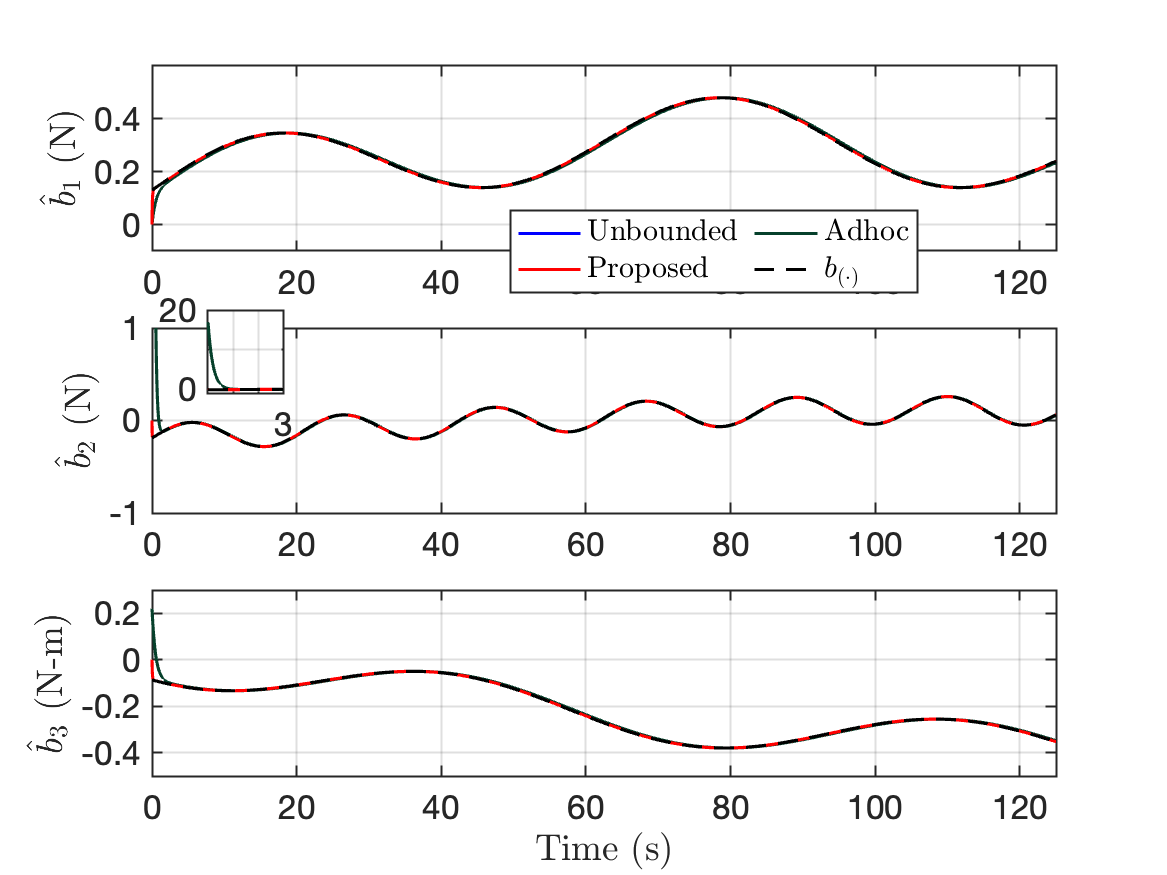}
			\caption{Estimation of disturbance vector.}
			\label{figa8comp:e2}
		\end{subfigure}
		\caption{Performance comparison of proposed controller against different methods for 8-shaped trajectory.}
		\label{figcase: a8comp}
	\end{figure*}
	\Cref{figcase: a8comp} shows the performance of the proposed controller against the commonly used methods in the literature. As seen from the \Cref{figa8comp:tau}, the control input demand goes manyfold higher than the available control input in the case of unbounded control design. And when ad-hoc input saturation is applied, then the tracking performance deteriorates, as evident from \Cref{figa8comp:eta}. It can be clearly seen that the USV is unable to track the desired heading. The same can also be inferred from the error plot shown in \Cref{figa8comp:e1}. The position and heading error go to zero quickly in case of unbounded input, but in reality, the actuator's capabilities are limited, questioning the practical applicability of that control law. However, the proposed control law does a better job of tracking while respecting the actuators' bounds, at the same time providing a mathematical guarantee for stability.
	%%==========input MRS==============================
	\subsection{Tracking with input magnitude and rate saturation constraint}
	In this subsection, we evaluate the tracking performance by the USV under the input magnitude as well as the input rate saturation constraint. The proposed control algorithm is given by \Cref{eqn:zeta_dot,eqn:tauc_dot,eqn:tau_d}, which is designed to satisfy the constraints on the input magnitude as well as its rate. Here also, we consider the same reference trajectories as in the previous subsections.
	\subsubsection{Tracking of an elliptical trajectory}
	The reference trajectory is described by the \Cref{eqn: ellipse}.
	\begin{figure*}[h!]
		\centering
		\begin{subfigure}{0.45\linewidth}
			\centering
			\includegraphics[width=\linewidth]{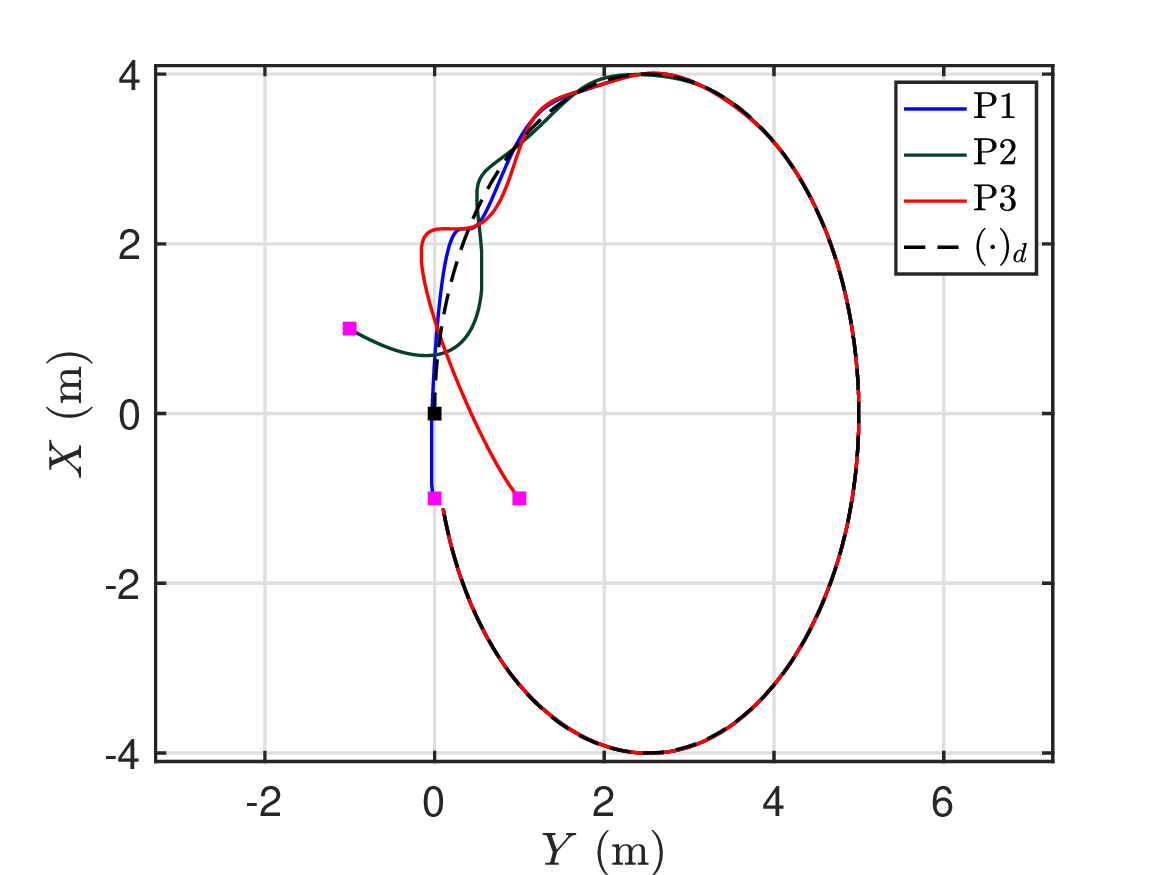}   
			\caption{Actual and desired trajectories.}
			\label{figMRSc:xy_plot}
		\end{subfigure}%
		\begin{subfigure}{0.45\linewidth}
			\centering
			\includegraphics[width=\linewidth]{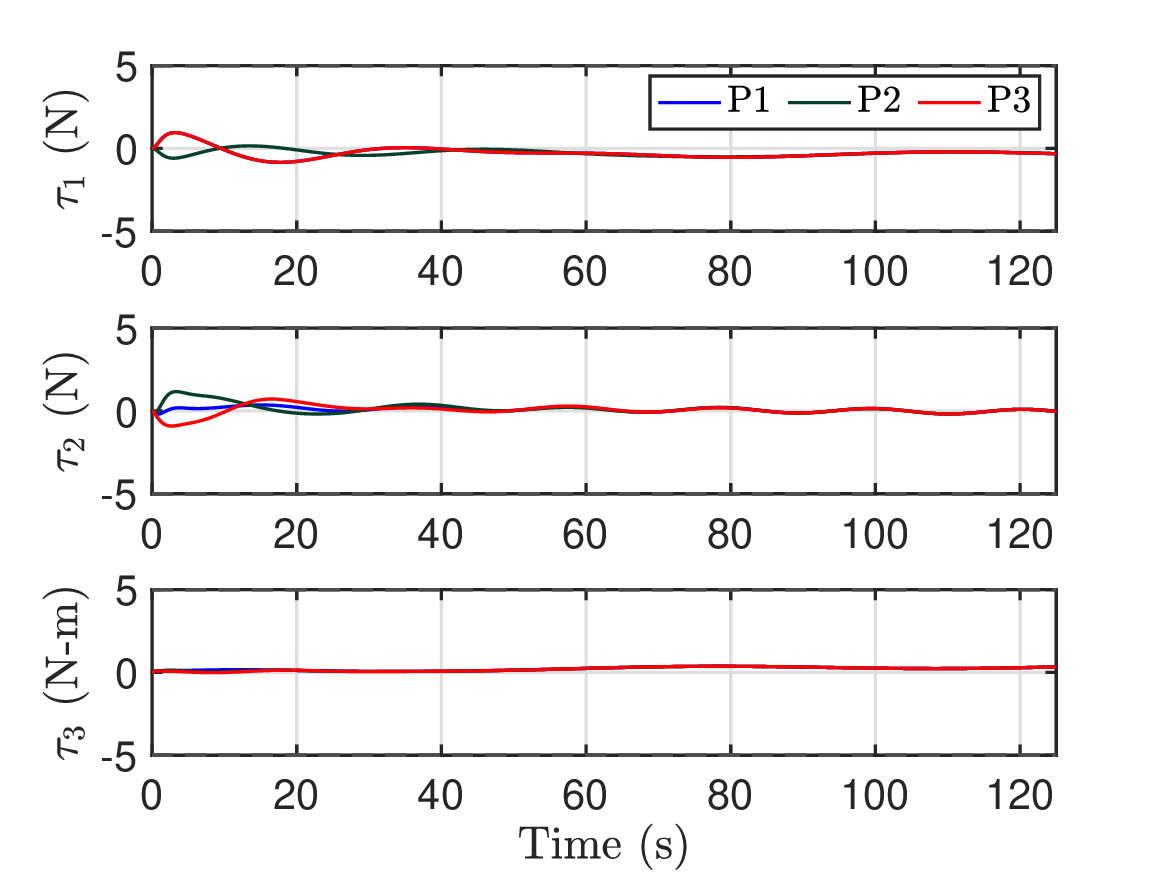}
			\caption{Control inputs.}
			\label{figMRSc:tau}
		\end{subfigure}
		\begin{subfigure}{0.45\linewidth}
			\centering
			\includegraphics[width=\linewidth]{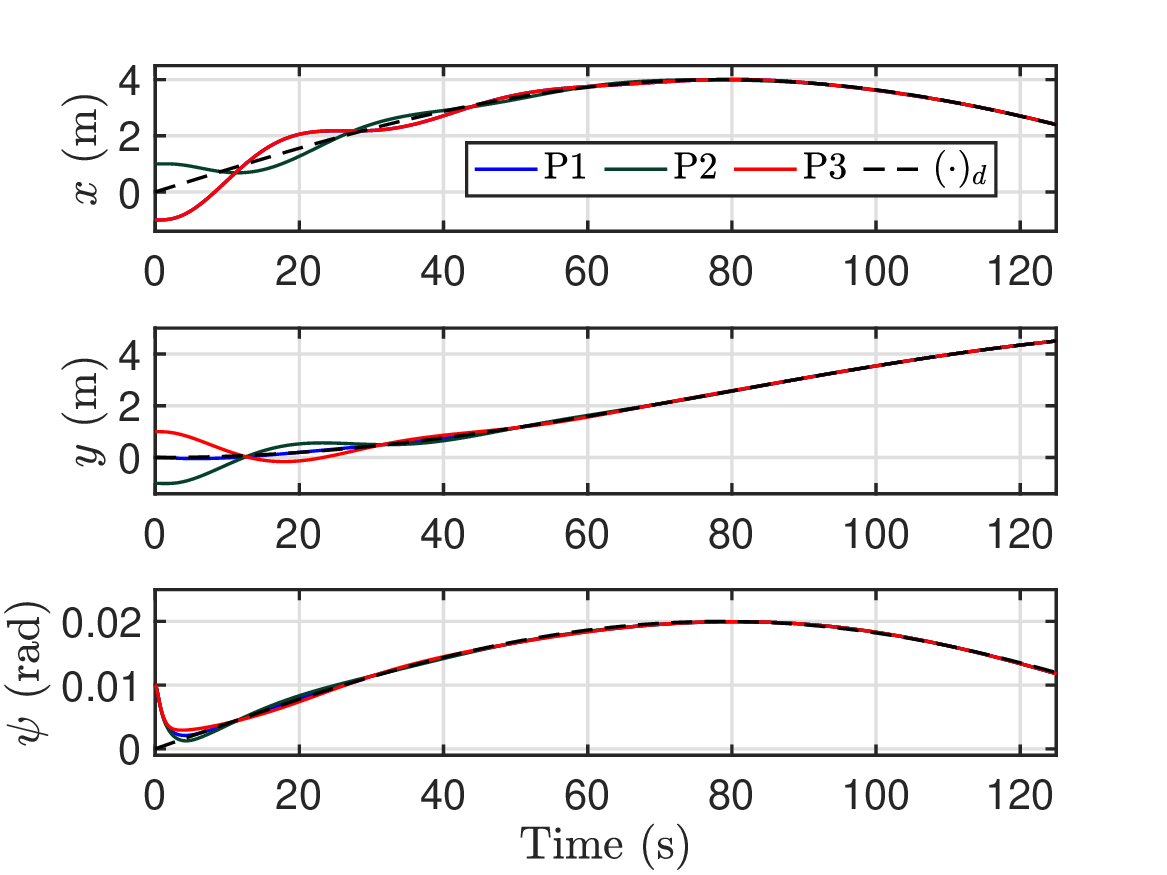}
			\caption{Position and heading of the vessel.}
			\label{figMRSc:eta}
		\end{subfigure}%
		\begin{subfigure}{0.45\linewidth}
			\centering
			\includegraphics[width=\linewidth]{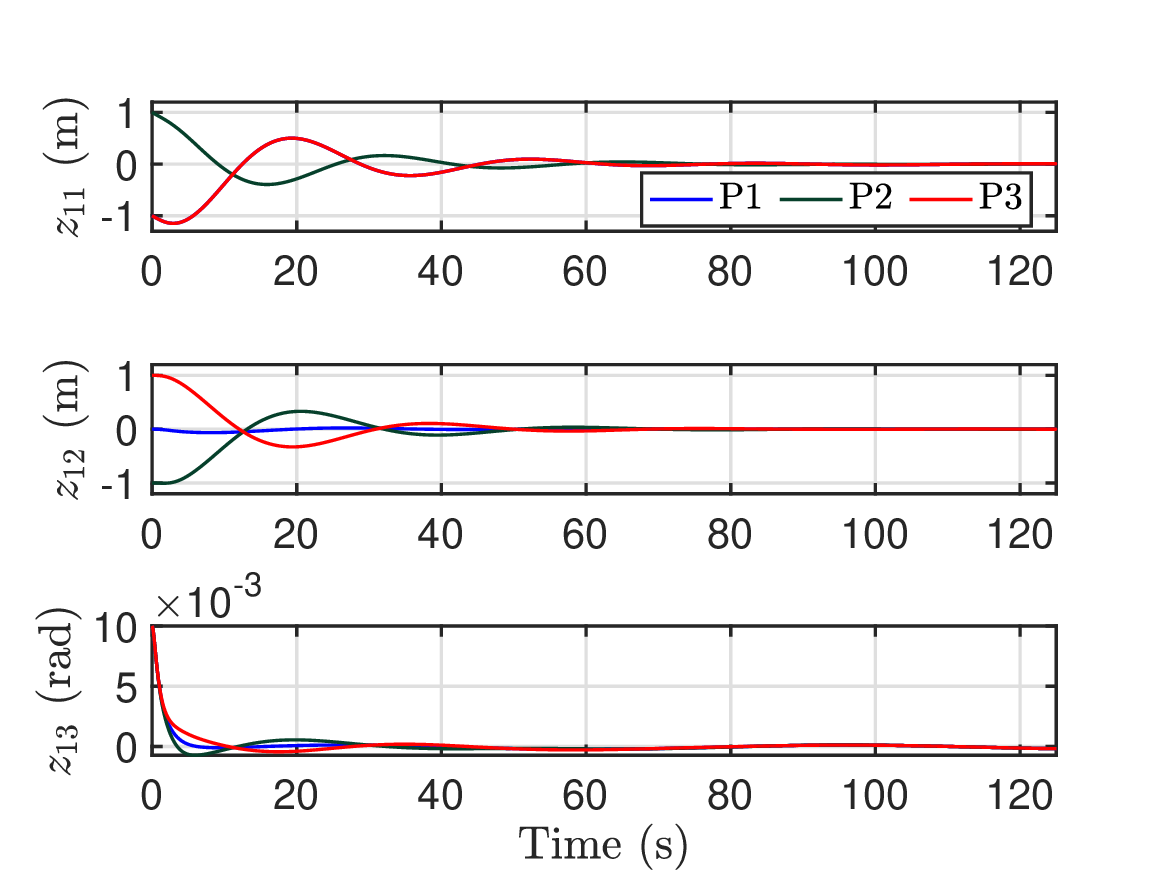}
			\caption{Errors in position and heading.}
			\label{figMRSc:e1}
		\end{subfigure}
		\begin{subfigure}{0.45\linewidth}
			\centering
			\includegraphics[width=\linewidth]{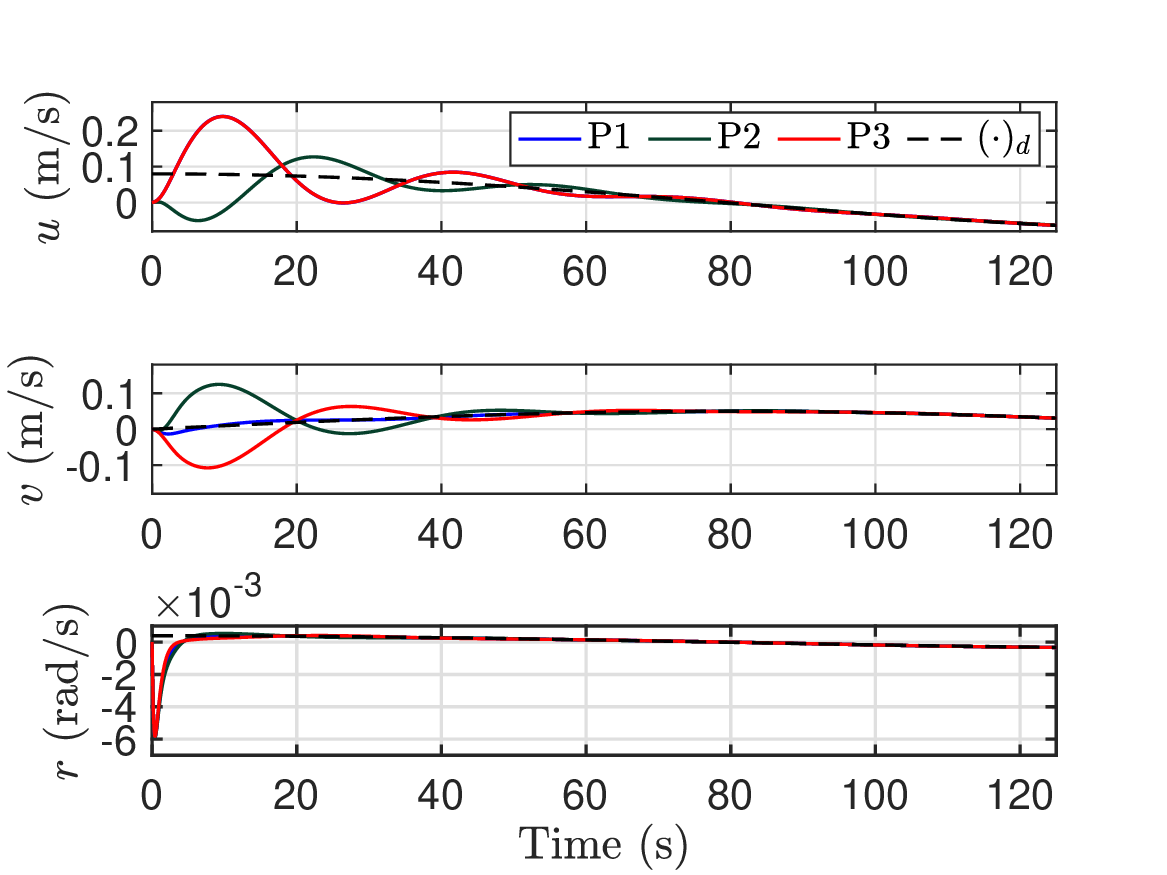}
			\caption{Surge, sway, and yaw rate.}
			\label{figMRSc:nu}
		\end{subfigure}%
		\begin{subfigure}{0.45\linewidth}
			\centering
			\includegraphics[width=\linewidth]{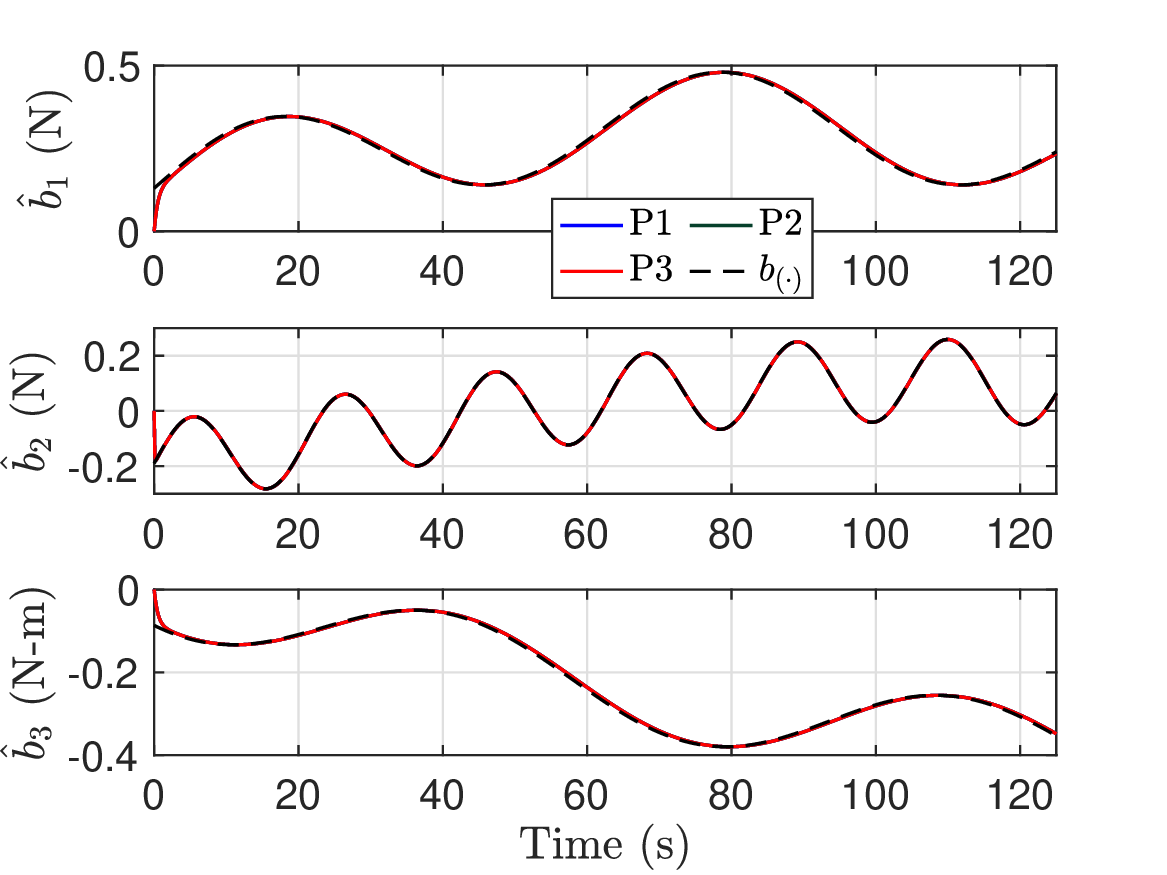}
			\caption{Estimation of disturbance vector.}
			\label{figMRSc:d}
		\end{subfigure}
		\caption{Performance validation of proposed controller for a USV under input magnitude and rate constraint, with elliptical trajectory.}
		\label{figcase: MRS_circle}
	\end{figure*}
	\cref{figcase: MRS_circle} illustrates the trajectory tracking performance while adhering to the constraints on the input and its rate. In \Cref{figMRSc:tau}, we can see that the control input never goes beyond the actuators' bounds, while the error in position and heading nears zero, as shown in \Cref{figMRSc:e1}. The disturbance vector is also being estimated as depicted in the \Cref{figMRSc:d}.
	\begin{figure}
		\centering
		\includegraphics[width=0.5\linewidth]{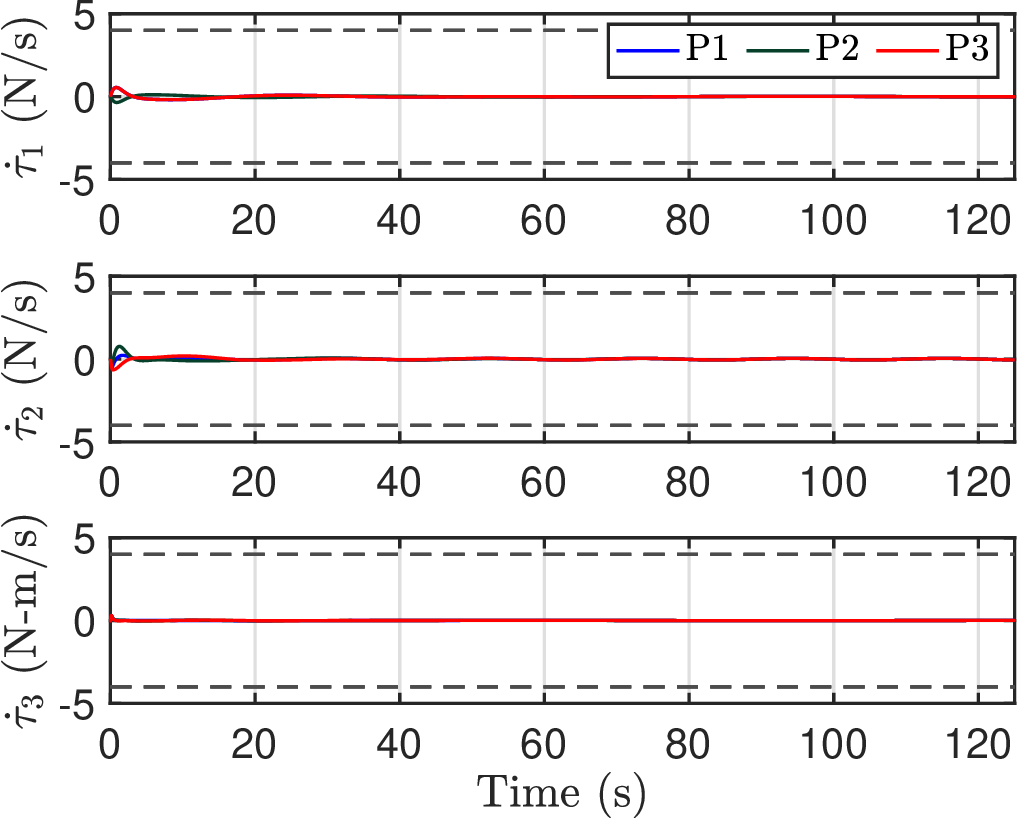}
		\caption{Demonstration of constraint satisfaction on the actuators' input rate for elliptical trajectory.}
		\label{figMRSc:tau_dot}
	\end{figure}
	Also, from \Cref{figMRSc:tau_dot}, it is evident the bounds on the input rate, that is, $\dot{\tau}_i ~(i \in \{1,2,3\})$, is always strictly respected and never crosses the bound $(-\tau_{dM}, \tau_{dM})$.
	\subsubsection{Tracking of an 8-shaped trajectory}
	To further attest to the performance of the proposed controller given by the \Cref{eqn:zeta_dot,eqn:tauc_dot,eqn:tau_d}, a more complex 8-shaped trajectory has been considered. The equation governing the 8-shaped trajectory is given by \Cref{eqn: 8-shape}. The USV is made to start from three different positions, namely P1, P2, and P3, with an initial heading as $0.01$ rad. The different initial positions are given by $(-1,0)$, $(1,-1)$ and $(-1,1)$. 
	\begin{figure*}
		\centering
		\begin{subfigure}{0.45\linewidth}
			\centering
			\includegraphics[width=\linewidth]{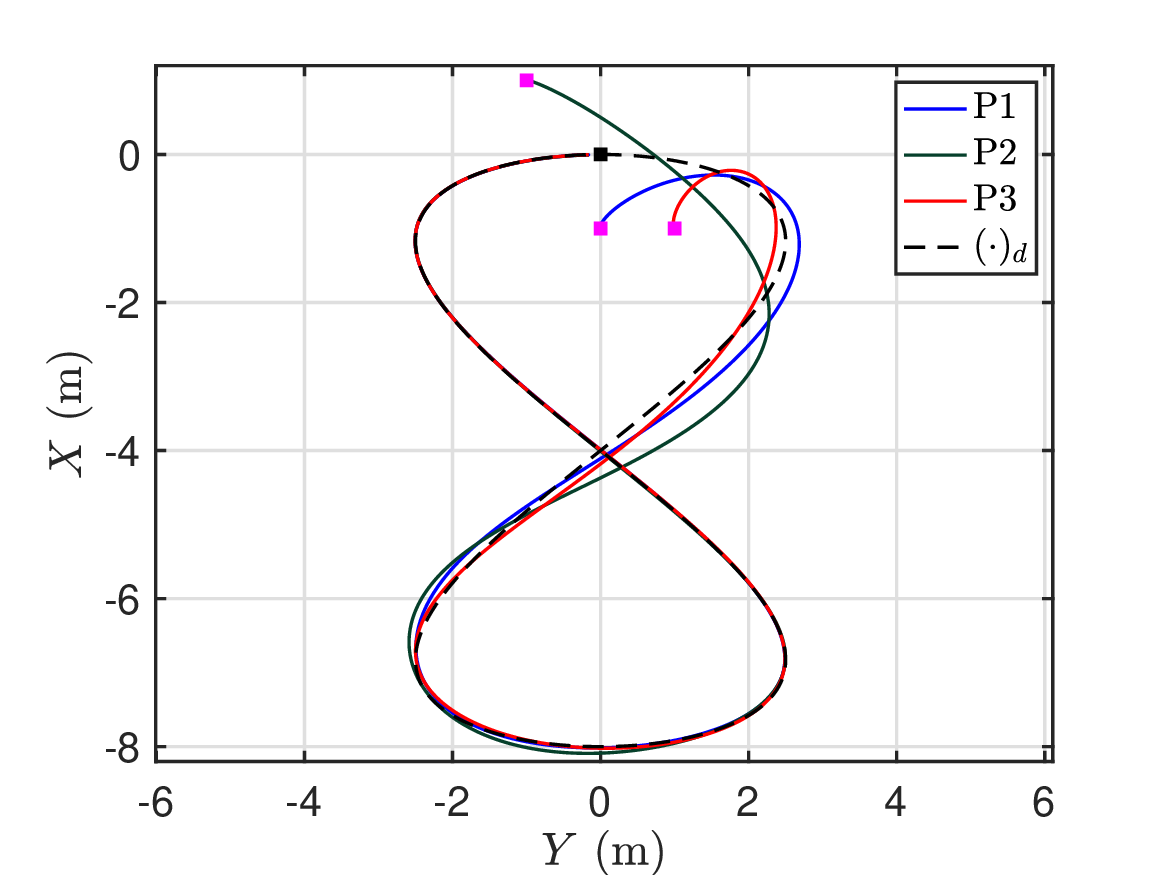}   
			\caption{Actual and desired trajectories.}
			\label{figMRS_8:xy_plot}
		\end{subfigure}%
		\begin{subfigure}{0.45\linewidth}
			% \centering
			\includegraphics[width=\linewidth]{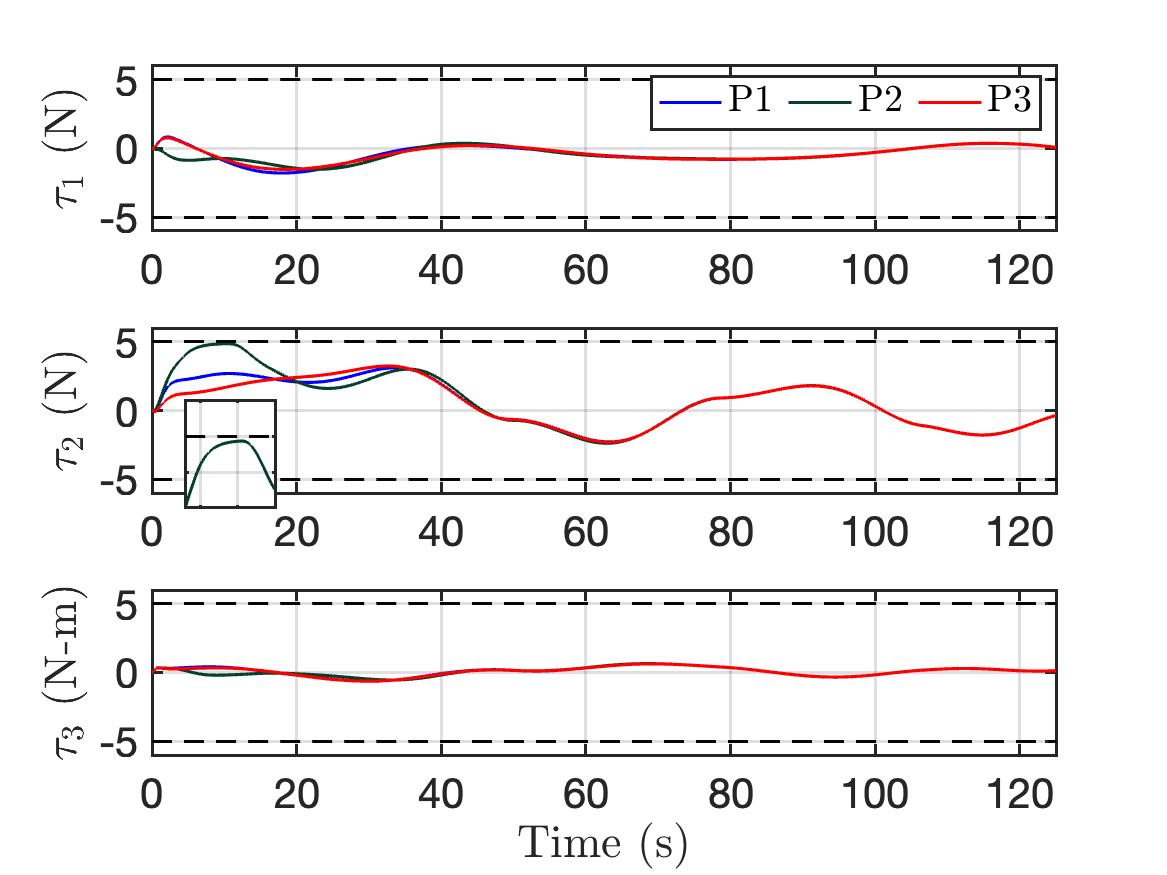}
			\caption{Control inputs.}
			\label{figMRS_8:tau}
		\end{subfigure}
		\begin{subfigure}{0.45\linewidth}
			\centering
			\includegraphics[width=\linewidth]{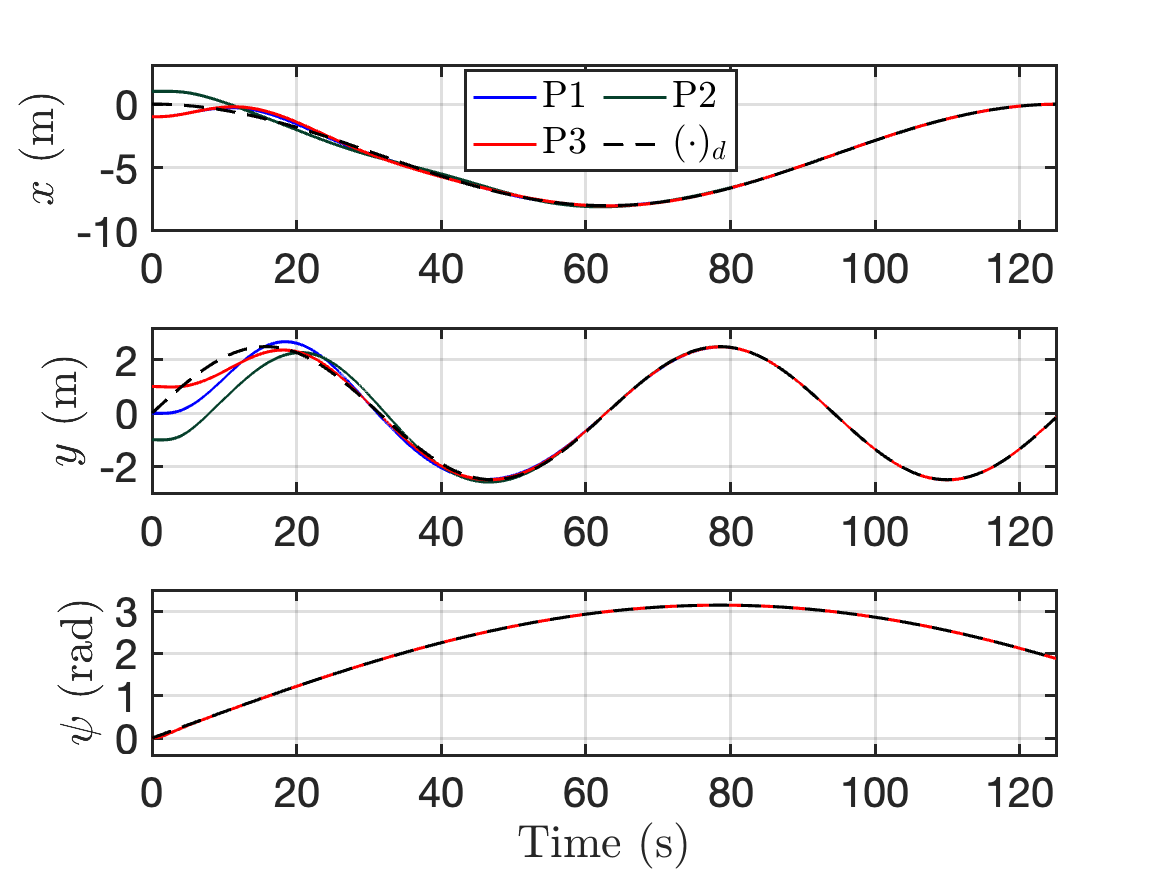}
			\caption{Position and heading of the vessel.}
			\label{figMRS_8:eta}
		\end{subfigure}%
		\begin{subfigure}{0.45\linewidth}
			% \centering
			\includegraphics[width=\linewidth]{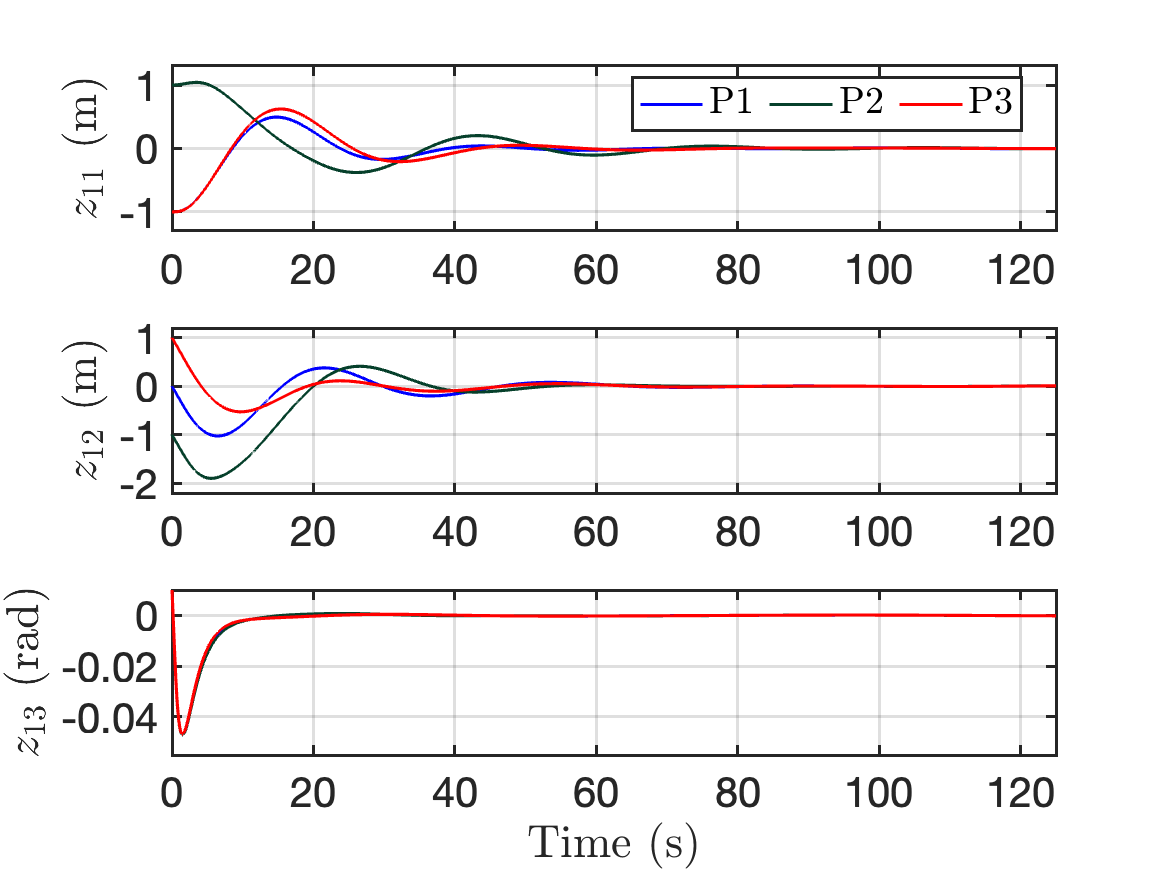}
			\caption{Error in position and heading.}
			\label{figMRS_8:e1}
		\end{subfigure}
		\begin{subfigure}{0.45\linewidth}
			\centering
			\includegraphics[width=\linewidth]{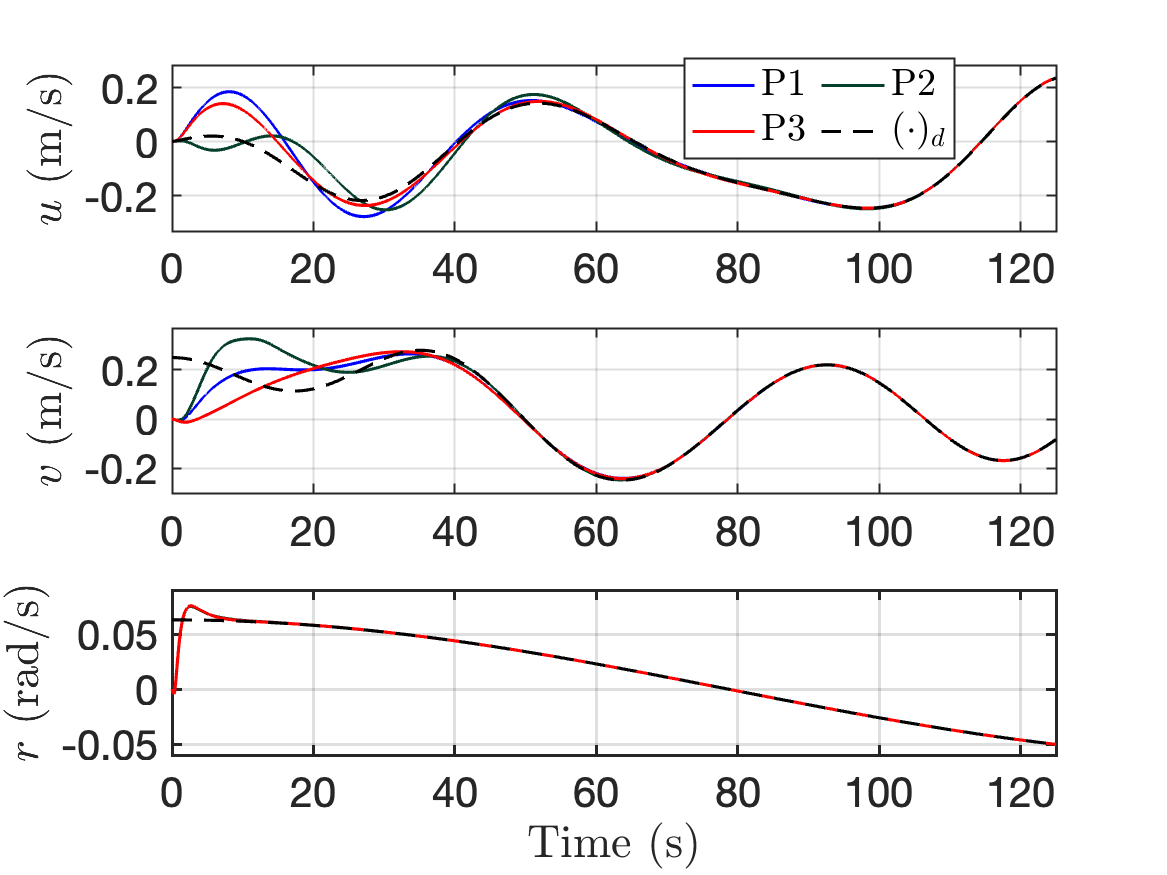}
			\caption{Surge, sway, and yaw rate.}
			\label{figMRS_8:nu}
		\end{subfigure}%
		\begin{subfigure}{0.45\linewidth}
			\centering
			\includegraphics[width=\linewidth]{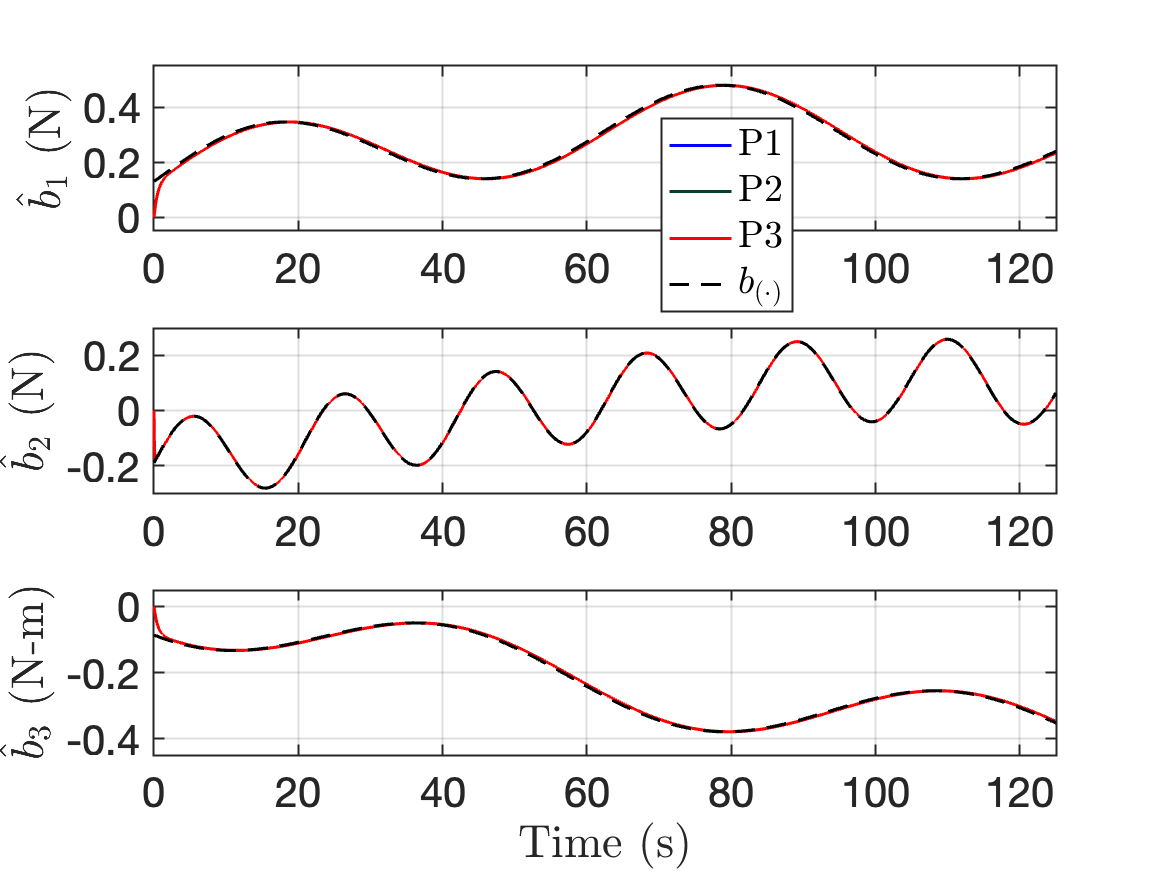}
			\caption{Estimation of disturbance vector.}
			\label{figMRS_8:d}
		\end{subfigure}
		\caption{Performance validation of proposed controller for a USV under input magnitude and rate constraint, with 8-shaped trajectory.}
		\label{figcase: MRS_8}
	\end{figure*}
	\cref{figcase: MRS_8} illustrates the tracking performance of the proposed algorithm while respecting the bounds on actuators' input magnitude as well as their rates. As seen from the \Cref{figMRS_8:xy_plot}, the USV follows the desired trajectory. The error in position and heading, along with their rates, can be seen going to zero in \Cref{figMRS_8:e1,figMRS_8:eta,figMRS_8:nu}. 
	\begin{figure}
		\centering
		\includegraphics[width=0.5\linewidth]{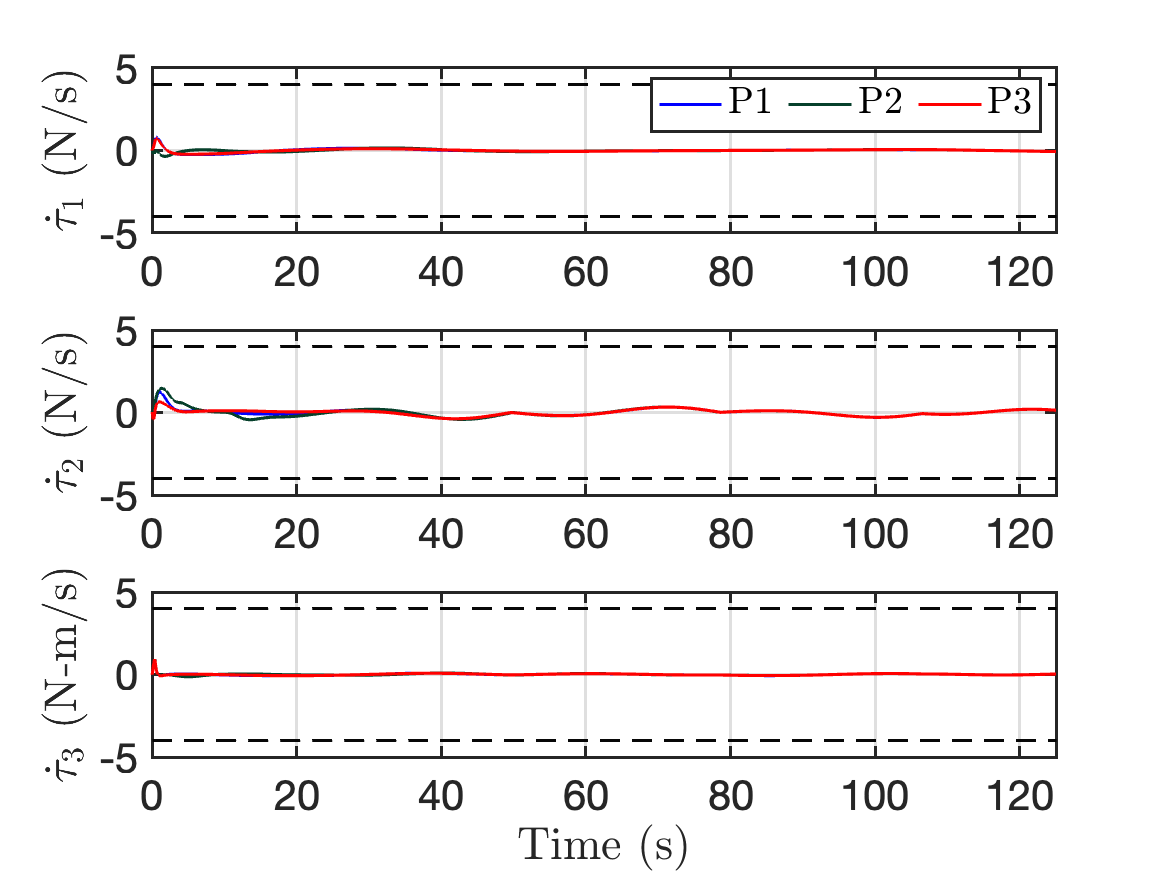}
		\caption{Demonstration of constraint satisfaction on the actuators' input rate for 8-shaped trajectory.}
		\label{fig:tau_dot_8}
	\end{figure}
	\cref{fig:tau_dot_8} shows that the rate of input magnitude always remains within the specified bounds.
	\subsubsection{Comparison across different methods for 8-shaped trajectory}
	In this section, we present the comparison of the performance of the proposed algorithm, which takes into account the actuator constraint on the input and their rate, with two other methods. The most commonly employed method is to design the control law ignoring the actuator bounds and then just clip the actuator input to its maximum or minimum value when the demand goes beyond the actuator capabilities. And the other method considers no bound on the actuator. The results from these three methods are denoted as ``Proposed'', ``Adhoc'', and ``Unbounded'', respectively, in the figures.
	\begin{figure*}
		\centering
		\begin{subfigure}{0.45\linewidth}
			\centering
			\includegraphics[width=\linewidth]{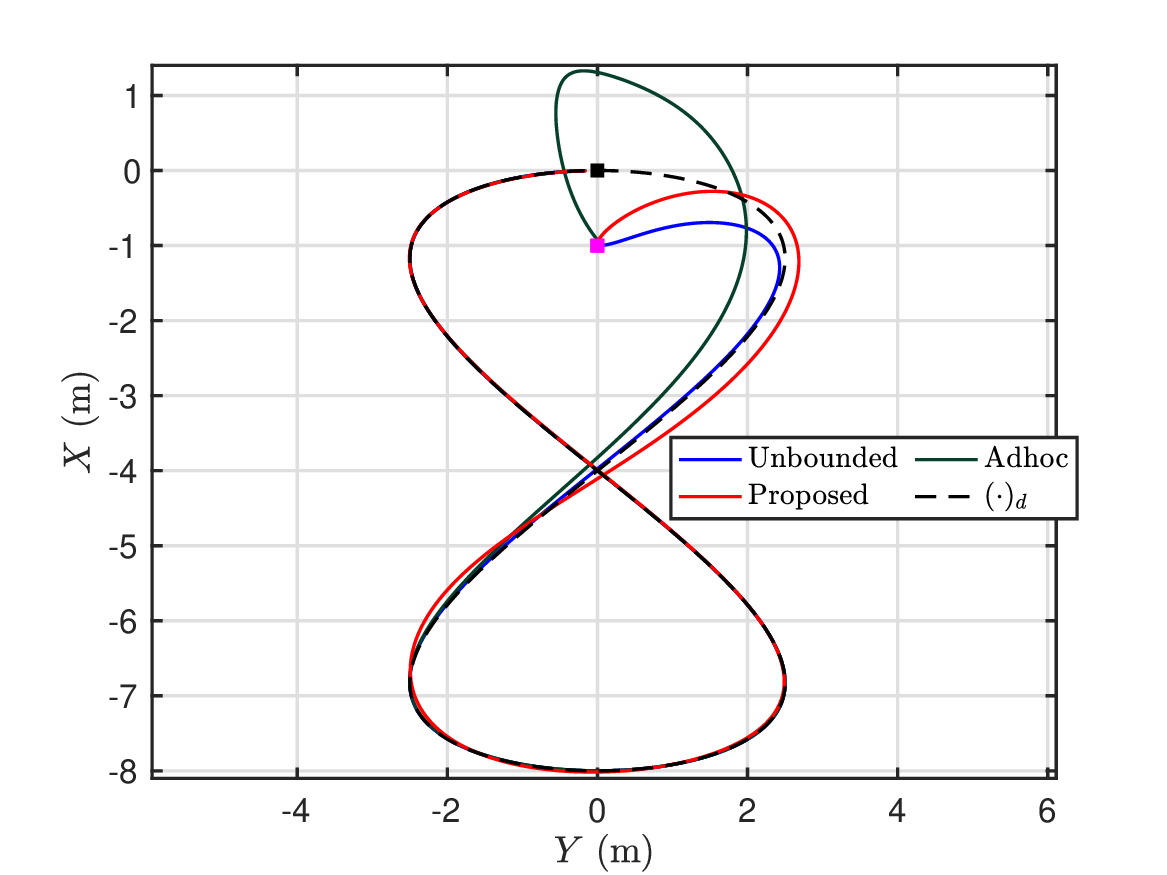}   
			\caption{Actual and desired trajectories.}
			\label{figa8comp_MRS:xy_plot}
		\end{subfigure}%
		\begin{subfigure}{0.45\linewidth}
			% \centering
			\includegraphics[width=\linewidth]{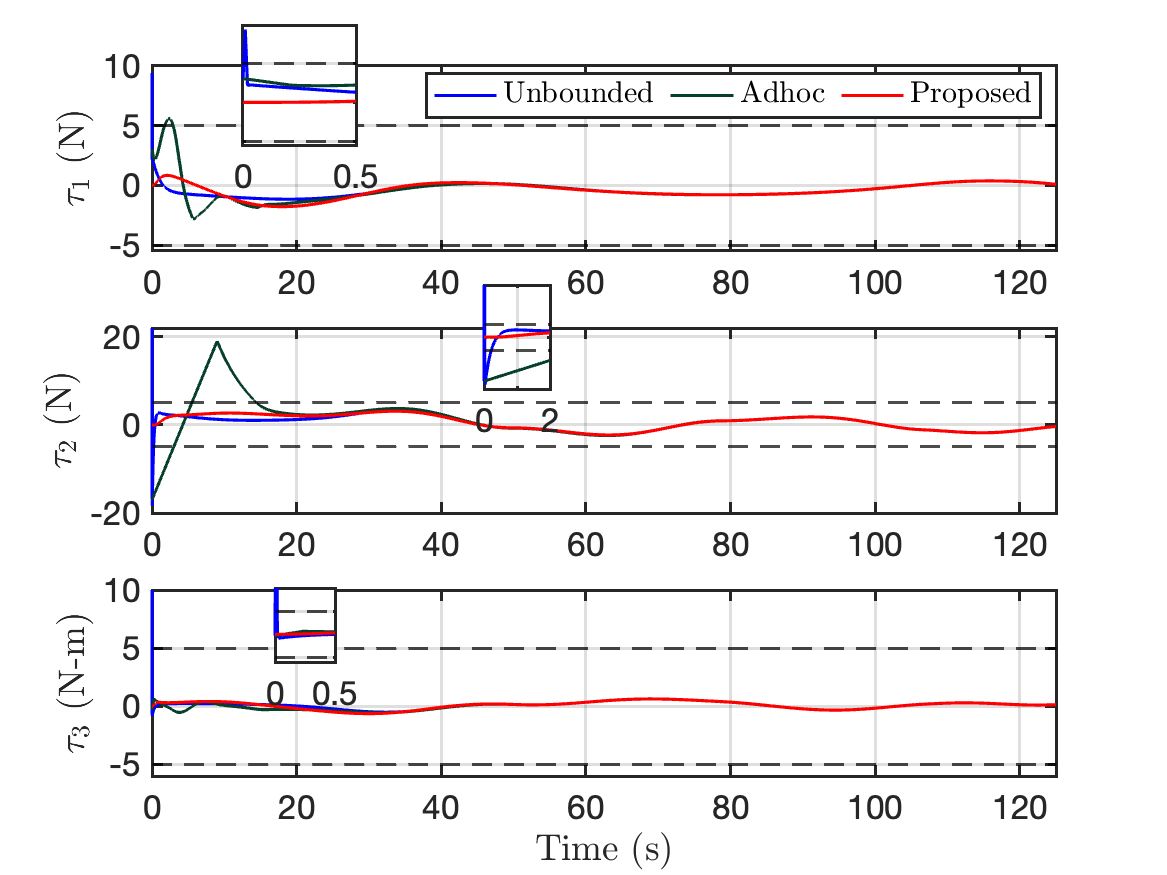}
			\caption{Control inputs.}
			\label{figa8comp_MRS:tau}
		\end{subfigure}
		\begin{subfigure}{0.45\linewidth}
			\centering
			\includegraphics[width=\linewidth]{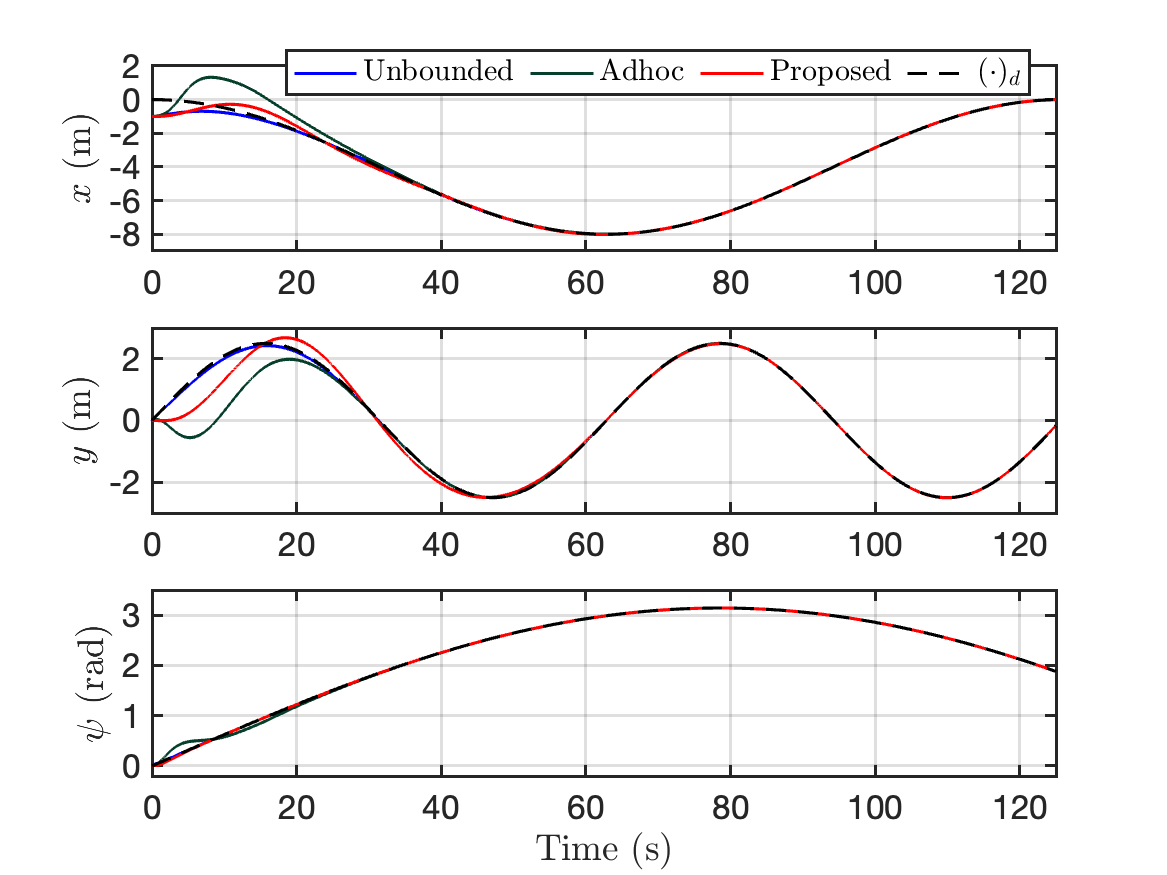}
			\caption{Position and heading of the vessel.}
			\label{figa8comp_MRS:eta}
		\end{subfigure}%
		\begin{subfigure}{0.45\linewidth}
			% \centering
			\includegraphics[width=\linewidth]{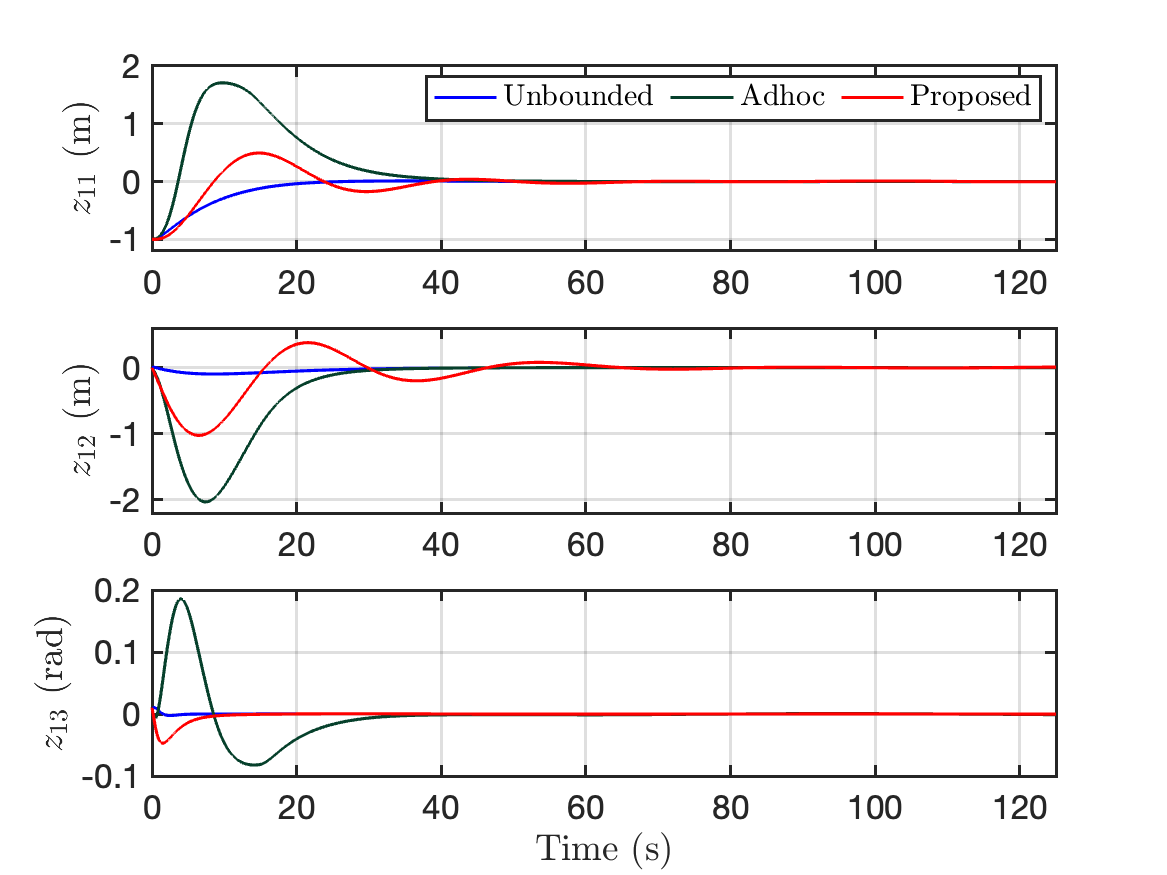}
			\caption{Error in position and heading.}
			\label{figa8comp_MRS:e1}
		\end{subfigure}
		\begin{subfigure}{0.45\linewidth}
			\centering
			\includegraphics[width=\linewidth]{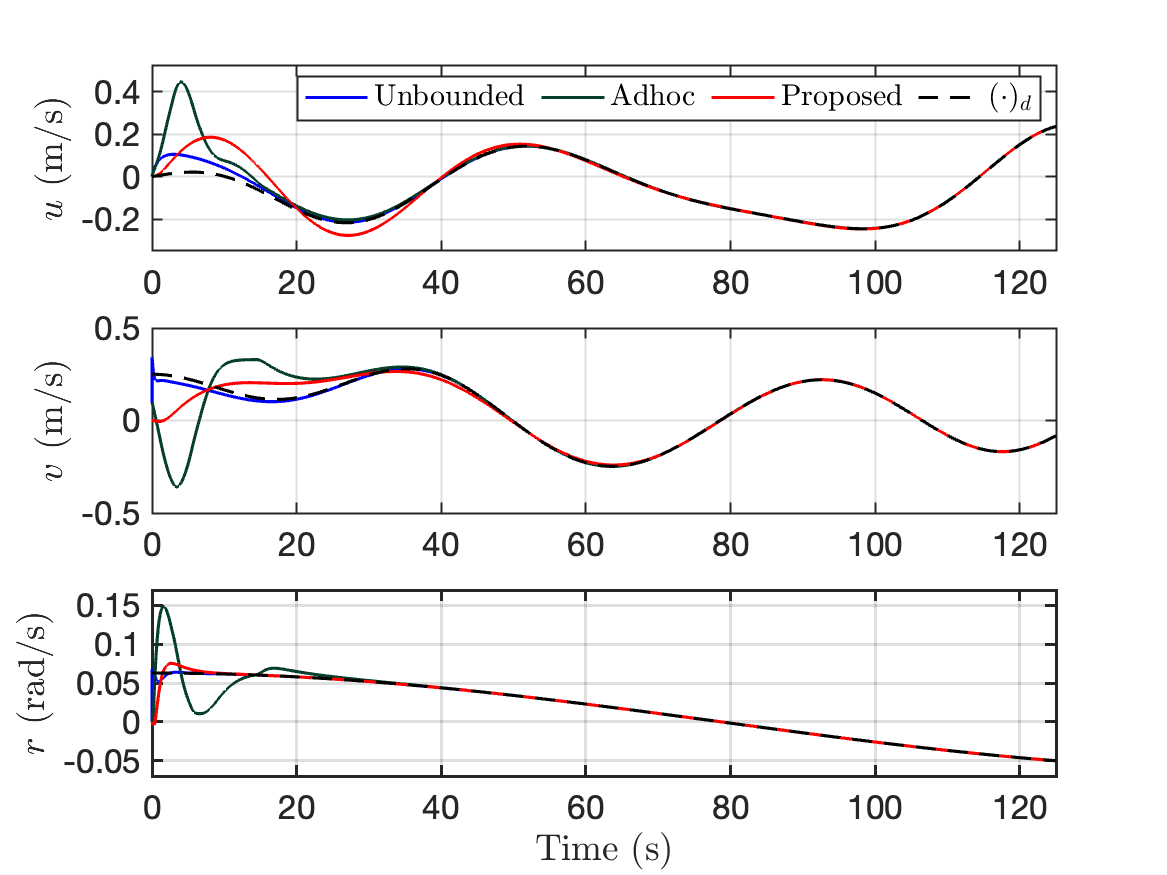}
			\caption{Surge, sway, and yaw rate.}
			\label{figa8comp_MRS:nu}
		\end{subfigure}%
		\begin{subfigure}{0.45\linewidth}
			\centering
			\includegraphics[width=\linewidth]{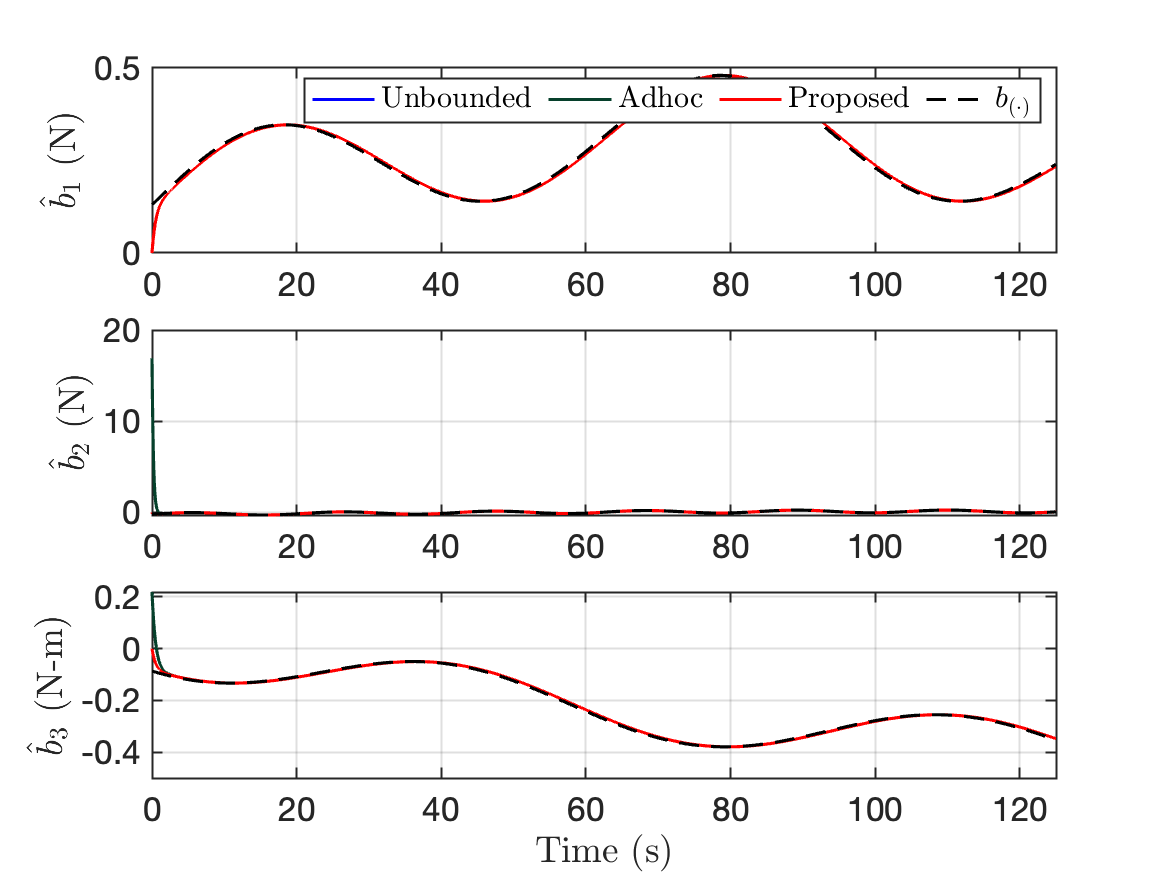}
			\caption{Estimation of disturbance vector.}
			\label{figa8comp_MRS:d}
		\end{subfigure}
		\caption{Performance validation of proposed controller against different methods for 8-shaped trajectory.}
		\label{figcase: a8comp_MRS}
	\end{figure*}
	\begin{figure}
		\centering
		\includegraphics[width=0.5\linewidth]{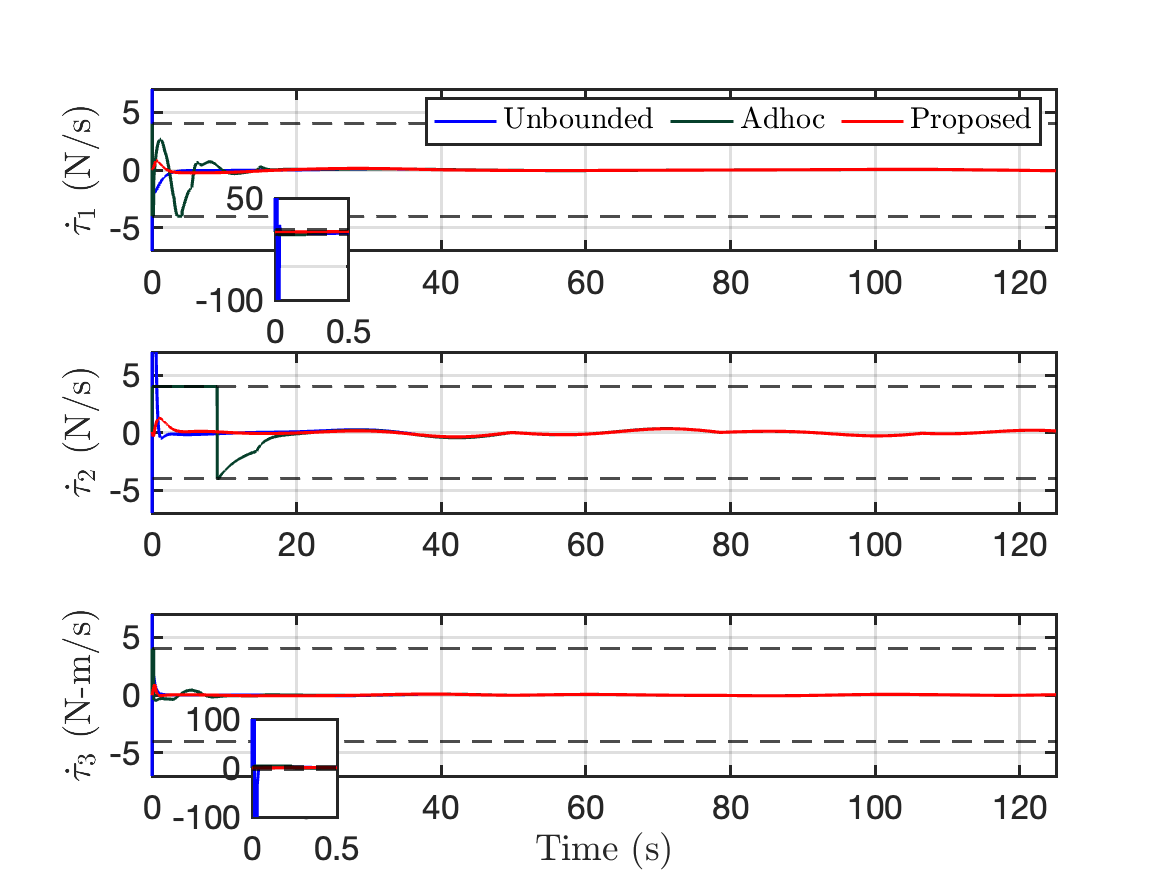}
		\caption{Comparison of input magnitude rate demand for different methods.}
		\label{figa8comp_MRS:tau_dot}
	\end{figure}
	\cref{figcase: a8comp_MRS} shows the comparison of the trajectory tracking algorithm. As seen from the \Cref{figa8comp_MRS:xy_plot}, the USV tracks the desired trajectory under all three methods. However,  the vehicle has to take a longer path when the actuator constraints are imposed in an ad-hoc manner than under the proposed method. \cref{figa8comp_MRS:tau} shows that the control input demand gets saturated in the case of the ad-hoc method, and in the case of the unbounded method, an unrealistic control demand is being made. From \Cref{figa8comp_MRS:eta,figa8comp_MRS:e1}, it can be seen that the error in position and heading approaches zero. Along with this, we can also verify from \Cref{figa8comp_MRS:nu}, that the surge, sway and yaw rate also tracks the desired value. Additionally, the estimates of the disturbance vector are depicted in the \Cref{figa8comp_MRS:d}. \cref{figa8comp_MRS:tau_dot} depicts that the rate of input magnitude constraints is satisfied while tracking the desired trajectory.
	\section{Conclusions}\label{sec:conclusion}
	This work proposed effective nonlinear control strategies for the trajectory tracking of marine surface vessels while accounting for the actuator constraints on the input magnitude and its rate. Unlike the common static nonlinear or hyperbolic tangent functions, this design employed a smooth input saturation model that is affine in the model input. This affine nature allowed seamless integration with the vessel's motion equations during controller design. Following an ad-hoc way of dealing with actuator constraints does not guarantee stability and also deteriorates the tracking performance, which can be seen in the simulation results.
	The system's stability with the proposed controller was established using the Lyapunov-based approach. As demonstrated via numerical simulations, the USV achieves the desired trajectory tracking even in the presence of unknown time-varying disturbances with the proposed controller, while the actuator input constraints are satisfied. Future research would explore parameter uncertainty in the model and control of underactuated vessels to enhance the model's applicability further. 
	\bibliography{references.bib}
\end{document}